\documentclass[a4paper, 12pt]{report}
\usepackage[english]{babel}
\usepackage{graphicx}
\usepackage{subfig}
\usepackage{qcircuit}
\usepackage{amsthm}
\usepackage{amsmath}
\usepackage{amssymb}
\usepackage{float}
\usepackage[font=small,labelfont=bf,indention=2em,belowskip=-8pt,aboveskip=4pt]{caption}
\usepackage{adjustbox}
\usepackage{multirow}
\usepackage{xcolor}
\usepackage{listings}
\usepackage{csquotes}
\usepackage[toc,page]{appendix}
\usepackage{hyperref}
\usepackage[backend=biber]{biblatex}
\DeclareMathOperator*{\argmax}{arg\,max}

\title{Towards the implementation of a quantum classifier}
\author{Lorenzo Confalonieri}
\date{}

\begin{document}

\begin{titlepage}
    \begin{figure}
        \centering
        \includegraphics[scale=0.93]{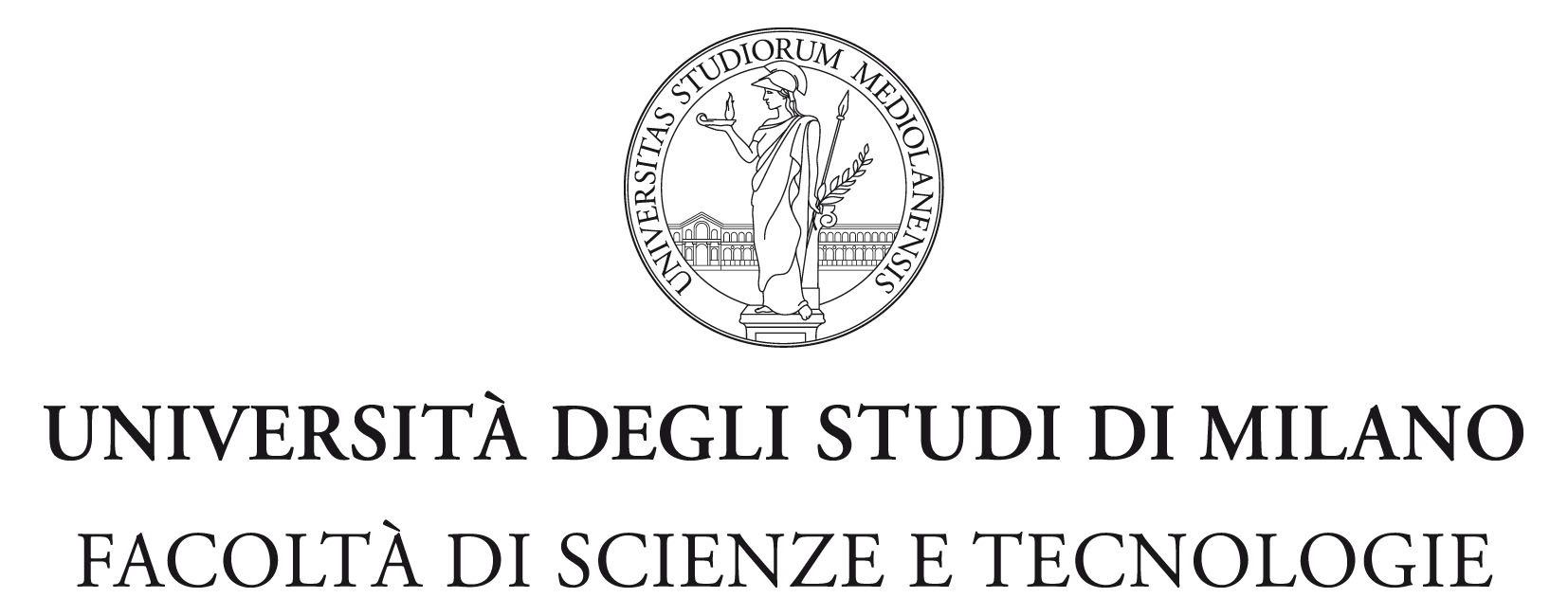}
    \end{figure}
    \begin{center}
       {\LARGE Corso di Laurea Triennale in Fisica}
    \end{center}
    \bigskip \bigskip \bigskip
    \begin{center}
        {\LARGE \textbf{Towards the implementation of a quantum classifier}}
    \end{center}
    \bigskip
    \begin{center}
        {\Large Lorenzo Confalonieri \\
        \bigskip
        \Large Relatore: Dr. Stefano Carrazza \\
        \bigskip
        \Large Correlatore: Adrián Pérez Salinas}
    \end{center}
    \vfill
    \begin{center}
        {\large Anno Accademico 2020/2021}
    \end{center}
\end{titlepage}

\begin{center}
    \LARGE
    \textbf{Acknowledgments}
\end{center}
\bigskip
Before presenting this thesis, I would like to focus on the people who helped bring these pages to life. 
\newline
I am extremely thankful to my thesis supervisor Dr. Stefano Carrazza, for his valuable suggestions and motivating guidance. He taught me to reflect on the right questions and encouraged me to discover the answers. I will always be grateful for this. 
\newline
I thank also my co-supervisor Adrián Pérez Salinas, for the precious feedbacks and the help in writing the thesis.
\newline
I want to mention also my colleague Nicole Zattarin, with whom I shared a part of this project.
\newline
I want to thank my classmates with whom I spent the last three years of university. In particular, I would like to mention Laura, Andrea, Michele, and Lorenzo; they were always there for me.
\newline
I would like to extend my sincere thanks to all my friends since high school. Especially I would like to thank father Luca, Andrea, Daniele, Giorgio and Alessandro.
\newline
My warm and heartfelt thanks to my family for teaching me strength, passion, devotion and the importance of taking a break now and then.
\newline
I would like to offer my special thanks to my sister Martina for helping me to write the thesis in English, and to my grandparents Irene, Mariadele, Aurelio and Celestino.
\newline
Last, but not least, my parents Marilena and Roberto, deserve endless gratitude for always supporting me. All my accomplishments are only because they always believed in me.

\tableofcontents

\chapter*{Introduction}
\label{chap:introduction}
The learning process is based on \textit{experience}.
Our brain can distinguish between multiple categories thanks to the big amount of information imported through the five senses: taste, sight, touch, smell, and sound.
However, our brain is not perfect; how can we create a mathematical model that can classify things better than our brain?

In the last decades, researchers studied what is known as \textit{artificial intelligence}. 
Due to the fast increase in speed of CPUs, the electronic circuits that execute instructions in a classical computer,  and its number of cores, in the last years a part of artificial intelligence called machine learning, has spread in a lot of different fields.
For ``machine learning" we refer to a model that can learn without being programmed for that specific situation.
For example, in medical physics have been studied applications of machine learning in radiology and radiotherapy \cite{https://doi.org/10.1002/mp.14088}.
Applications outside the world of physics are, for example, the prediction of stock prices \cite{ChowdhuryReaz2020Ptsp} and the study of engagement during a game-based task or non-game-based task with facial emotion detection \cite{NinausManuel2019Ieei}.

Machine learning can be defined into three main categories.

\begin{enumerate}
    \item \textit{Supervised learning.} The model learns from a given data set called training data set where data is labeled. Regression and classification problems belong to this section.
    \item \textit{Unsupervised learning.} The model tries to understand an information from the given data without a given label. In this category, there are, for example, clustering problems, density estimation, and autoencoders.
    \item \textit{Reinforcement learning.} The model learns how to react to the environment through penalties and rewards. Automatic driving and gaming bots belong to this category. This is the most advanced and complex model that can be implemented. 
\end{enumerate}

In this work, we focus on supervised learning, in particular, on the issue of binary classification using a new paradigm, \textit{Quantum Computing}.

The aim of this thesis is to show that is possible to build a binary classifier with a quantum computer. We will call this model, binary quantum classifier.
In order to achieve our goal, we have implemented a library with the support of Qibo \cite{qibo}, an open-source framework for quantum computing, simulation, and hardware control.

We test our quantum classifier on two data sets.
The first one is a reduced version of the MNIST data set where we take only handwritten images of zeros and ones.
The second one represents two categories of particle physics processes that can occur at high energy colliders, such as LHC at CERN \cite{ButterworthJonathanM2008Jsaa}.

We show that our classifier learns from these data sets and we highlight its performances and limitations in comparison with classical technologies.

We start by describing the fundamentals of quantum circuits: qubits and gates.
Then we discuss how to design a quantum circuit for the classification of variables and images.
Finally, we show and discuss the results obtained from the two data sets previously described and we compare metrics with the ones obtained from neural networks.

\chapter{Quantum computing}
\label{chap:qc}

In this chapter we are going to define the fundamentals subjects of quantum computing: qubits and gates and then we
explain what is quantum computing today and why researchers are studying this new approach to information.

\section{Qubits and gates}
A qubit, or quantum bit, is the basic unit of quantum information as a bit is for a classical computer. Qubits live in a Hilbert vector space with basis $|0\rangle, |1\rangle$. The general form is the following:
\begin{equation}
    |\psi\rangle = a|0\rangle + \sqrt{1-|a|^2} e^{i\phi}|1\rangle ,
\end{equation}
where $a$ and $\phi$ are real values.

When we measure the qubit we find the state $|0\rangle$ with probability $|a|^2$ and the state $|1\rangle$ with probability $1-|a|^2$.

If we rewrite the qubit as $|\psi\rangle = \alpha |0\rangle + \beta |1\rangle$, where $\alpha$ and $\beta$ are complex numbers,  we can use a very useful vector representation:
\begin{equation}
    |\psi\rangle = 
    \begin{pmatrix}
        \alpha  \\
        \beta
    \end{pmatrix}.
\end{equation}

A qubit in a quantum computer can be realized in many different ways: by a single atom, a single electron, a single photon, or a cold superconducting electrical circuit.

In a classical computer, the most basic unit of information is a bit. A bit is a logic state with only two possible values: true/false, 1/0, yes/no, on/off.
As in classical computing we define logic gates that are applied to bits, in quantum computing we define quantum gates that are applied to qubits.
A quantum gate is an operator that can be applied to a single qubit or multiple qubits.
A set of qubits and gates define a computation model known as, quantum circuit.
This kind of approach to quantum computing has been developed by Google \cite{google}, IBM \cite{ibm}, Rigetti \cite{rigetti} and Intel \cite{intel}.

We can represent a single-qubit gate in the basis $|0\rangle , |1\rangle$ with a general matrix $M$ where $a, b, c, d$ are complex numbers.
\begin{equation}
    M = 
    \begin{pmatrix}
        a & b  \\
        c & d
    \end{pmatrix}.
\end{equation}
$M$ satisfies the unitary condition:
\begin{equation} 
    \begin{pmatrix}
        a & b  \\
        c & d
    \end{pmatrix}
    \begin{pmatrix}
        \overline{a} & \overline{c}  \\
        \overline{b} & \overline{d}
    \end{pmatrix}
    =
    \begin{pmatrix}
        1 & 0  \\
        0 & 1
    \end{pmatrix},
\end{equation}
where $\overline{a}, \overline{b}, \overline{c}, \overline{d}$ are the conjugates of $a, b, c, d$.

The action on a qubit is then:
\begin{equation}
    \begin{pmatrix}
        a & b  \\
        c & d
    \end{pmatrix}
    \begin{pmatrix}
        \alpha  \\
        \beta
    \end{pmatrix}
    = 
    \begin{pmatrix}
        a \alpha + b \beta \\
        c \alpha + d \beta
    \end{pmatrix}.
\end{equation}

Common single-qubit gates and multi-qubit gates used to build a quantum circuit are presented in Appendix~\ref{chap:gates}.

When we pass from a single qubit system to a multi-qubit system is not just increasing the size of the model, as what happens in classical computing.
In quantum computing is possible to create entangled states.
A quantum entangled state of qubits is a state in which each qubit cannot be described independently from the state of the other qubits.
We can write a general quantum state as:
\begin{equation}
    |\psi\rangle = \sum_{i=0}^{2^n - 1} \alpha_i |i\rangle,
\end{equation}
where $n$ is the number of qubits of the state.
We can note that the number of parameters that describe the system increases exponentially with the number of qubits and this is a consequence of entanglement.

Bell states are very simple entangled states and they represent a basis for the four-dimensional Hilbert space of two qubits.
\begin{equation}
    \begin{split}
        |\psi_{1}\rangle = \frac{1}{\sqrt{2}} (|00\rangle + |11\rangle),\quad
        |\psi_{2}\rangle = \frac{1}{\sqrt{2}} (|00\rangle - |11\rangle), \\
        |\psi_{3}\rangle = \frac{1}{\sqrt{2}} (|01\rangle + |10\rangle),\quad
        |\psi_{4}\rangle = \frac{1}{\sqrt{2}} (|01\rangle - |10\rangle).
    \end{split}
    \label{bell_eq}
\end{equation}
The creation of the four Bell states is described in Appendix~\ref{sec:bell}.

\section{Quantum computing applications}
In this section, we motivate the use of quantum computing in real-world applications
introducing what is known today to give a quantum advantage.

\subsection{Quantum computing today}
The reason that drives the research in quantum computing is that many algorithms, involving quantum circuits, are known to give an advantage over classical algorithms \cite{cern_quantum_seminary}.
Grover's algorithm \cite{GroverLov1996Afqm}, for example, is used to find an element that satisfies a certain condition in an unsorted list of $N$ elements.
Any classical algorithm needs $O(N)$ queries to the list in the worst case. Grover's algorithm can find the element with $O(\sqrt{N})$ queries.
Another important example is Shor's algorithm \cite{ShorP.W1994Afqc}, it finds a factor of an n-bit integer in time $O(n^{2}(\log{n})(\log{\log{n}}))$ where the best classical algorithm needs $O(\exp{(cn^{\frac{1}{3}}(\log{n})^{\frac{2}{3}})})$.
Other important applications of quantum computing that have a scientific interest are the simulation of electronic structures and chemical bonds \cite{McCaskeyAlexanderJ2019Qcaa}.
All these algorithms are tested on Noisy Intermediate-Scale Quantum (NISQ) devices \cite{Preskill_2018}.
IBM has several NISQ devices with a number of qubits from 1 to 65 \cite{ibm}.
Rigetti \cite{rigetti} has available 6 quantum computers with a number of qubits from 8 to 31. 

In this work, the quantum circuits created are simulated with Qibo \cite{qibo},
an open-source framework for quantum computing \cite{efthymiou2020qibo}.
The key features of Qibo are the definition of a standard language for the construction and execution of quantum circuits,
efficient simulation backends with CPU, multi-thread CPU, GPU and multi-GPU, and quantum hardware control.
\subsection{Quantum machine learning}
In this work, we study a near-term application of NISQ devices, quantum machine learning.
Classical input data is embedded in a trainable quantum circuit, called variational quantum circuit (VQC) \cite{Havl_ek_2019}, assisted by a classical computer.
This kind of approach is studied because we can exploit a complex Hilbert space that grows exponentially with the number of qubits.
This can lead to an advantage over classical machine learning models where the representational ability grows only linearly with the number of classical bits.
It has been shown that a single qubit is sufficient to build a universal quantum classifier.
Using multiple qubits is possible to introduce entanglement between qubits using two-qubits gates (Appendix~\ref{chap:gates}) and reduce the number of gates to use in the circuit.
In this thesis, we explore the possibility to use a VQC with multiple qubits to solve a binary classification problem.

\chapter{Design a binary quantum classifier}
\label{chap:binary_classifier}

\begin{figure}
    \centering
    \includegraphics[scale=0.20]{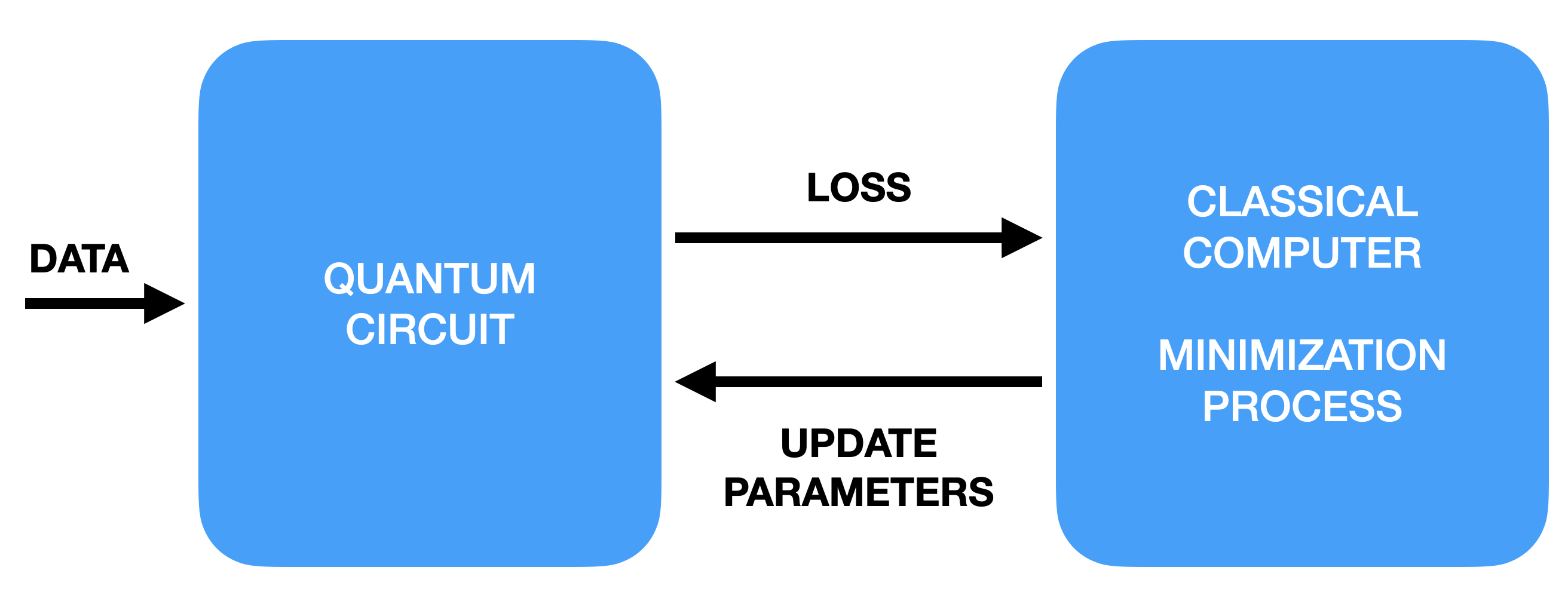}
    \caption{Simple diagram that shows how the hybrid method works.}
    \label{fig:diagram}
\end{figure}
In machine learning, a classifier is a model that allows discrimination between multiple categories.
The aim of this work is to build a binary classifier, a specific classifier that can discriminate data in two categories, labeled as 0 and 1.
For this purpose we need a procedure to input data in the circuit, a set of gates that are trainable, a loss function, that measure how good our training is, and a minimizer to train parameters.
In this chapter we are about to show how to design a quantum binary classifier for two different types of data:
\begin{enumerate}
    \item small arrays of data, in other words, a few variables
    \item images
\end{enumerate}
We use a hybrid strategy which consists of creating a quantum circuit, where we input data, with a set of gates that can be trained, like rotation gates.
We evaluate the loss function and then a classical computer decides how to change the parameters of the quantum circuit.
Then we re-execute the quantum circuit until the minimization process of the loss function ends.
A schematic representation of this approach is shown in Figure~\ref{fig:diagram}.

After we input the data in the circuit (for example an image of the data set) we set the parameters of the gates.
Measuring the first qubit, we get $|0\rangle$ or $|1\rangle$. If we obtain $|0\rangle$ most of the time we say that the predicted label is 0, otherwise, the predicted label is 1.
We do this operation for each element of the data set and we evaluate the loss function.
The minimizer now decides how to change the parameters of the circuit and all the process restart.

The goodness of the classification is measured by accuracy $\mathcal{A}$: the number of elements in the data set that are classified correctly, divided by the size of the data set.
We can write this definition as:
\begin{equation}
    \mathcal{A}=\frac{1}{N}\sum_{i=1}^{N}{\delta{(L_{i}^{predicted}-L_{i}^{real})}},
\end{equation}
where $N$ is the total number of elements in the data set, $L_{i}^{predicted}$ is the predicted label for the i-element, and $L_{i}^{real}$ is the real label for the i-element.

\section{How to input data in the circuit}
We designed two different methods to input data in the circuit: one for small arrays and the other for images.

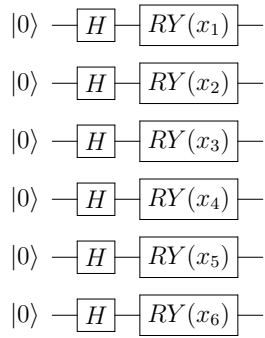
\begin{figure}[b]
    \centering
    \adjustbox{scale=0.80}{
        \subfloat{
            \Qcircuit @C=1.em @R=.7em{
                & \lstick{|0\rangle} & \gate{H} & \gate{RY(x_1)} & \qw\\
                & \lstick{|0\rangle} & \gate{H} & \gate{RY(x_2)} & \qw\\
                & \lstick{|0\rangle} & \gate{H} & \gate{RY(x_3)} & \qw\\
                & \lstick{|0\rangle} & \gate{H} & \gate{RY(x_4)} & \qw\\
                & \lstick{|0\rangle} & \gate{H} & \gate{RY(x_5)} & \qw\\
                & \lstick{|0\rangle} & \gate{H} & \gate{RY(x_6)} & \qw
            }
        }
    }
    \caption{Input layer of a classifier circuit of 6 qubits for a small array $\vec{x}$.}
    \label{fig:inputlayer}
\end{figure}
\subsubsection{Small arrays}
Given the number of data in the array, $L_{array}$, we create a quantum circuit with the number of qubits equal to the length of the array.
This procedure is perfect for small arrays, up to $10 \sim 20$ elements, because we only need a few numbers of qubits.
Firstly, we divide each element of the array by its maximum value and we multiple for $\pi$.
We create the circuit by adding a first layer of Hadamard gates and a layer of RY rotations with $\theta$ equal to the values in the array. We refer to this as the input layer.
In Figure~\ref{fig:inputlayer} we show an example of the input layer for a circuit of 6 qubits.

\subsubsection{Images}
Images could be very large. An image of size 32x32 with 1 color channel and 1byte of color depth contains 1024 numbers between 0 and 255.
If we use the input layer showed before we would need 1024 qubits. For this reason, we designed another method, specific to images.
The strategy consists of using the basis of an n-qubits system that is composed of $2^{n}$ states. For a 10-qubits circuit, we have exactly 1024 states.
\begin{equation}
    |\psi\rangle = a_{00...0}|00...0\rangle + ... + a_{11...1}|11...1\rangle
    \label{eq:base}
\end{equation} 
We identify two steps:
\begin{enumerate}
    \item Transform the image in an array. This could be done in different ways. The easiest way is to concatenate each row of the image.
    Another possibility consists of dividing the image into 4 squares and for each of them concatenate columns.
    \item Input the array in the circuit. We normalize the array dividing by its norm and then, we set the parameters of Equation~\ref{eq:base} to the values of the normalized array.
\end{enumerate}
Once we have created the array representing the image, we set the initial state of the quantum circuit.

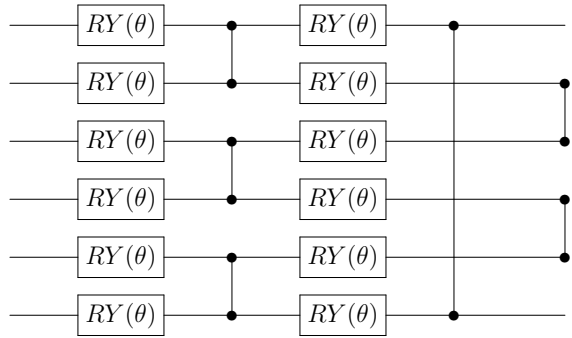
\begin{figure}[b]
    \centering
    \adjustbox{scale=0.80}{
        \subfloat{
            \Qcircuit @C=1em @R=.7em @!C {
                & \gate{RY(\theta)} & \ctrl{1} & \gate{RY(\theta)}    & \ctrl{5} & \qw \\
                & \gate{RY(\theta)} & \ctrl{0} & \gate{RY(\theta)}    & \qw & \ctrl{1} \\
                & \gate{RY(\theta)} & \ctrl{1} & \gate{RY(\theta)}    & \qw & \ctrl{0} \\
                & \gate{RY(\theta)} & \ctrl{0} & \gate{RY(\theta)}    & \qw & \ctrl{1} \\
                & \gate{RY(\theta)} & \ctrl{1} & \gate{RY(\theta)}    & \qw & \ctrl{0} \\
                & \gate{RY(\theta)} & \ctrl{0} & \gate{RY(\theta)}    & \ctrl{0}  & \qw 
            }
        }
    }
    \caption{Ansatz for a 6 qubits classifier. Rotations RY($\theta$) are trainable.}
    \label{fig:Ansatz}
\end{figure}
\section{Ansatz and Measurement}

Once we have data in the circuit we design an Ansatz with rotation gates that we train and two-qubits gates that connect qubits.
For rotations, we choose RY rotations with initial $\theta$ random between $0$ and $2\pi$, and for two-qubits gates we choose CZ gates.
The Ansatz for a 6-qubits classifier is shown in Figure~\ref{fig:Ansatz}.
It is possible to concatenate multiple layers of Ansatz and study what happens by increasing this number.
We call the number of layers of Ansatz, to make it easier, number of layers.

At the end of the circuit, we add a final column of rotations RY and a measurement gate to the first qubit.

\section{Loss functions}
We implemented three different loss functions to evaluate the proper functioning of training.
We need to evaluate the loss function for all the elements in the data set used for training.
Each loss function evaluation is based on the measure of the first qubit of the quantum classifier circuit.
For each element we execute the circuit multiple times; we call this amount number of shots (nshots).
We also need to store the frequency of predicted label 0 and predicted label 1.
We define a function \textit{prediction} that takes as input an image or an array and returns these two frequencies.

\subsubsection{Mean squared error ``square"}
We compute the loss $\mathcal{L}$ by taking take the frequency of finding the target label of the i-element of the data set, $F_i$, and doing a mean squared error:
\begin{equation}
    \mathcal{L} =  \frac{1}{N}\sum_{i=1}^{N}{(1-F_i)^{2}}.
\end{equation}
A possible pseudo-code is the following.
\begin{lstlisting}
    for image in images:
        # compute the frequency to find that "image"
        # is 0 and the frequency that is 1.
        predict = prediction(image)
        loss += ((1 - predict[target])**2) / len(images)
\end{lstlisting}

\subsubsection{Cross entropy ``ce"}
Instead of applying a mean square error to $F_i$, we can apply a $-\ln{}$ function, which in information theory it is a measure of uncertain:
\begin{equation}
    \mathcal{L} =  \frac{1}{N}\sum_{i=1}^{N}{-\ln{F_i}}.
\end{equation}
In this case, the algorithm would look like this:
\begin{lstlisting}
    for image in images:
        # compute the frequency to find that "image"
        # is 0 and the frequency that is 1.
        predict = prediction(image)
        loss -= ln(predict[target]) / len(images)
\end{lstlisting}

\subsubsection{Limited cross-entropy ``lce"}
We consider only the elements that we are classifying with a wrong predicted label.
This may help to increase the classification ability on the images with a lot of noise.
If $F_i$ is lower than 0.55 we apply $-\ln{}$ function:
\begin{equation}
    \mathcal{L} = \frac{1}{N}\sum_{i=1}^{N}{-\ln{F_{i}^{<0.55}}}.
\end{equation}
In pseudo-code this means:
\begin{lstlisting}
    for image in images: 
        # compute the frequency to find that "image"
        # is 0 and the frequency that is 1.
        predict = prediction(image)
        if predict[target] < 0.55:
            loss -= ln(predict[target]) / len(images)
\end{lstlisting}
\section{Minimization process and minimizers}

We implemented multiple minimizers to train the circuit. Most of them are taken from scipy minimizers \cite{scipy}.

\subsubsection{Powell}
Powell minimizer is a free-derivative method that performs sequential one-dimensional minimizations along each vector of the directions set.
The initial directions set is ${\mathbf{s}_{1}, ..., \mathbf{s}_{N}}$, where $N$ is the number of variables of the function to minimize and correspond to the normals aligned to each axis.
Given the initial point $\pmb{\theta}_{0}$ we find a set of scalars $\alpha_{i}$ that minimize the function evaluations of $\pmb{\theta}_{0}+\alpha_{1}\mathbf{s}_{1}$, $\pmb{\theta}_{0}+\sum_{i=1}^{2}{\alpha_{i} \mathbf{s}_{i}}$, ..., $\pmb{\theta}_{0}+\sum_{i=1}^{N}{\alpha_{i} \mathbf{s}_{i}}$.
The new point is then $\pmb{\theta}_{1} = \pmb{\theta}_{0}+\sum_{i=1}^{N}{\alpha_{i}\mathbf{s}_{i}}$.
We add $\sum_{i=1}^{N}{\alpha_{i}\mathbf{s}_{i}}$ to the directions set and we remove the vector $\mathbf{s}_{i}$ that contributed the most to the new point: $\mathbf{s}_{removed}=\argmax_{i=1}^{N}{|\alpha_{i}|||\mathbf{s}_{i}||}$
The algorithm iterates until no improvement is made.
Bounds are implemented and, in our case, each $\theta$ must be in the range $(0, 2\pi)$.

\subsubsection{COBYLA (Constrained Optimization BY Linear Approximation)}
The algorithm is based on linear approximations to the objective function and each constraint.
The algorithm evaluates the objective function and the constraints at the vertices of a trust region.
The number of vertices of this region is $N+1$, where $N$ is the number of variables of the function to minimize.
A linear polynomial is used to create an interpolation of the objective function and the constraints.
At each step, the size of the trust region is modified by the algorithm and it is decreased as the convergence is achieved.
Bounds are not implemented and we did not use any constraints during the minimization process.

\subsubsection{Other scipy optimizers}
Other minimizers implemented are CG, BFGS, L-BFGS-B, Newton-CG, TNC, trust-constr.
Most of them use approximations of the gradient by evaluating the function to minimize near the initial point.

\subsubsection{Genetic algorithm}
Genetic algorithms are inspired by the evolution of a specific generation in nature.
We start defining the generation randomly and then, at each step, we perform mutations of genes and crossover between pairs of chromosomes.
The parameters are the size of the population, the number of generations to create, the probability of crossover, and the probability of mutation.
This kind of approach may work in situations where the derivative is not computable or is useless.

\subsubsection{Barren plateaus}
The main problem in quantum machine learning when we use a hybrid approach is barren plateaus.
The landscape of trainable parameters in classical machine learning typically descends slowly.
When we train a variational quantum circuit we find what is known as barren plateaus, the landscape of the parameters is mainly flat therefore gradient, hessian and higher-order derivatives vanish exponentially \cite{CerezoM2021Hodo}.
This issue makes the minimization process harder.
Free-derivative minimizers, like Powell and COBYLA, could work better than algorithms based on gradients but they are affected too.
In addition, recently has been proved that barren plateaus effects also free-derivative methods because the differences in loss function evaluations vanish exponentially when we increase the number of qubits \cite{arrasmith2020effect}.
\chapter{Digits classification}
\label{chap:digits_classification}

\begin{figure}
    \centering
    \includegraphics[scale=0.3]{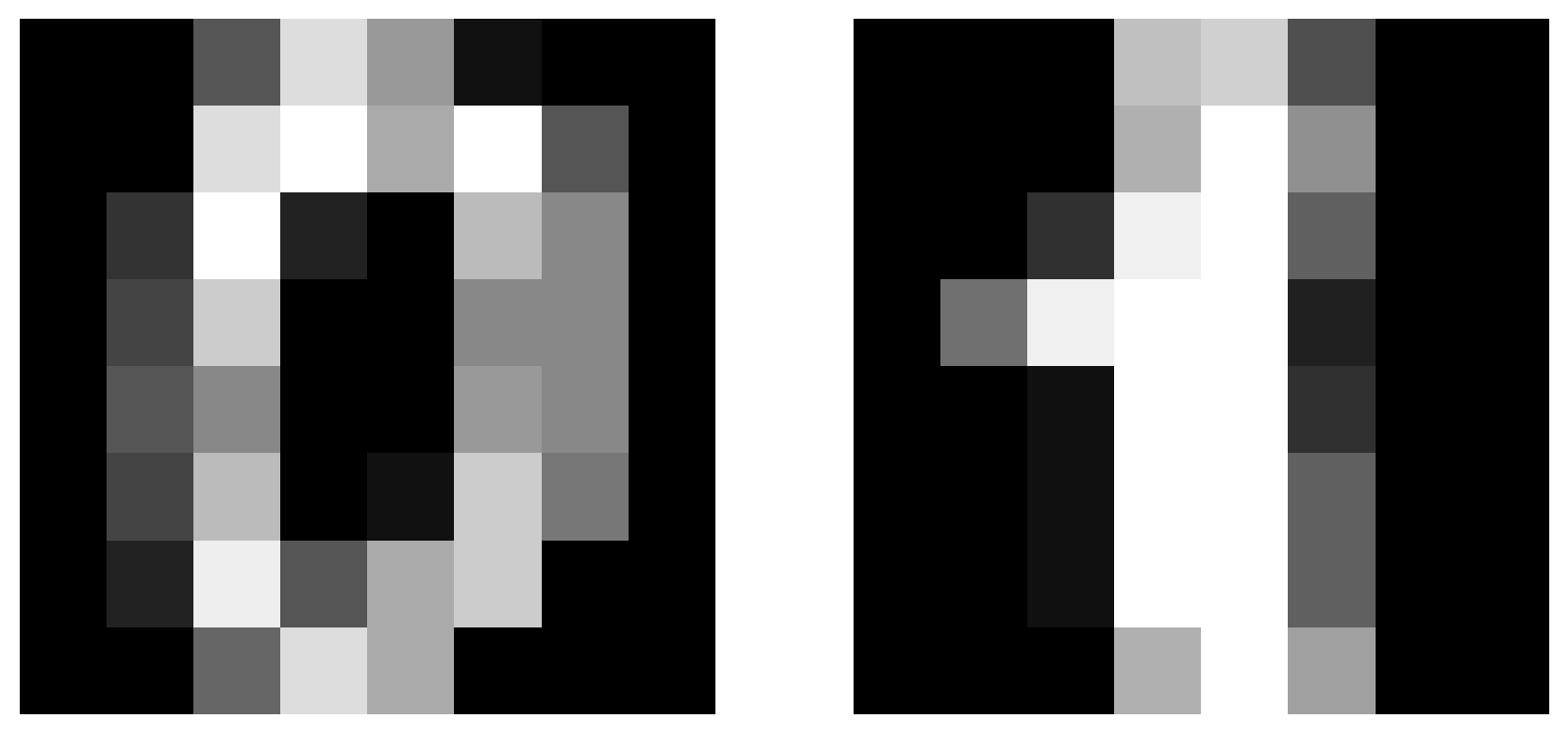}
    \caption{Samples from the digits data set.
    On the left, we have a handwritten zero and on the right a handwritten one.
    A darker pixel stands for a lower value.}
    \label{fig:samples_digits}
\end{figure}
\begin{figure}
    \centering
    \adjustbox{scale=0.7}{
        \subfloat{
            \Qcircuit @C=0.7em @R=.5em @!C {
                & \gate{RY(\theta)} & \ctrl{1} & \gate{RY(\theta)}    & \ctrl{5} & \qw & \lstick{...} & \gate{RY(\theta)} & \meter\\
                & \gate{RY(\theta)} & \ctrl{0} & \gate{RY(\theta)}    & \qw & \ctrl{1} & \lstick{...} & \gate{RY(\theta)} & \qw\\
                & \gate{RY(\theta)} & \ctrl{1} & \gate{RY(\theta)}    & \qw & \ctrl{0} & \lstick{...} & \gate{RY(\theta)} & \qw\\
                & \gate{RY(\theta)} & \ctrl{0} & \gate{RY(\theta)}    & \qw & \ctrl{1} & \lstick{...} & \gate{RY(\theta)} & \qw\\
                & \gate{RY(\theta)} & \ctrl{1} & \gate{RY(\theta)}    & \qw & \ctrl{0} & \lstick{...} & \gate{RY(\theta)} & \qw\\
                & \gate{RY(\theta)} & \ctrl{0} & \gate{RY(\theta)}    & \ctrl{0}  & \qw & \lstick{...} & \gate{RY(\theta)} & \qw
                \gategroup{1}{2}{6}{6}{2.em}{--}
                \gategroup{1}{8}{6}{8}{.9em}{--}
                \inputgroupv{1}{6}{1em}{5.1em}{|\psi_{input}\rangle \hspace{7mm}}
            }
        }
    }
    \caption{Quantum classifier circuit for the classification of 8x8 images.
    The first dashed box is the Ansatz and the second one represents the final rotations. Rotations RY$(\theta)$ are trainable.
    Dashed horizontal lines between Ansatz and final rotations indicate the possibility to concatenate multiple Ansätze.}
    \label{fig:circuit_digits}
\end{figure}
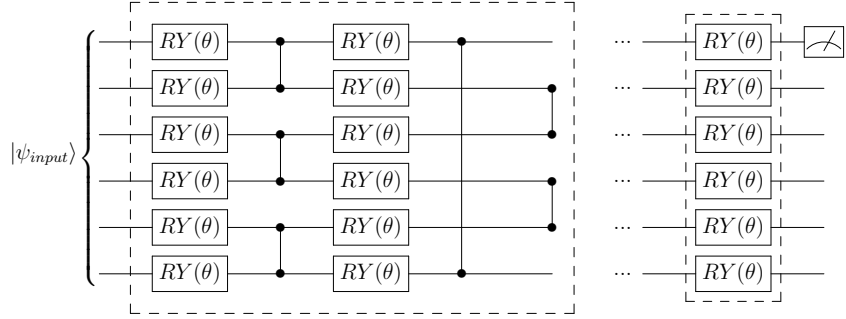
\section{Data set and Ansatz}
When a student starts a course about machine learning one of the first examples that he studies is the classification of handwritten digits belonging to the MNIST data set \cite{mnist}.
This problem is solvable in machine learning, for example, with a CNN (convolutional neural network).

Here we want to study if it is possible to discriminate between a handwritten zero and a handwritten one with our quantum classifier.

The data set has 360 images. The size of each picture is 8x8. The MNIST data set is taken from \textit{scikit-learn} \cite{digits} and are considered only handwritten zeros and handwritten ones.
A sample of the images is shown in Figure~\ref{fig:samples_digits}.

We use an Ansatz of 6 qubits because we can store a number of pixels equal to $2^{6}$, that is exactly the number of pixels in our 8x8 images. The corresponding circuit is illustrated in Figure~\ref{fig:circuit_digits}.

\begin{figure}
    \centering
    \includegraphics[scale=0.23]{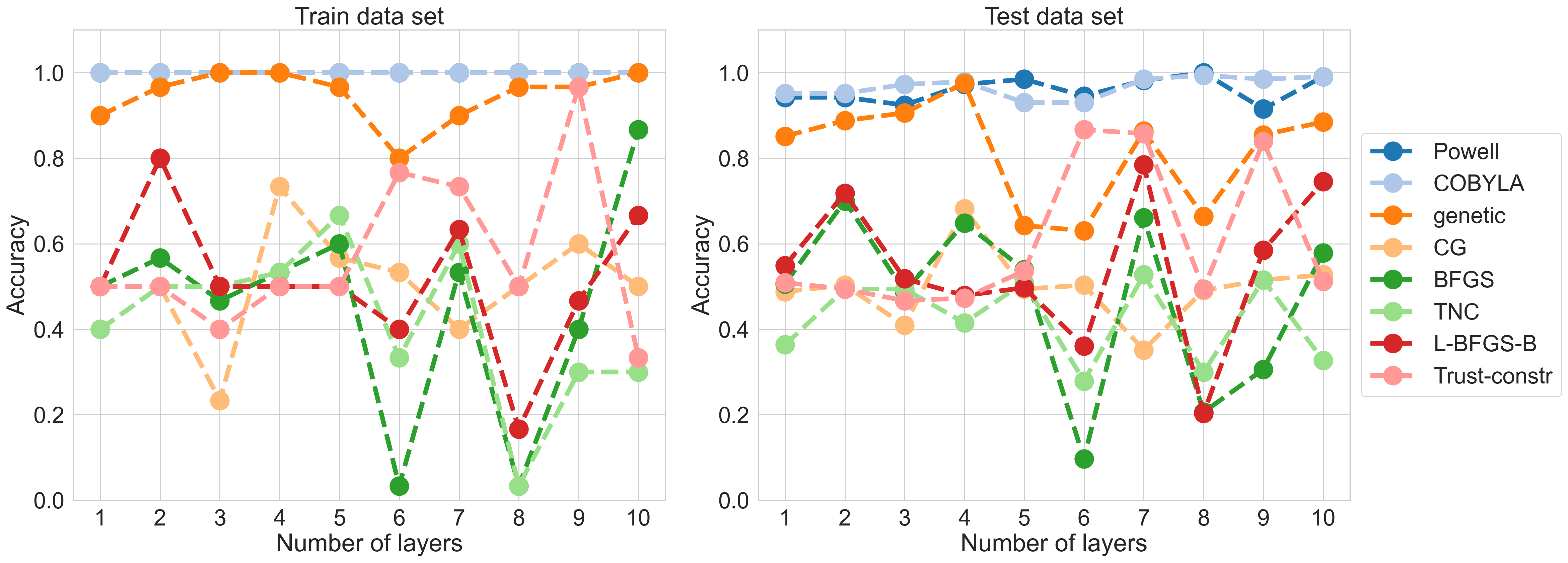}
    \caption{Accuracy of the train set (on the left) and the test set (on the right) for different minimizers and by increasing the number of layers.
    The loss function used is ``square".
    The quantum classifier circuit has 6 qubits and it is illustrated in Figure~\ref{fig:circuit_digits}. The number of shots is 1000.}
    \label{fig:result_digits_square}
\end{figure}
\begin{figure}
    \centering
    \includegraphics[scale=0.23]{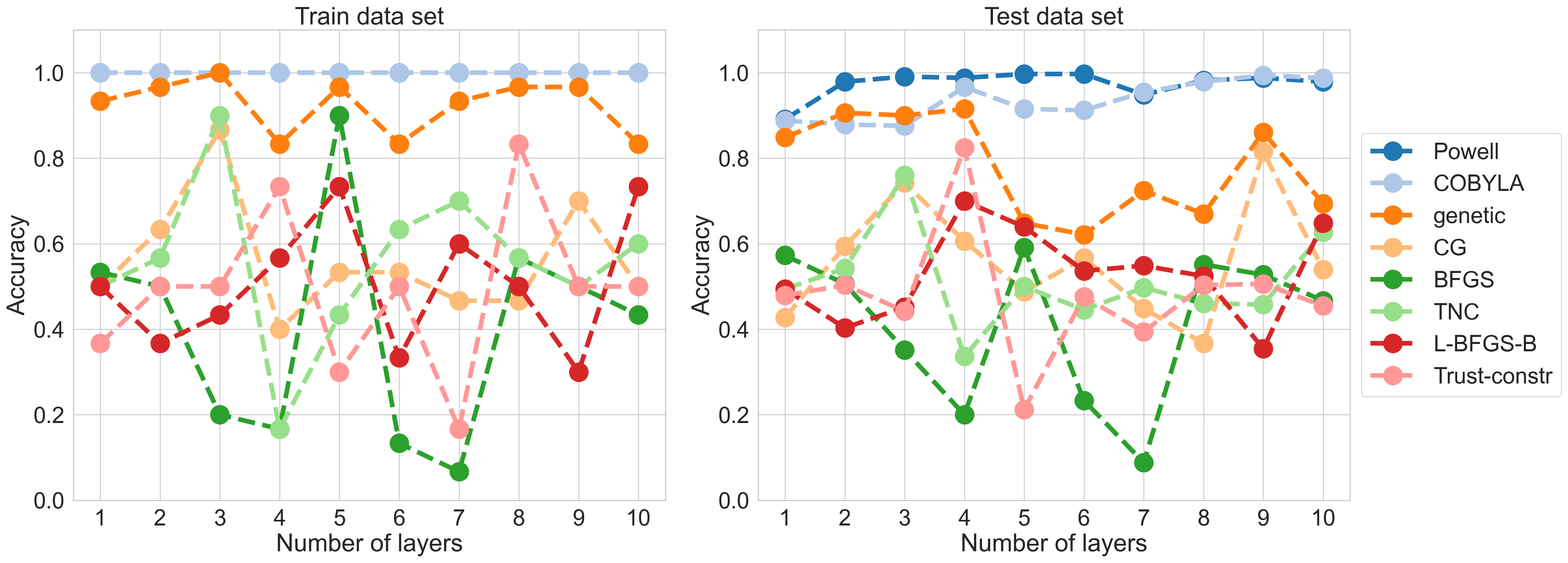}
    \caption{Accuracy of the train set (on the left) and the test set (on the right) for different minimizers and by increasing the number of layers.
    The loss function used is ``ce".
    The quantum classifier circuit has 6 qubits and it is illustrated in Figure~\ref{fig:circuit_digits}. The number of shots is 1000.}
    \label{fig:result_digits_ce}
\end{figure}
\section{Comparison between minimizers}
To understand what is the best optimizer for this kind of classification we compare the different minimizers introduced in Chapter~\ref{chap:binary_classifier}.
We used as train set 30 images and as test set the other 330 images available in the data set.
The number of shots is 1000.
We increased the number of layers and computed accuracy.

We tried with ``square" (Figure~\ref{fig:result_digits_square}) and ``ce" (Figure~\ref{fig:result_digits_ce}) loss function obtaining similar results.

The best optimizers are Powell and COBYLA.
The genetic algorithm works well, but only with a small number of layers. When this number exceeds 4, the number of parameters becomes too big for this minimizer and the performances are deprecated.

We can deduce that algorithms that use approximations of gradient do not work. This result is correlated to the presence of barren plateaus.
Algorithms like Powell, COBYLA, and genetic are the only ones that we tested able to train the model because they are gradient-free.
We expect the same behavior from the result of the next data set, the jet data set.

\section{Scoring results and comments}

Given the results of the previous section, we decide to use COBYLA optimizer with the ``square" loss function and a train set of size 30.
The classifier circuit has 3 layers corresponding to 42 trainable parameters.
The number of shots used in this section is 2000.

We want now to split the data set in order to do the training on different parts of it.
To do that we divide the data set into a fixed test set of size 100 and a fixed train set of size 200.
We take randomly 30 images for training from the train set, the other part becomes the validation set. We repeat this last operation 5 times.
For each training we calculate the accuracy of the test set and eventually we get an average value; results are exposed in Table~\ref{table:test_digits}.
We compute also the confusion matrix of the validation set (Table~\ref{table:conf_digits}), which represents the correct rate of classification of a specific label.
We can confirm that the quantum classifier discriminate perfectly the two classes belonging to the validation set.
All the handwritten zeros infact are predicted as zeros; the same is for the handwritten ones.
The same conclusion comes from the test accuracy that is almost $\sim$100\%.

\begin{table}
    \centering
    \begin{tabular}{c|c}
        Data set & Test accuracy \\ 
        \hline \hline
        Digits & $0.990 \pm 0.013$ \\ 
        \hline
    \end{tabular}
    \caption{Accuracy of test set averaged over 5 different trainings of 30 images. The test set has a size of 100 images and it is fixed through all different trainings.
    The quantum classifier used has 3 layers of Ansatz. We used COBYLA optimizer to minimize the ``square" loss function. The number of shots is 2000.}
    \label{table:test_digits}
\end{table}
\begin{table}
    \centering
    \begin{tabular}{c|cc}
        \multirow{2}{*}{Real label} & \multicolumn{2}{c}{Predicted label}\\
        & 0 & 1 \\ \cline{2-3}
        0 & $100 \pm 0.0 \% $ & $0.0 \pm 0.0 \% $ \\ 
        1 & $0.0 \pm 0.0 \% $ & $100.0 \pm 0.0 \% $ \\ 
        \hline
    \end{tabular}
    \caption{Confusion matrix for digits data set. The average is computed over 5 different trainings of 30 images.
    Values are calculated on the validation set, the part of the train set not used for training, of size 170.
    The quantum classifier used has 3 layers of Ansatz. We used COBYLA optimizer to minimize the ``square" loss function.
    The number of shots is 2000.}
    \label{table:conf_digits}
\end{table}

\chapter{Jet tagging}
\label{chap:jet_tagging}

\section{Data sets and Ansatz}

In high energy physics is frequent the problem of classification of two different jets, this problem is known as jet tagging.
In this work, we want to study if it is possible to do this kind of binary classification with our quantum classifier and to tell how well can we do this classification in terms of accuracy (how many jets are classified correctly).

Collisions at LHC collider occur at such high energies that there is enough energy to produce massive particles with enough velocity to decay in products that are collimated. This is the case of the following interaction:
\begin{equation}
	pp \rightarrow W^{+}W^{-} \rightarrow qqqq.
\end{equation}
The interaction $W \rightarrow qq' $ was generated with a center of mass energy of $\sqrt{s}$=14TeV.
The two pairs of quarks generated are collimated in one single jet.
Pairs of quarks and gluons can also be generated from these processes:
\begin{equation}
	pp \rightarrow qq, qg, gg.
\end{equation}
We call this second class ``0” and the first one ``1”.
Additionally, in a real situation, we can have proton-proton in-time interactions known as ``pile-up”.
In this work we study both cases, \textit{“without pile-up”} and \textit{“with pile-up”}.
These data sets are taken from Ref. \cite{baldi}.

\begin{figure}
    \centering
    \includegraphics[scale=0.33]{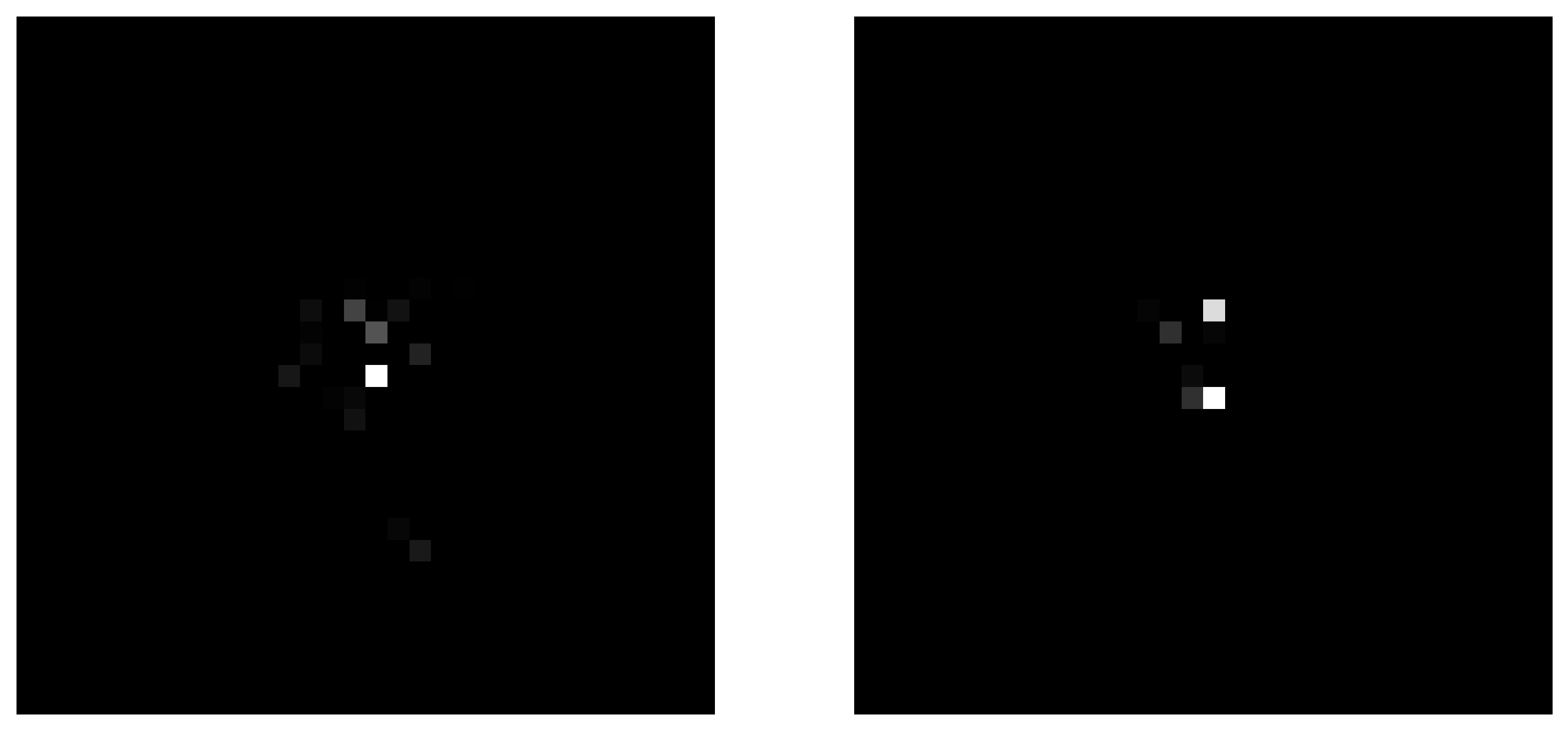}
    \caption{The picture on the left shows an example of the class ``0"; on the right an example of class ``1".  Both images belong to the ``without pile-up" data set. A darker pixel stands for a lower value.}
    \label{fig:sample_jet_nopile}
\end{figure}
\begin{figure}
    \centering
    \includegraphics[scale=0.33]{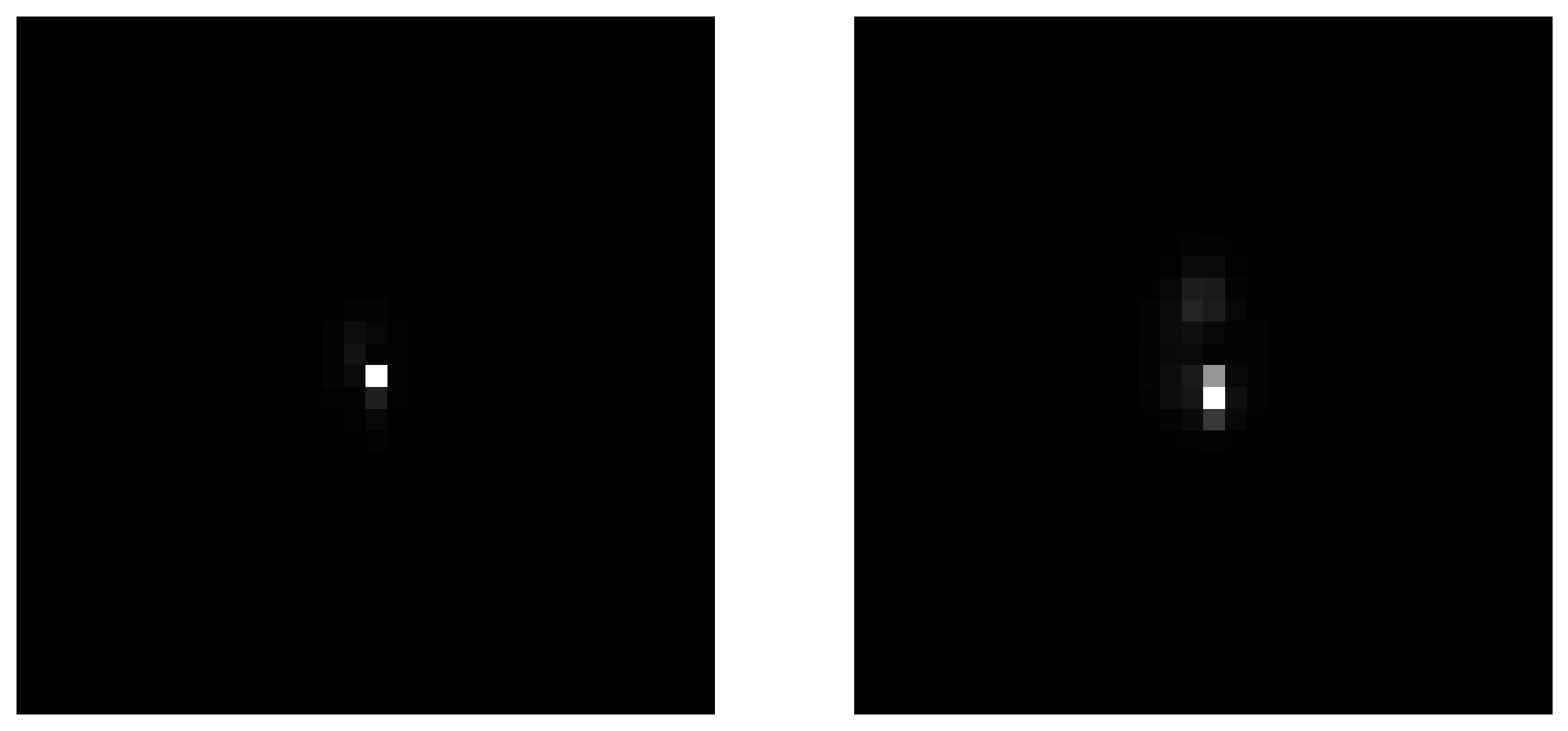}
    \caption{On the left it is shown the sum of all images from class ``0"; on the right the sum for the class ``1". Both images belong to the ``without pile-up" data set. A darker pixel stands for a lower value.}
    \label{fig:sample_jetsum_nopile}
\end{figure}
\begin{figure}
    \centering
    \includegraphics[scale=0.33]{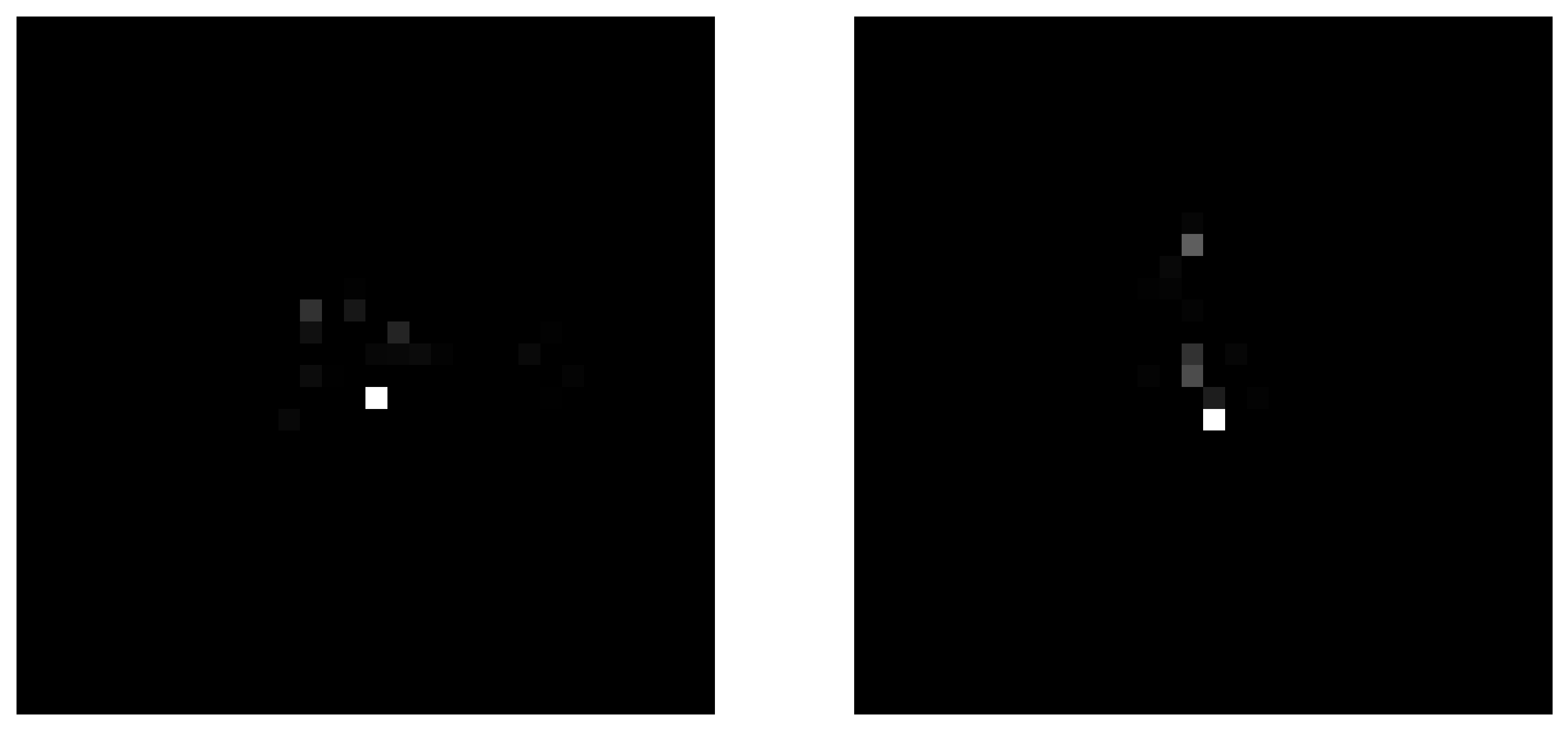}
    \caption{The picture on the left shows an example of the class ``0"; on the right an example of class ``1". Both images belong to the ``with pile-up" data set. A darker pixel stands for a lower value.}
    \label{fig:sample_jet_pile}
\end{figure}
\begin{figure}
    \centering
    \includegraphics[scale=0.33]{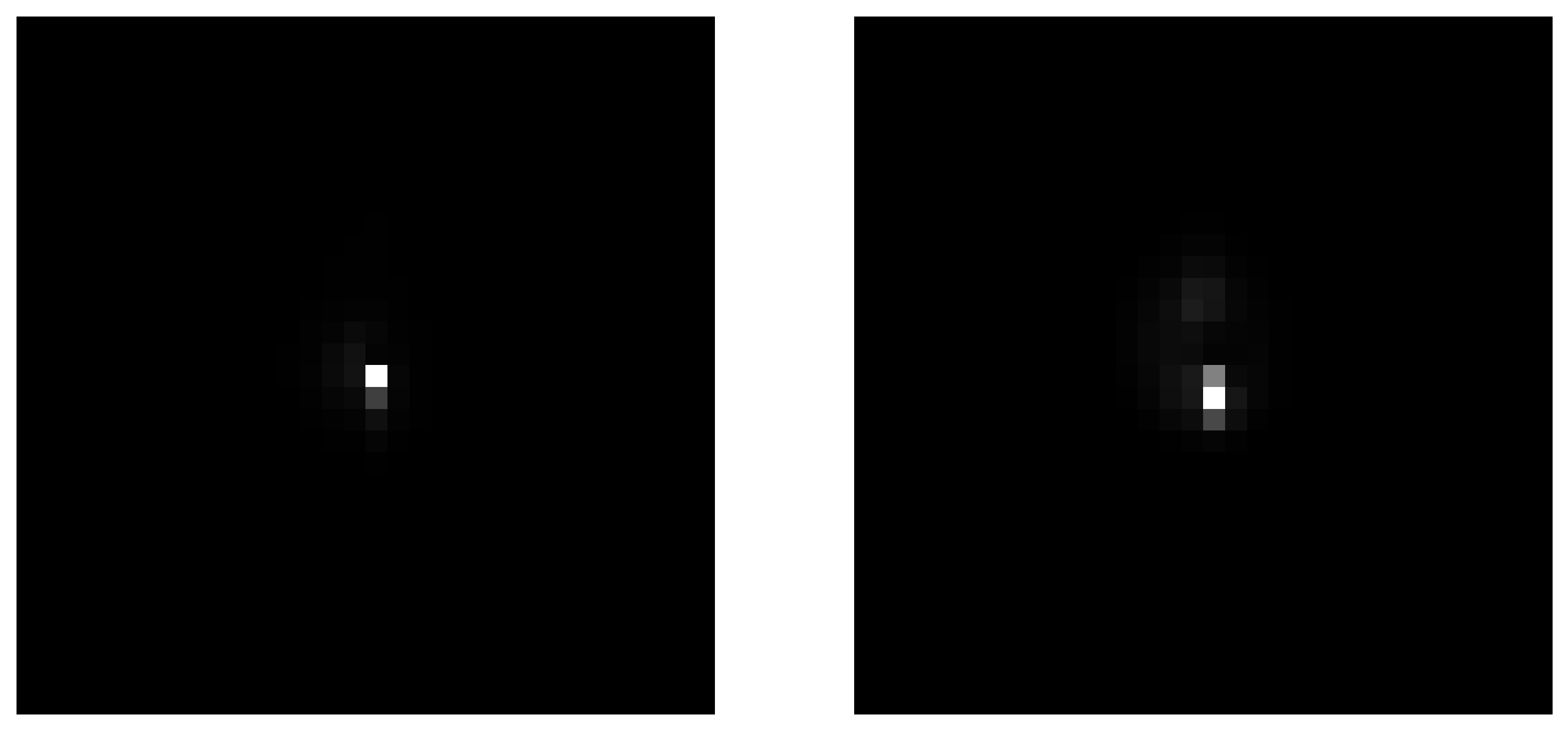}
    \caption{On the left it is shown the sum of all images from class ``0"; on the right the sum for the class ``1". Both images belong to the ``with pile-up" data set. A darker pixel stands for a lower value.}
    \label{fig:sample_jetsum_pile}
\end{figure}
\subsection{Images}

To do jet tagging we can use images that represent the amount of energy deposited at different points of the cylindrical calorimeter surface.
Images are pre-processed and converted into 32x32 images with values from 0 to 255.

Examples are illustrated for both data sets ``without pile-up" (Figure~\ref{fig:sample_jet_nopile} and Figure~\ref{fig:sample_jetsum_nopile}) and ``with pile-up" (Figure~\ref{fig:sample_jet_pile} and Figure~\ref{fig:sample_jetsum_pile}).

\begin{figure}
    \centering
    \adjustbox{scale=0.80}{
        \subfloat{
            \Qcircuit @C=1.4em @R=.7em {
                & \gate{RY(\theta)} & \ctrl{1} & \gate{RY(\theta)}    & \ctrl{9} & \qw & \rstick{   } & \lstick{...} & \gate{RY(\theta)} & \meter\\
                & \gate{RY(\theta)} & \ctrl{0} & \gate{RY(\theta)}    & \qw & \ctrl{1} & \rstick{   } & \lstick{...} & \gate{RY(\theta)} & \qw\\
                & \gate{RY(\theta)} & \ctrl{1} & \gate{RY(\theta)}    & \qw & \ctrl{0} & \rstick{   } & \lstick{...} & \gate{RY(\theta)} & \qw\\
                & \gate{RY(\theta)} & \ctrl{0} & \gate{RY(\theta)}    & \qw & \ctrl{1} & \rstick{   } & \lstick{...} & \gate{RY(\theta)} & \qw\\
                & \gate{RY(\theta)} & \ctrl{1} & \gate{RY(\theta)} & \qw & \ctrl{0} & \rstick{   } & \lstick{...} & \gate{RY(\theta)} & \qw\\
                & \gate{RY(\theta)} & \ctrl{0} & \gate{RY(\theta)}    & \qw & \ctrl{1} & \rstick{   } & \lstick{...} & \gate{RY(\theta)} & \qw\\
                & \gate{RY(\theta)} & \ctrl{1} & \gate{RY(\theta)}    & \qw  & \ctrl{0} & \rstick{   } & \lstick{...} & \gate{RY(\theta)} & \qw\\
                & \gate{RY(\theta)} & \ctrl{0} & \gate{RY(\theta)}    & \qw & \ctrl{1} & \rstick{   } & \lstick{...} & \gate{RY(\theta)} & \qw\\
                & \gate{RY(\theta)} & \ctrl{1} & \gate{RY(\theta)}    & \qw & \ctrl{0} & \rstick{   } & \lstick{...} & \gate{RY(\theta)} & \qw\\
                & \gate{RY(\theta)} & \ctrl{0} & \gate{RY(\theta)}    & \ctrl{0}  & \qw & \rstick{   } & \lstick{...} & \gate{RY(\theta)} & \qw
                \gategroup{1}{2}{10}{6}{2.em}{--}
                \gategroup{1}{9}{10}{9}{.9em}{--}
                \inputgroupv{1}{10}{1em}{10.4em}{|\psi_{input}\rangle \hspace{7mm}}
            }
        }
    }
    \caption{Quantum classifier circuit for classification of 32x32 images.
    The first dashed box is the Ansatz and the second one represents the final rotations.
    Dashed horizontal lines between Ansatz and final rotations indicate the possibility to concatenate multiple Ansätze.
    Rotations RY$(\theta)$ are trained during the minimization process.}
    \label{fig:circuit_jet_images}
\end{figure}
\begin{figure}
    \centering
    \adjustbox{scale=0.80}{
        \subfloat{
            \Qcircuit @C=1.4em @R=.7em {
                & \lstick{|0\rangle} & \gate{H} & \gate{RY(x_1)} & \qw & \gate{RY(\theta)} & \ctrl{1} & \gate{RY(\theta)} & \ctrl{5} & \qw & \rstick{   } & \lstick{...} & \gate{RY(\theta)} & \meter \\
                & \lstick{|0\rangle} & \gate{H} & \gate{RY(x_2)} & \qw & \gate{RY(\theta)} & \ctrl{0} & \gate{RY(\theta)} & \qw &\ctrl{1} & \rstick{   } & \lstick{...} & \gate{RY(\theta)} & \qw\\
                & \lstick{|0\rangle} & \gate{H} & \gate{RY(x_3)} & \qw & \gate{RY(\theta)} & \ctrl{1} & \gate{RY(\theta)} & \qw & \ctrl{0} & \rstick{   } & \lstick{...} & \gate{RY(\theta)} & \qw\\
                & \lstick{|0\rangle} & \gate{H} & \gate{RY(x_4)} & \qw & \gate{RY(\theta)} & \ctrl{0} & \gate{RY(\theta)} & \qw & \ctrl{1} & \rstick{   } & \lstick{...} & \gate{RY(\theta)} & \qw\\
                & \lstick{|0\rangle} & \gate{H} & \gate{RY(x_5)} & \qw & \gate{RY(\theta)} & \ctrl{1} & \gate{RY(\theta)} & \qw & \ctrl{0} & \rstick{   } & \lstick{...} & \gate{RY(\theta)} & \qw\\
                & \lstick{|0\rangle} & \gate{H} & \gate{RY(x_6)} & \qw & \gate{RY(\theta)} & \ctrl{0} & \gate{RY(\theta)} & \ctrl{0} & \qw & \rstick{   } & \lstick{...} & \gate{RY(\theta)} & \qw
                \gategroup{1}{3}{6}{4}{.9em}{--}
                \gategroup{1}{6}{6}{10}{1.7em}{--}
                \gategroup{1}{13}{6}{13}{.9em}{--}
            }
        }
    }
    \caption{Quantum classifier circuit used for jet features.
    The first dashed box is the input layer, the second is the Ansatz and the last one is the final rotations.
    Dashed horizontal lines between Ansatz and final rotations indicate the possibility to concatenate multiple Ansätze.
    Rotations RY$(\theta)$ are trained during the minimization process.}
    \label{fig:circuit_jet_features}
\end{figure}
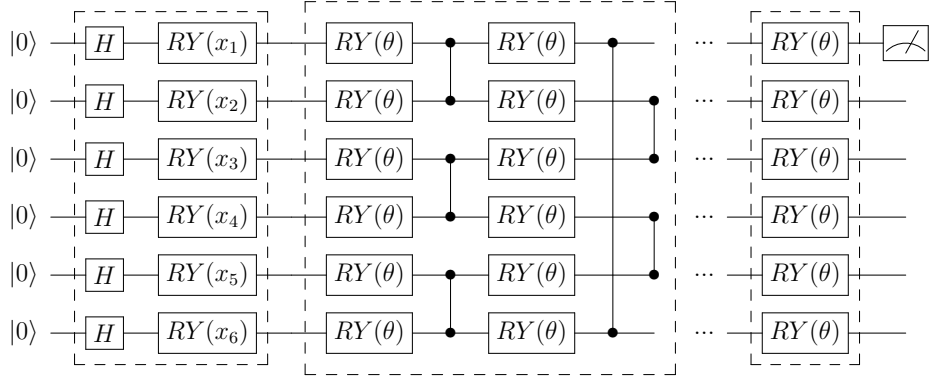
\subsubsection{Ansatz}
Images of size 32x32 contain 1024 pixels, therefore we need a 10-qubits classifier as the one illustrated in Figure~\ref{fig:circuit_jet_images}.

In this work, we also investigate the possibility of resizing these images from 32x32 to 16x16 using an 8-qubits quantum classifier.
By reducing the number of qubits, the effects of barren plateaus may be reduced.

\subsection{Features}

We can also approach the classification with six high-level jet variables: the invariant mass of the trimmed jet, N-subjettiness $\tau_{21}^{\beta=1}$, and the energy correlation functions $C_{2}^{\beta=1}$, $C_{2}^{\beta=2}$, $D_{2}^{\beta=1}$, $D_{2}^{\beta=2}$.
We call these variables features.

In this case, we need a classifier circuit for a small number of variables. We use a quantum classifier of 6 qubits because we have 6 features.
In Figure~\ref{fig:circuit_jet_features} we show the classifier circuit needed.

\newpage
\section{Comparison between minimizers}
Once we have defined the quantum classifier circuit, we need to find a minimizer to train the model.
In this section, we are about to study the different minimizers introduced in Chapter~\ref{chap:binary_classifier} with different loss functions and by increasing the size of the train set.
For the test set, we use 6000 images, not used for training.
The number of shots used is 2000.

\subsubsection{Features}
In Appendix~\ref{chap:pic_jet_features} we can see what happens by increasing the number of layers in the classifier quantum circuit.
There is no evidence of a best optimizer for this certain situation.
Increasing the size of the train data set does not make a difference.
We can also see that this behavior persists even with different loss functions.

It is plausible that there is not enough flexibility in the model to use the information contained in the 6 features.

\begin{figure}
    \centering
    \includegraphics[scale=0.33]{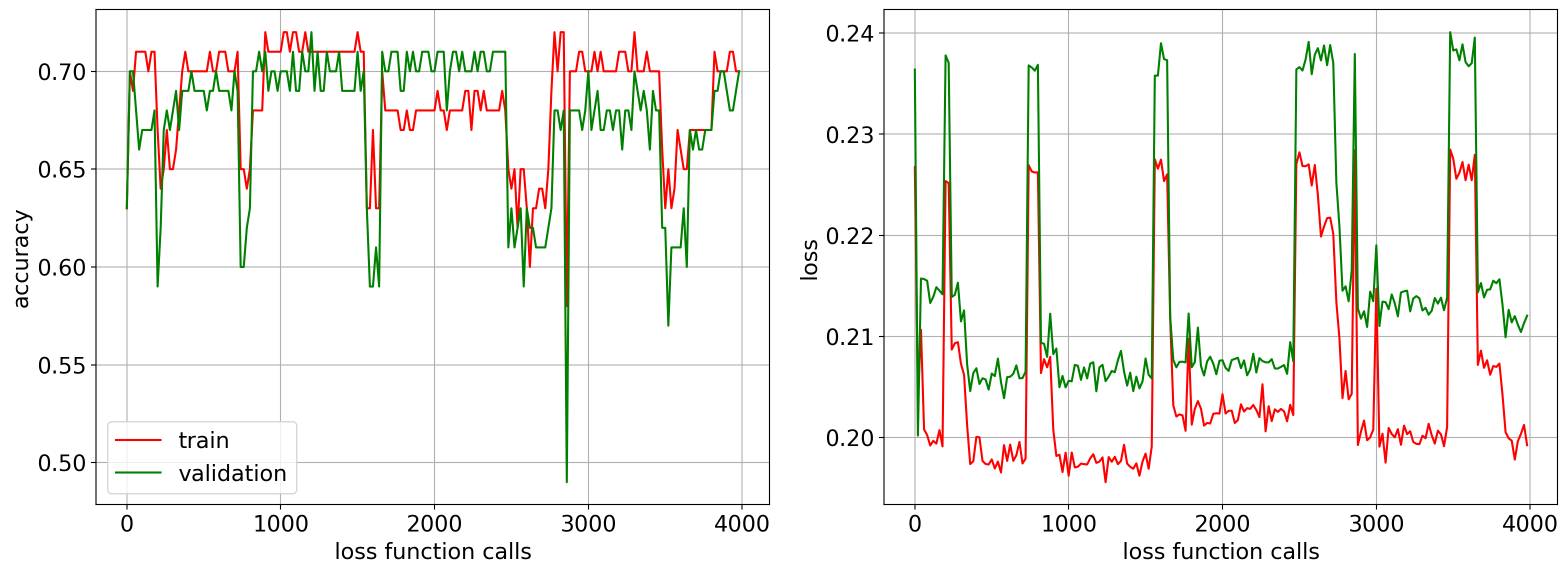}
    \caption{Value of accuracy (on the left) and loss function (on the right) during minimization with COBYLA.
    In red we see values for train set and in green for validation set.
    Both sets have the same size, which is 100. It has been used ``with pile-up" data set.
    The quantum classifier circuit has 10 qubits and 6 layers of Ansatz. The number of shots is 2000.}
    \label{fig:history_loss_jet}
\end{figure}
\subsubsection{Images}
In Appendix~\ref{chap:pic_jet_images32} we can see what happens by increasing the number of layers in the classifier quantum circuit using 32x32 images.
It is evident that only three optimizers are able to train the circuit: Powell, COBYLA, and genetic.

Increasing the size of the train data set, the gap between the training accuracy and test accuracy decreases.

Increasing the number of layers does not help to increase the accuracy. It seems like only the first layer of Ansatz is relevant for the classification.

We also studied how the loss function changes during the minimization process.
We used COBYLA optimizer with a train set of 100 images and a validation set of 100 different images.
Instead of using CZ gates, we used CRZ gates to give the model the ability to maintain the performance of the previous configuration after we have increased the number of layers.
We obtain the plot shown in Figure~\ref{fig:history_loss_jet} for the data set ``with pile-up". The vertical picks correspond to the moment when we increased the number of layers and then the optimizer started exploring the parameters space.
There is no evidence of an improvement in terms of accuracy and loss function when we increase the number of layers.
There is also no evidence of overlearning. This indicated that the model has not enough expressiveness.

We also tried to reduce the size of images from 32x32 to 16x16 (results are shown in Appendix~\ref{chap:pic_jet_images16}).
Results show that when we resize the images we lost too much information and the accuracy decreases. 

Given all these informations, we continue to use only COBYLA optimizer with the ``square" loss function and a train data set of 100 images with original size 32x32.

\begin{table}
    \centering
    \begin{tabular}{c|cc}
        \multirow{2}{*}{Real label} & \multicolumn{2}{c}{Predicted label}\\
        & 0 & 1 \\ \cline{2-3}
        0 & $75.2 \pm 3.0 \% $ & $24.8 \pm 3.0 \% $ \\ 
        1 & $33.6 \pm 3.4 \% $ & $66.4 \pm 3.4 \% $ \\ 
        \hline
    \end{tabular}
    \caption{Confusion matrix for ``without pile-up" data set. The average is computed over 5 different trainings on the validation sets.
    The classifier circuit is a 10-qubits circuit with 6 layers of Ansatz.
    The model is trained with COBYLA optimizer and the ``square" loss function. The number of shots is 2000.}
    \label{fig:confusion_nopile}
\end{table}
\begin{table}
    \centering
    \begin{tabular}{c|cc}
        \multirow{2}{*}{Real label} & \multicolumn{2}{c}{Predicted label}\\
        & 0 & 1 \\ \cline{2-3}
        0 & $67.2 \pm 4.2 \% $ & $32.8 \pm 4.2 \% $ \\ 
        1 & $30.4 \pm 4.0 \% $ & $69.6 \pm 4.0 \% $ \\ 
        \hline
    \end{tabular}
    \caption{Confusion matrix for ``with pile-up" data set. The average is computed over 5 different trainings on the validation sets.
    The classifier circuit is a 10-qubits circuit with 6 layers of Ansatz.
    The model is trained with COBYLA optimizer and the ``square" loss function. The number of shots is 2000.}
    \label{fig:confusion_pile}
\end{table}
\section{Scoring results}
In this section, with the aim of understanding how well we classified jet images, we focus on four metrics often used in machine learning and, in particular, in classification problems.

\textit{Confusion matrix} represents the rate of predicted zeros that are real zeros, the rate of predicted ones that are real zeros, etc.

\textit{ROC curve} shows true positive rate vs. false positive rate. In an ideal situation, we would have only one point at the top left of the plot meaning that we are classifying correctly all the given images.

\textit{AUC score} is the area under the ROC curve. Ideally, it would be 1.

\textit{Test accuracy} is the accuracy of the test set of size 6000.

In order to compute these four metrics we do a \textit{shuffle split}: a given set of 200 images is divided into two sets of the same size, train set, and validation set.
In our case, we do this shuffle split five times. Every time we train the classifier circuit with 6 layers and we compute the confusion matrix, ROC curve, and AUC score on the validation set.
The test accuracy is computed on the test set.

All of these metrics are computed on both data sets: ``without pile-up" and ``with pile-up".
As we wrote in the previous section we use COBYLA optimizer and ``square" loss function.
The number of shots used is 2000.

\subsubsection{Confusion matrices}
What we see by comparing confusion matrices of ``without pile-up" (Table~\ref{fig:confusion_nopile}) and ``with pile-up" (Table~\ref{fig:confusion_pile}), is that we have more elements on the diagonal of the first matrix.
Pile-up is a sort of noise for our training that decreases the ability to classify images.

\begin{figure}
    \centering
    \includegraphics[scale=0.57]{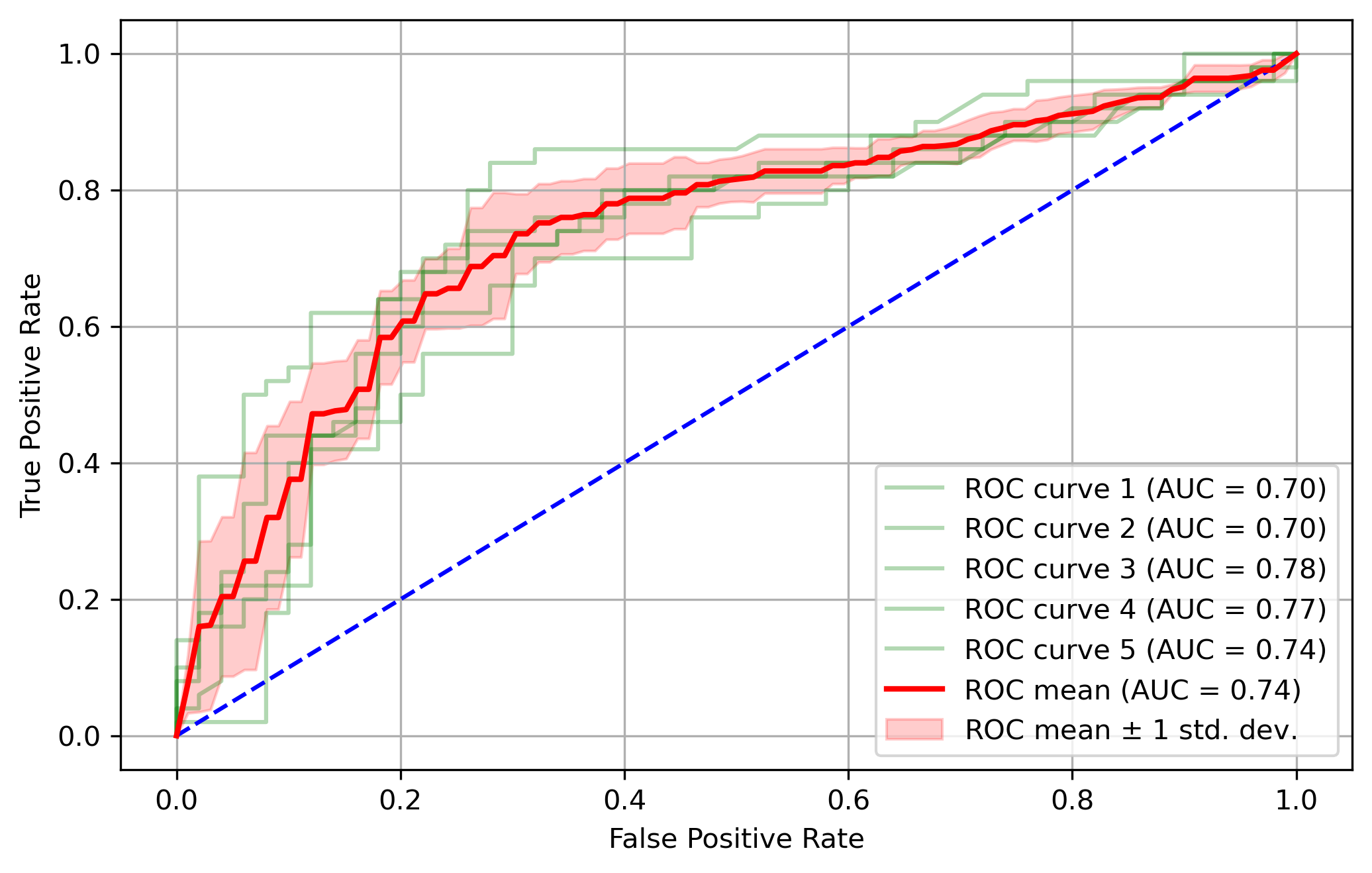}
    \caption{ROC curve and AUC score for the five validation sets in green.
    In red is shown the average ROC and AUC. The data set is ``without pile-up".
    The classifier circuit is a 10-qubits circuit with 6 layers of Ansatz.
    The model is trained with COBYLA optimizer and the ``square" loss function.
    The number of shots is 2000.}
    \label{fig:roc_nopile}
\end{figure}
\begin{figure}
    \centering
    \includegraphics[scale=0.57]{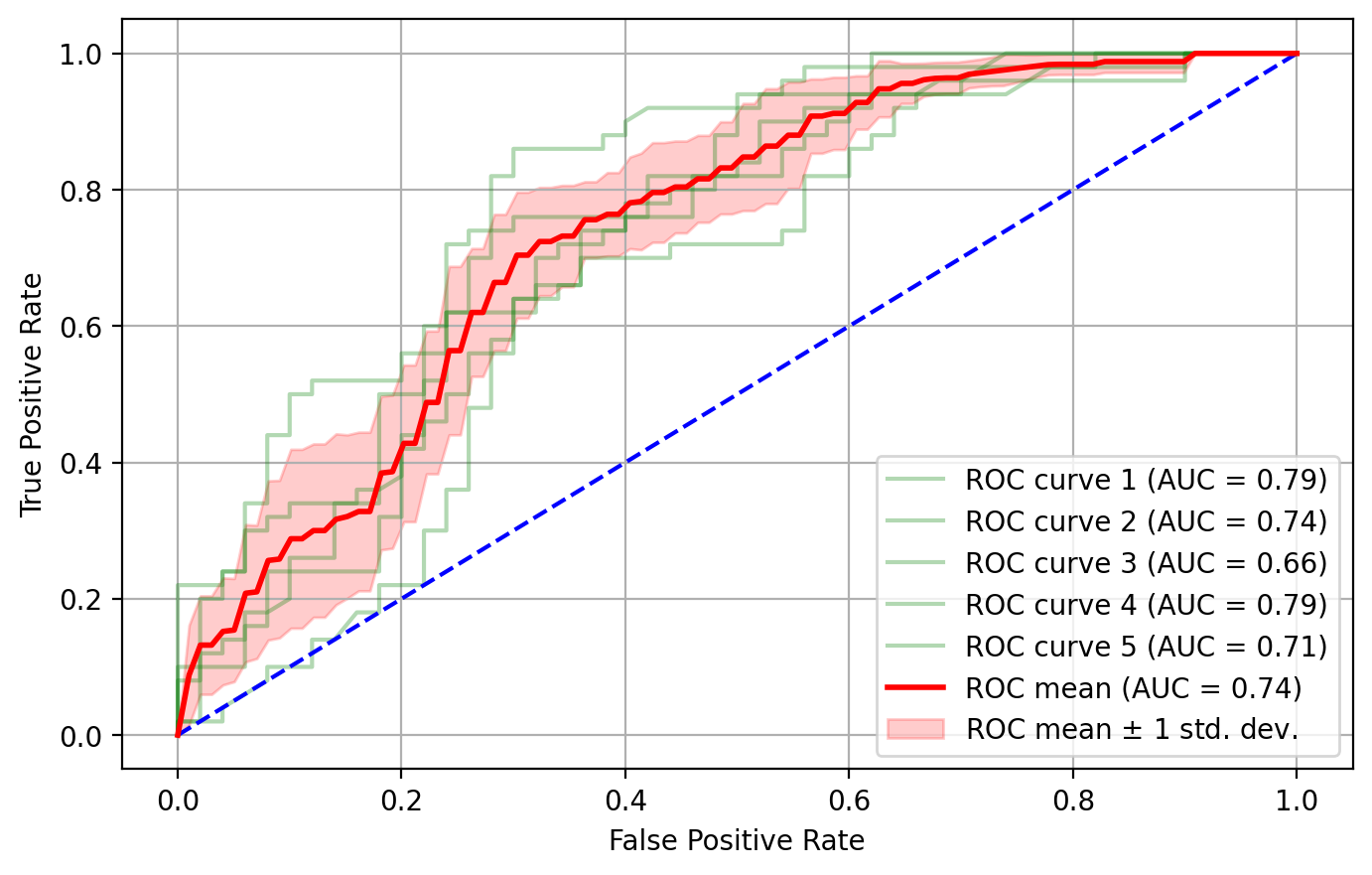}
    \caption{ROC curve and AUC score for the five validation sets in green.
    In red is shown the average ROC and AUC. The data set is ``with pile-up".
    The classifier circuit is a 10-qubits circuit with 6 layers of Ansatz.
    The model is trained with COBYLA optimizer and the ``square" loss function.
    The number of shots is 2000.}
    \label{fig:roc_pile}
\end{figure}
\subsubsection{ROC curve and AUC score}
If we compare ROC curves between the two data sets (Figure~\ref{fig:roc_nopile} and Figure~\ref{fig:roc_pile}), we see that on average the AUC score is the same, but the standard deviation is higher in the case ``with pile-up".
This is strongly correlated to the fact that pile-up is toughness on the classification.
\subsubsection{Test accuracy}
Finally, we show test accuracy in the column "Quantum classifier" of Table~\ref{table:cnn_qml}. The results obtained before with confusion matrices, ROC curve, and AUC score are summarized here.
The accuracy for the data set with pile-up is slightly lower and the standard deviation is higher.
\begin{table}
    \centering
    \begin{tabular}{c|c|c}
        \multicolumn{3}{c}{Test accuracy} \\
        \hline \hline
        Data set & CNN & Quantum classifier \\ 
        \hline
        Without pile-up & $0.688 \pm 0.028$ & $0.757 \pm 0.003$\\ 
        With pile-up & $0.678 \pm 0.019$ & $0.710 \pm 0.010$\\ 
        \hline
    \end{tabular}
    \caption{Comparison between the accuracy of test set computed with a CNN and a quantum classifier.
    Averages are calculated over 5 different trainings.
    The CNN has a convolutional layer with 1 filter of kernel size (2,2) and activation function ``relu",
    a MaxPooling layer of size (4,4), a Flatten layer, and a Dense layer with one final neuron.
    The total number of trainable parameters is 55.
    The network is trained with the ``Adam" optimizer and the loss function is binary cross-entropy.
    The classifier circuit is a 10-qubits circuit with 6 layers of Ansatz.
    The model is trained with COBYLA optimizer and the ``square" loss function.
    The number of shots is 2000.}
    \label{table:cnn_qml}
\end{table}
\section{Comparison with a neural network}

In paper \cite{baldi} the authors used a very wide and deep neural network to classify images.
They tested different networks with dimensions, in terms of the number of trainable parameters, around 1 million.
In particular, they discovered that a CNN (convolutional neural network) works better than a DNN (dense neural network).

Our quantum classifier, in comparison, is smaller: we have only 30 trainable parameters for one layer and 50 parameters for 2 layers.
This means that it is not possible to compare their results with ours.

With the intention of doing a fair comparison, we designed a very small CNN (with only 55 trainable parameters) and we computed accuracy for the test set.
We trained our network with 100 images with the same method described in the previous section.
The CNN is made of the following parts.

$\bullet$\textit{Convolutional layer} with 1 filter of kernel size(2, 2). This means that the output shape is (31, 31, 1). The activation function is ``relu".

$\bullet$\textit{MaxPooling layer} of pool size (4, 4). The output shape is (7, 7, 1).

$\bullet$\textit{Flatten layer} with output shape (49) and a \textit{Dense layer} with one final neuron.

We used as optimizer ``Adam" with binary cross-entropy as loss function.

The results obtained in terms of accuracy of the test set are compared with the ones obtained from the quantum classifier in Table~\ref{table:cnn_qml}.

Despite the results with the neural network are slightly worse, we cannot assert that we have a real gain from the quantum classifier.
On the other hand, we can confirm that is possible to do a classification of images with this kind of quantum circuit, reaching the results of small neural networks.

\chapter{Conclusion}
\label{chap:conclusion}
In this work, we started exploring the possibility of classifying two categories with a quantum circuit.
We designed a circuit with trainable RY rotations and CZ gates to connect two qubits.
We implemented the chance to use different minimizers and we measured performances mostly in terms of accuracy.

As a first test, we used a very known data set in machine learning: the MNIST data set.
We used a reduced version with only handwritten digits of zeros and ones.
We found that Powell and COBYLA minimizers give the model the possibility to reach an accuracy of almost $100\%$.
In particular, training a train set of 30 images, we obtained an accuracy of the test set (100 images) of $99.0 \pm 1.3 \%$.

Then we changed the target of our classification to two different jets generated in high-energy colliders as LHC.
Due to proton-proton in-time interaction known as pile-up, we used two different sets: ``without pile-up" and ``with pile-up".
First, we used 6 high-level variables representing jet proprieties called features.
In this situation, the accuracy of both train set and test set was around $\sim 50\%$.
The main reason behind this result is probably that our model has not enough expressiveness to extrapolate information from that small amount of data.

Next, we changed the data set to images representing energy deposits of the two jets in a calorimeter.
By incrementing the number of layers, we did not see an improvement in terms of accuracy and loss function.
The reason is that the structure of our circuit has not enough flexibility to represent the given images.
One of the indicators of this problem is that we do not have overlearning even if we train the model with 6 layers.
Another possible explanation is that minimizer is not able to solve barren plateaus well.

Eventually, we compared the results with the ones computed with a small CNN.
We present results from the ``without pile-up" data set and from the ``with pile-up" data set in Table~\ref{table:cnn_qml}.

Results from the CNN are slightly worse than the ones obtained with the quantum classifier, but this does not represent a real gain.
On the other hand, we established that this kind of approach is comparable with a small neural network.

What do we need to take the next step forward?
One game-changer of machine learning was the discovery of back-propagation.
In quantum machine learning, we need some kind of algorithm to solve barren plateaus.

Another level that needs to be reached, is to find a new kind of structure of the circuit able to extrapolate information from an image as a CNN does.

To conclude, we can also say that all these results in the context of the quantum classifier are simulated on a classical computer.
We need to push forward quantum computers in terms of the number of qubits and reliability to understand the behavior of this kind of approach on a real quantum computer.

\appendix

\chapter{Common gates}
\label{chap:gates}

In this chapter, we describe common single-qubit gates and two-qubits gates used in quantum circuits.

\section{Single-qubit gates}

\subsubsection{X Gate}
\begin{figure}
    \centering
    \subfloat{
        \Qcircuit @C=1.em @R=.7em{
            & \lstick{|0\rangle} & \gate{X} & \qw & \rstick{|1\rangle}\\
            & \lstick{|1\rangle} & \gate{X} & \qw & \rstick{|0\rangle}
        }
    }
    \caption{Effect of X gate on the states $|0\rangle$ and $|1\rangle$.}
    \label{fig:xgate}
\end{figure}
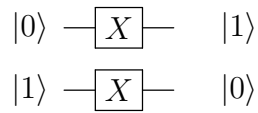

This gate is described by the unitary matrix:
\begin{equation}
    X = 
    \begin{pmatrix}
        0 & 1  \\
        1 & 0
    \end{pmatrix}.
\end{equation}
This behaves like a NOT gate. Infact $|0\rangle$ is mapped to $|1\rangle$ and $|1\rangle$ to $|0\rangle$ as shown in Figure~\ref{fig:xgate}.

\subsubsection{Z Gate}
This gate is described by the unitary matrix:
\begin{equation}
    Z = 
    \begin{pmatrix}
        1 & 0  \\
        0 & -1
    \end{pmatrix}.
\end{equation}
In a circuit we can represent the action on states $|0\rangle$ and $|1\rangle$ as illustrated in Figure~\ref{fig:zgate}.
\begin{figure}
    \centering
    \subfloat{
        \Qcircuit @C=1.em @R=.7em{
            & \lstick{|0\rangle} & \gate{Z} & \qw & \rstick{|0\rangle}\\
            & \lstick{|1\rangle} & \gate{Z} & \qw & \rstick{-|1\rangle}
        }
    }
    \caption{Action of Z gate on the states $|0\rangle$ and $|1\rangle$.}
    \label{fig:zgate}
\end{figure}

\subsubsection{H Gate (Hadamard gate)}
This gate is described by the unitary matrix:
\begin{equation}
    H = \frac{1}{\sqrt{2}}
    \begin{pmatrix}
        1 & 1  \\
        1 & -1
    \end{pmatrix}.
\end{equation}
We can define two new qubits:
\begin{equation}
    \begin{split}
        |+\rangle =& \frac{1}{\sqrt{2}}(|0\rangle+|1\rangle),
        \\
        |-\rangle =& \frac{1}{\sqrt{2}}(|0\rangle-|1\rangle).
    \end{split}
\end{equation}
The action of $H$ gate on the basis $|0\rangle$, $|1\rangle$ is shown in Figure~\ref{fig:hgate}.
\begin{figure}
    \centering
    \subfloat{
        \Qcircuit @C=1.em @R=.7em{
            & \lstick{|0\rangle} & \gate{H} & \qw & \rstick{|+\rangle}\\
            & \lstick{|1\rangle} & \gate{H} & \qw & \rstick{|-\rangle}
        }
    }
    \caption{Action of H gate on states $|0\rangle$ and $|1\rangle$.}
    \label{fig:hgate}
\end{figure}

\subsubsection{Y Gate}
This gate is described by the unitary matrix:
\begin{equation}
    Y =
    \begin{pmatrix}
        0 & -i  \\
        i & 0
    \end{pmatrix}.
\end{equation}
Figure~\ref{fig:ygate} show the circuit representation of the Y gate.
\begin{figure}
    \centering
    \subfloat{
        \Qcircuit @C=1em @R=.7em {
            & \gate{Y} & \qw
        }
    }
    \caption{Circuit representation of Y gate.}
    \label{fig:ygate}
\end{figure}

Note that X, Y, Z gates with identity are also called \textit{Pauli gates} because their representation corresponds to Pauli matrices.
\begin{equation}
    \begin{pmatrix}
        0 & 1  \\
        1 & 0
    \end{pmatrix} ; 
    \begin{pmatrix}
        0 & -i  \\
        i & 0
    \end{pmatrix} ; 
    \begin{pmatrix}
        1 & 0  \\
        0 & -1
    \end{pmatrix} ; 
    \begin{pmatrix}
        1 & 0  \\
        0 & 1
    \end{pmatrix}.
\end{equation}

\subsubsection{Rotation Gates}
\begin{figure}
    \centering
    \subfloat{
        \Qcircuit @C=1em @R=.7em {
            & \gate{RX(\theta)} & \qw\\
            & \gate{RY(\theta)} & \qw\\
            & \gate{RZ(\theta)} & \qw\\
        }
    }
    \caption{Rotation gates illustration in a quantum circuit.}
    \label{fig:rot}
\end{figure}
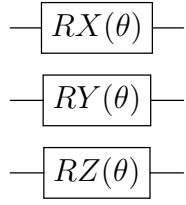

We can define three rotation gates $R_{X}, R_{Y}, R_{Z}$ as follows:
\begin{equation}
    R_{X}(\theta) = \exp{-i\frac{\theta}{2}X} = 
    \begin{pmatrix}
        \cos{\frac{\theta}{2}} & -i\sin{\frac{\theta}{2}}  \\
        -i\sin{\frac{\theta}{2}} & \cos{\frac{\theta}{2}}
    \end{pmatrix},
\end{equation}
\begin{equation}
    R_{Y}(\theta) = \exp{-i\frac{\theta}{2}Y} = 
    \begin{pmatrix}
        \cos{\frac{\theta}{2}} & -\sin{\frac{\theta}{2}}  \\
        \sin{\frac{\theta}{2}} & \cos{\frac{\theta}{2}}
    \end{pmatrix},
\end{equation}
\begin{equation}
    R_{Z}(\theta) = \exp{-i\frac{\theta}{2}Z} = 
    \begin{pmatrix}
        1 & 0  \\
        0 & \exp{i \theta}
    \end{pmatrix}.
\end{equation}
In a quantum circuit, we represent these rotation gates as shown in Figure~\ref{fig:rot}.

It has been shown that exist $\alpha, \beta, \gamma$ which allow us to write any one-qubit unitary gate $U$ as:
\begin{equation}
    U = R_{Z}(\alpha)R_{Y}(\beta)R_{Z}(\gamma).
\end{equation}

\section{Multiple-qubits gates}
Multiple-qubits gates allow connecting qubits. In this section, we are about to introduce four useful two-qubits gates, CNOT, CZ, CRX, and CRZ.
\subsubsection{CNOT Gate}
\begin{figure}
    \centering
    \subfloat{
        \Qcircuit @C=1em @R=.7em {
            & \ctrl{1} & \qw\\
            & \targ & \qw
        }
    }
    \caption{Representation of CNOT gate in a quantum circuit.}
    \label{fig:cnot}
\end{figure}
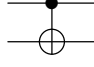

CNOT gate, also known as CX gate, applies an X gate to the second qubit, called \textit{target qubit}, if the first one, called \textit{control qubit}, is $|1\rangle$.
We can represent the gate with the following matrix acting on the basis $|00\rangle, |01\rangle, |10\rangle, |11\rangle$.
\begin{equation}
    CNOT =
    \begin{pmatrix}
        1 & 0 & 0 & 0 \\
        0 & 1 & 0 & 0 \\
        0 & 0 & 0 & 1 \\
        0 & 0 & 1 & 0 \\
    \end{pmatrix}
    .
\end{equation}
Figure~\ref{fig:cnot} illustrates the circuit representation of the CNOT gate.

\subsubsection{CZ Gate}
CZ gate is also called CPhase gate because it changes the phase of the second qubit, controlled by the state of the first qubit.
The matrix representation acting on the basis $|00\rangle, |01\rangle, |10\rangle, |11\rangle$ is:
\begin{equation}
    CZ =
    \begin{pmatrix}
        1 & 0 & 0 & 0 \\
        0 & 1 & 0 & 0 \\
        0 & 0 & 1 & 0 \\
        0 & 0 & 0 & -1 \\
    \end{pmatrix}
    .
\end{equation}
This gate is also reversible on the controlled qubit and the target qubit. Then we can represent it in a circuit as in Figure~\ref{fig:cz}:
\begin{figure}
    \centering
    \subfloat{
        \Qcircuit @C=1em @R=.7em {
            & \ctrl{1} & \qw\\
            & \ctrl{0}& \qw
        }
    }
    \caption{Representation of CZ gate in a quantum circuit.}
    \label{fig:cz}
\end{figure}
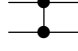

\subsubsection{CRX gate}
CRX gate is a controlled RX rotation of the second qubit by the first qubit.
The unitary matrix representation is:
\begin{equation}
    CRX =
    \begin{pmatrix}
        1 & 0 & 0 & 0 \\
        0 & 1 & 0 & 0 \\
        0 & 0 & \cos{\frac{\theta}{2}} & -\sin{\frac{\theta}{2}} \\
        0 & 0 & \sin{\frac{\theta}{2}} & \cos{\frac{\theta}{2}} \\
    \end{pmatrix}
    .
\end{equation}

\subsubsection{CRZ gate}
CRZ gate is a controlled RZ rotation of the second qubit by the first qubit.
The unitary matrix representation is:
\begin{equation}
    CRZ =
    \begin{pmatrix}
        1 & 0 & 0 & 0 \\
        0 & 1 & 0 & 0 \\
        0 & 0 & \exp{-i\frac{\theta}{2}} & 0 \\
        0 & 0 & 0 & \exp{i\frac{\theta}{2}} \\
    \end{pmatrix}
    .
\end{equation}

\newpage
\section{How to create Bell states with Qibo}
\label{sec:bell}
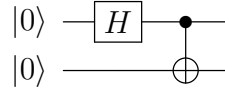
\begin{figure}
    \centering
    \subfloat{
        \Qcircuit @C=1em @R=.5em {
            & \lstick{|0\rangle} & \gate{H} & \ctrl{1} & \qw\\
            & \lstick{|0\rangle} & \qw & \targ & \qw
        }
    }
    \caption{Quantum Circuit necessary to create the first Bell state.}
    \label{fig:bell1}
\end{figure}
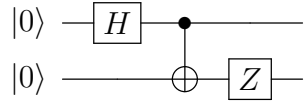
\begin{figure}
    \centering
    \subfloat{
        \Qcircuit @C=1em @R=.5em {
            & \lstick{|0\rangle} & \gate{H} & \ctrl{1} & \qw & \qw\\
            & \lstick{|0\rangle} & \qw & \targ & \gate{Z} & \qw
        }
    }
    \caption{Quantum circuit necessary to create the second Bell state.}
    \label{fig:bell2}
\end{figure}
In this section, we show how to write a circuit that prepares the four Bell states (Equation~\ref{bell_eq}).
To create the first Bell state we can use the following procedure illustrated in Figure~\ref{fig:bell1}:
\begin{enumerate}
    \item Prepare the system in the state $|00\rangle$.
    \item Apply Hadamard gate to the first qubit, obtaining the state $\frac{1}{\sqrt{2}}(|00\rangle + |10\rangle)$.
    \item Apply CNOT gate with the first qubit as control qubit and the second qubit as target qubit.
\end{enumerate}
To do that in Qibo we need to import from Qibo modules \textit{Circuit} and \textit{gates}.
\begin{lstlisting}
    from qibo.models import Circuit
    from qibo import gates
\end{lstlisting}
Then we need to define a circuit of 2 qubits.
\begin{lstlisting}
    circ = Circuit(2)
\end{lstlisting}
We add now the gates: H and CNOT.
\begin{lstlisting}
    circ.add(gates.H(0))
    circ.add(gates.CNOT(0, 1))
\end{lstlisting}
In the end we add a measure gate to both qubits.
\begin{lstlisting}
    circ.add(gates.M(0, 1))
\end{lstlisting}

If we created the second Bell state, we would apply the following gates to the first Bell state as shown in Figure~\ref{fig:bell2}:
\begin{lstlisting}
    circ.add(gates.Z(1))
\end{lstlisting}

If we created the third Bell state, we would apply the following gates to the first Bell state as illustrated in Figure~\ref{fig:bell3}:
\begin{lstlisting}
    circ.add(gates.X(1))
\end{lstlisting}
\begin{figure}
    \centering
    \subfloat{
        \Qcircuit @C=1em @R=.5em {
            & \lstick{|0\rangle} & \gate{H} & \ctrl{1} & \qw & \qw\\
            & \lstick{|0\rangle} & \qw & \targ & \gate{X} & \qw
        }
    }
    \caption{Quantum circuit necessary to create the third Bell state.}
    \label{fig:bell3}
\end{figure}
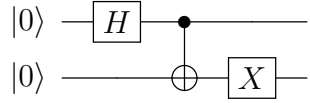

Eventually, if we created the fourth Bell state, we would apply the following gates to the first Bell state as shown in Figure~\ref{fig:bell4}:
\begin{lstlisting}
    circ.add(gates.X(1))
    circ.add(gates.Z(1))
\end{lstlisting}
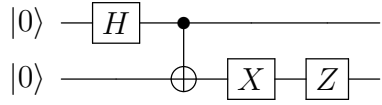
\begin{figure}
    \centering
    \subfloat{
        \Qcircuit @C=1em @R=.5em {
            & \lstick{|0\rangle} & \gate{H} & \ctrl{1} & \qw & \qw & \qw\\
            & \lstick{|0\rangle} & \qw & \targ & \gate{X} & \gate{Z} & \qw
        }
    }
    \caption{Quantum circuit necessary to create the fourth Bell state.}
    \label{fig:bell4}
\end{figure}
We can execute the circuit multiple times, for example, 1000, and obtain the frequencies of each state by writing the following code.
\begin{lstlisting}
    results = circ(nshots=1000)
    results.frequencies()
\end{lstlisting}
If we would do this operation to the first Bell state we would find $\sim 50\%$ of the time the state $|00\rangle$ and $\sim 50\%$ of the time the state $|11\rangle$.
\chapter{Results from training of jet images of size 32x32}
\label{chap:pic_jet_images32}

In this Appendix, we show the accuracies obtained for the train set and the test set with the quantum classifier
by increasing the size of the train set, the number of layers, and by changing the minimizer and the loss function.
For the training process, we use original images (of size 32x32) from both data sets, with and without pile-up.
The quantum classifier circuit has 10 qubits and it is illustrated in Figure~\ref{fig:circuit_jet_images}.

\begin{figure}
    \centering
    \includegraphics[scale=0.23]{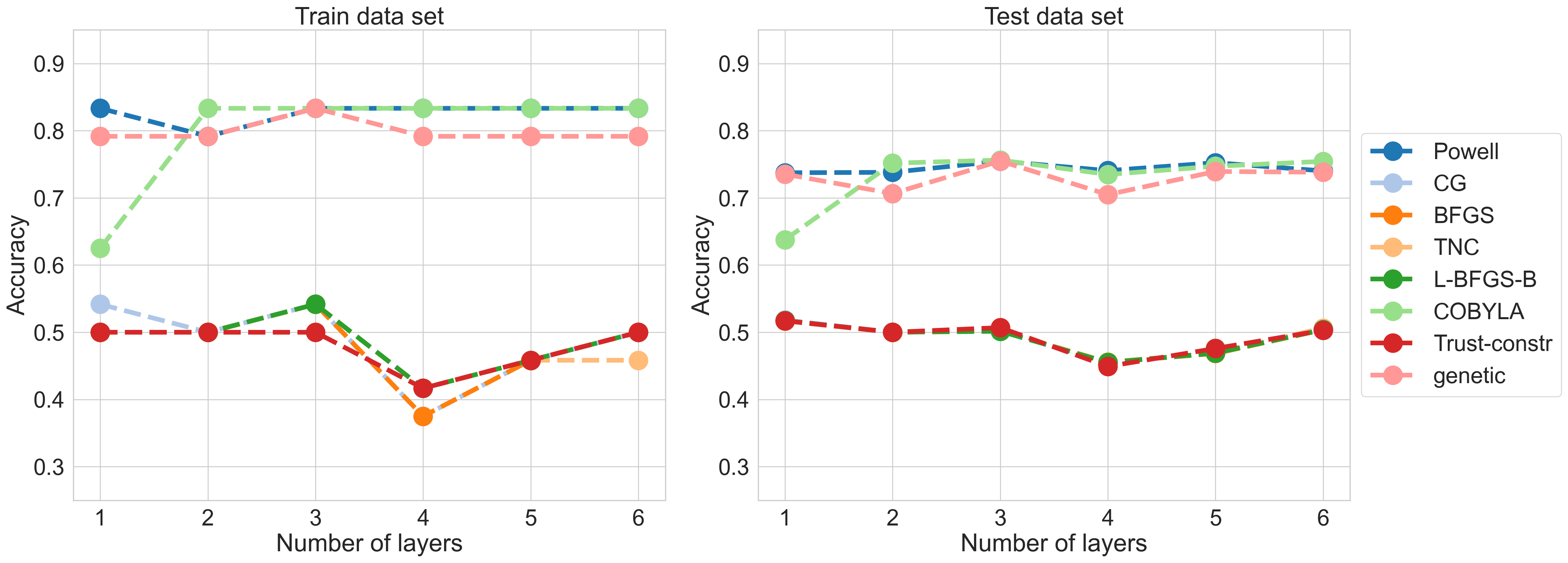}
    \caption{The plots show the accuracy of the train set on the left and the accuracy of the test set on the right as a function of the number of layers and minimizer used.
    The quantum classifier is a 10-qubits circuit for the classification of 32x32 images.
    The data set is the ``without pile-up" data set and the size of the train set is 25.
    The size of the test set is 6000.
    The loss function is ``square". The number of shots is 2000.}
    \label{fig:noresize_squarescipy_jet_25_pileFalse}
\end{figure}

\begin{figure}
    \centering
    \includegraphics[scale=0.23]{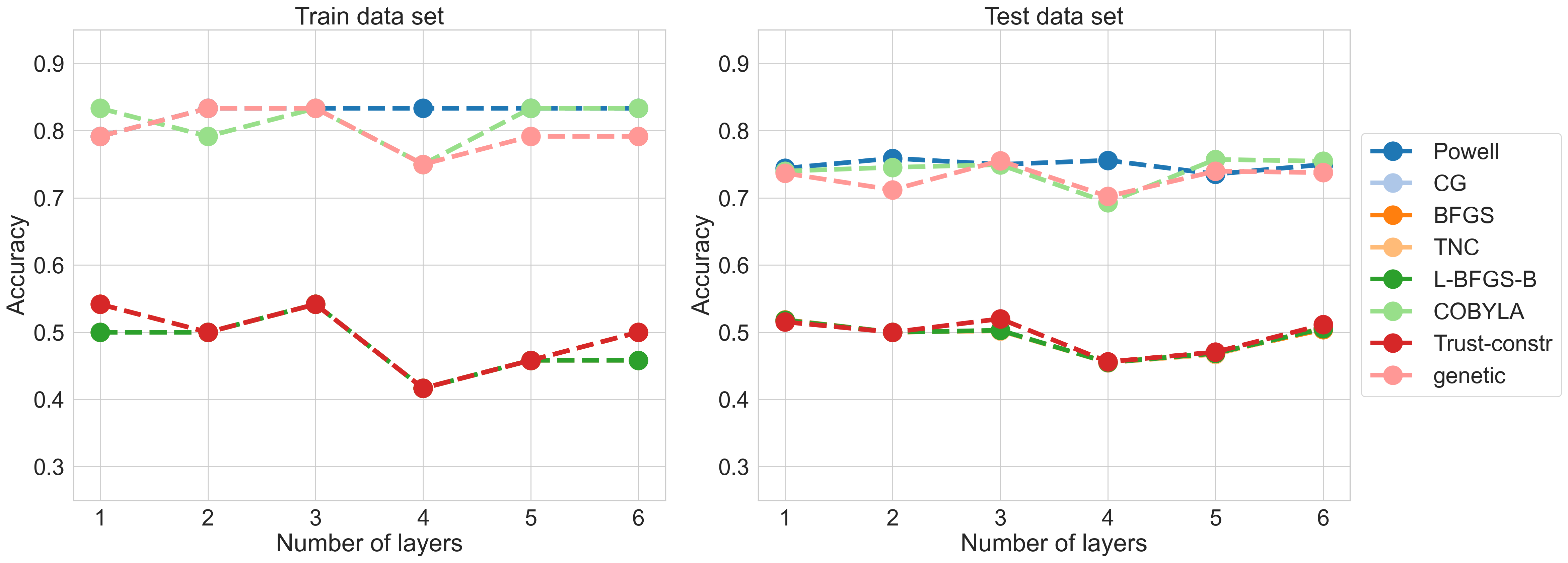}
    \caption{The plots show the accuracy of the train set on the left and the accuracy of the test set on the right as a function of the number of layers and minimizer used.
    The quantum classifier is a 10-qubits circuit for the classification of 32x32 images.
    The data set is the ``without pile-up" data set and the size of the train set is 25.
    The size of the test set is 6000.
    The loss function is ``ce". The number of shots is 2000.}
    \label{fig:noresize_cescipy_jet_25_pileFalse}
\end{figure}

\begin{figure}
    \centering
    \includegraphics[scale=0.23]{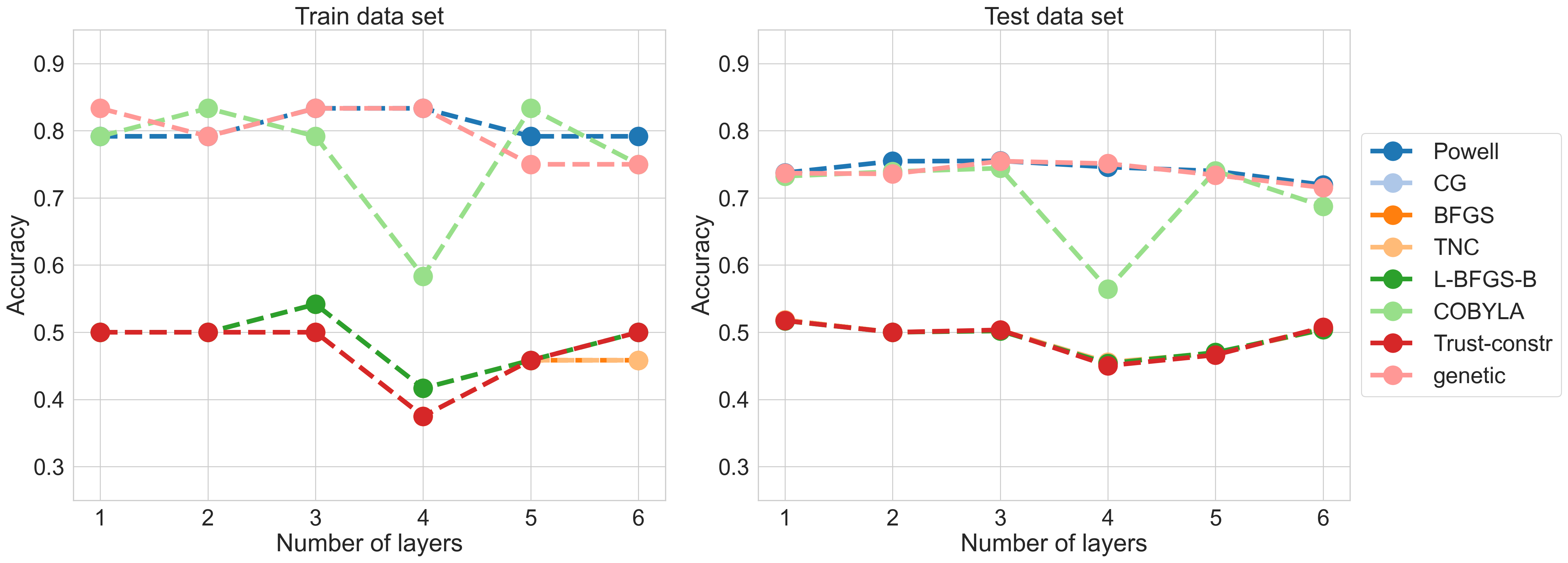}
    \caption{The plots show the accuracy of the train set on the left and the accuracy of the test set on the right as a function of the number of layers and minimizer used.
    The quantum classifier is a 10-qubits circuit for the classification of 32x32 images.
    The data set is the ``without pile-up" data set and the size of the train set is 25.
    The size of the test set is 6000.
    The loss function is ``lce". The number of shots is 2000.}
    \label{fig:noresize_lcescipy_jet_25_pileFalse}
\end{figure}

\begin{figure}
    \centering
    \includegraphics[scale=0.23]{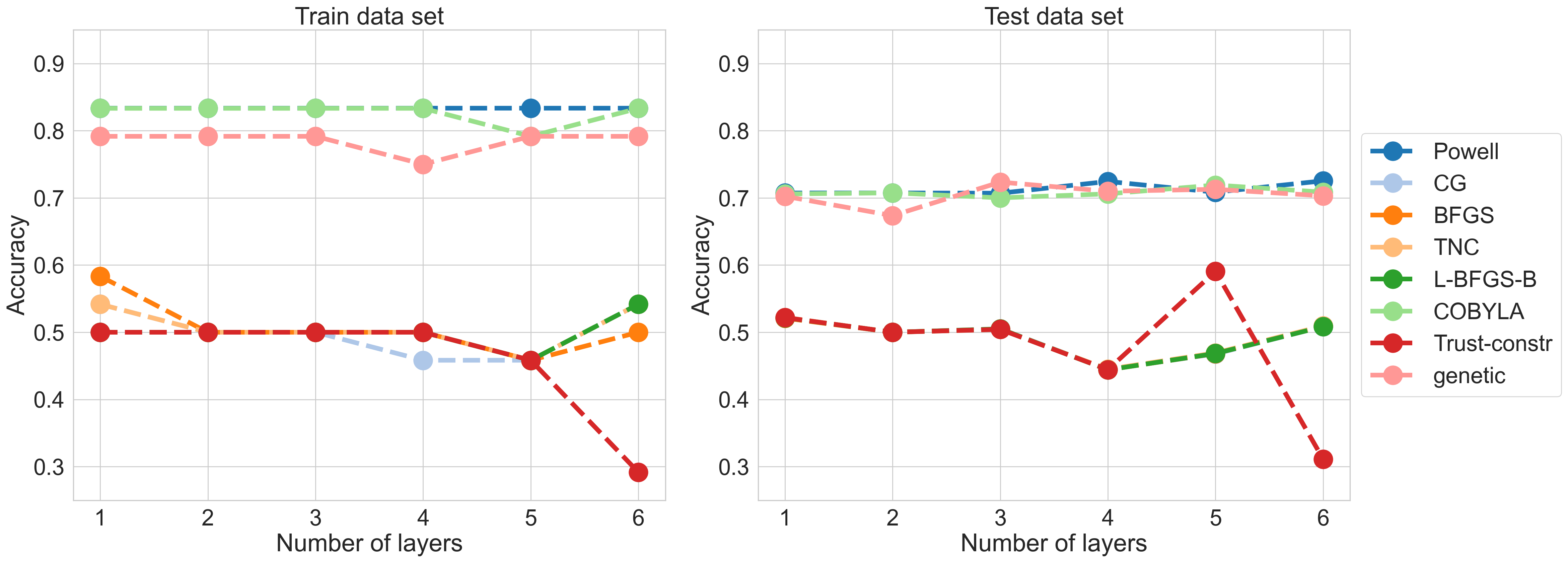}
    \caption{The plots show the accuracy of the train set on the left and the accuracy of the test set on the right as a function of the number of layers and minimizer used.
    The quantum classifier is a 10-qubits circuit for the classification of 32x32 images.
    The data set is the ``with pile-up" data set and the size of the train set is 25.
    The size of the test set is 6000.
    The loss function is ``square". The number of shots is 2000.}
    \label{fig:noresize_squarescipy_jet_25_pileTrue}
\end{figure}

\begin{figure}
    \centering
    \includegraphics[scale=0.23]{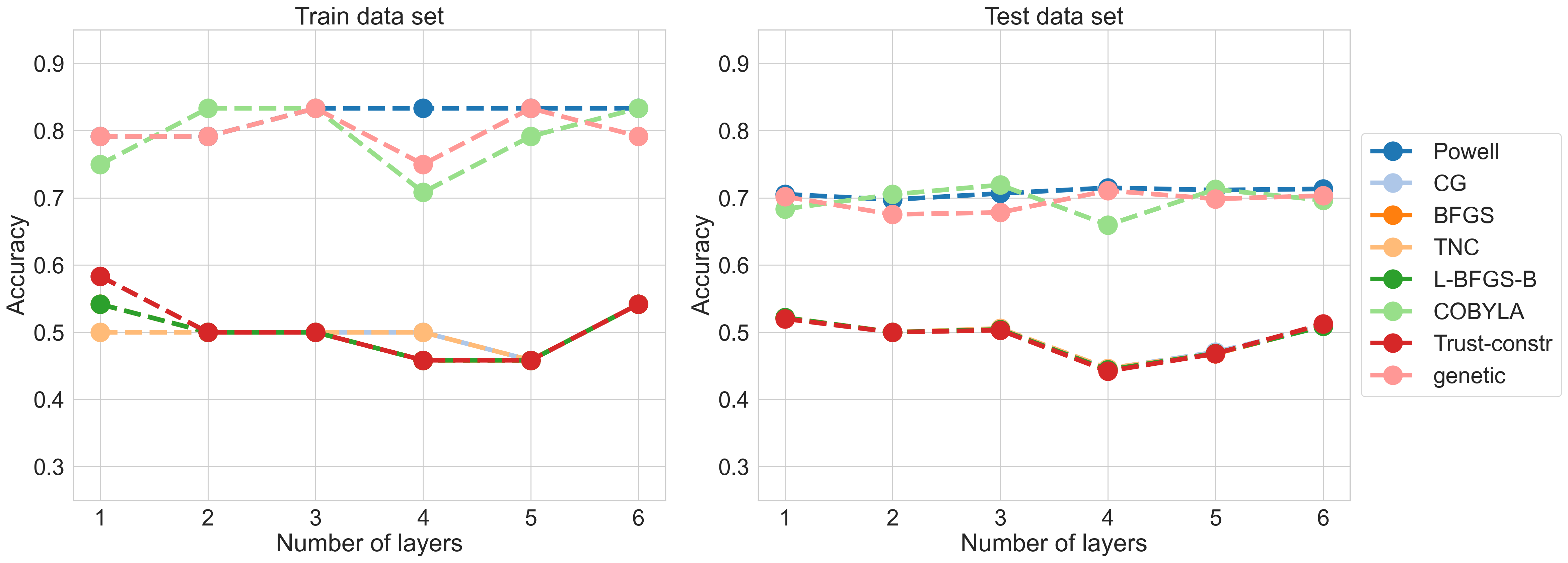}
    \caption{The plots show the accuracy of the train set on the left and the accuracy of the test set on the right as a function of the number of layers and minimizer used.
    The quantum classifier is a 10-qubits circuit for the classification of 32x32 images.
    The data set is the ``with pile-up" data set and the size of the train set is 25.
    The size of the test set is 6000.
    The loss function is ``ce". The number of shots is 2000.}
    \label{fig:noresize_cescipy_jet_25_pileTrue}
\end{figure}

\begin{figure}
    \centering
    \includegraphics[scale=0.23]{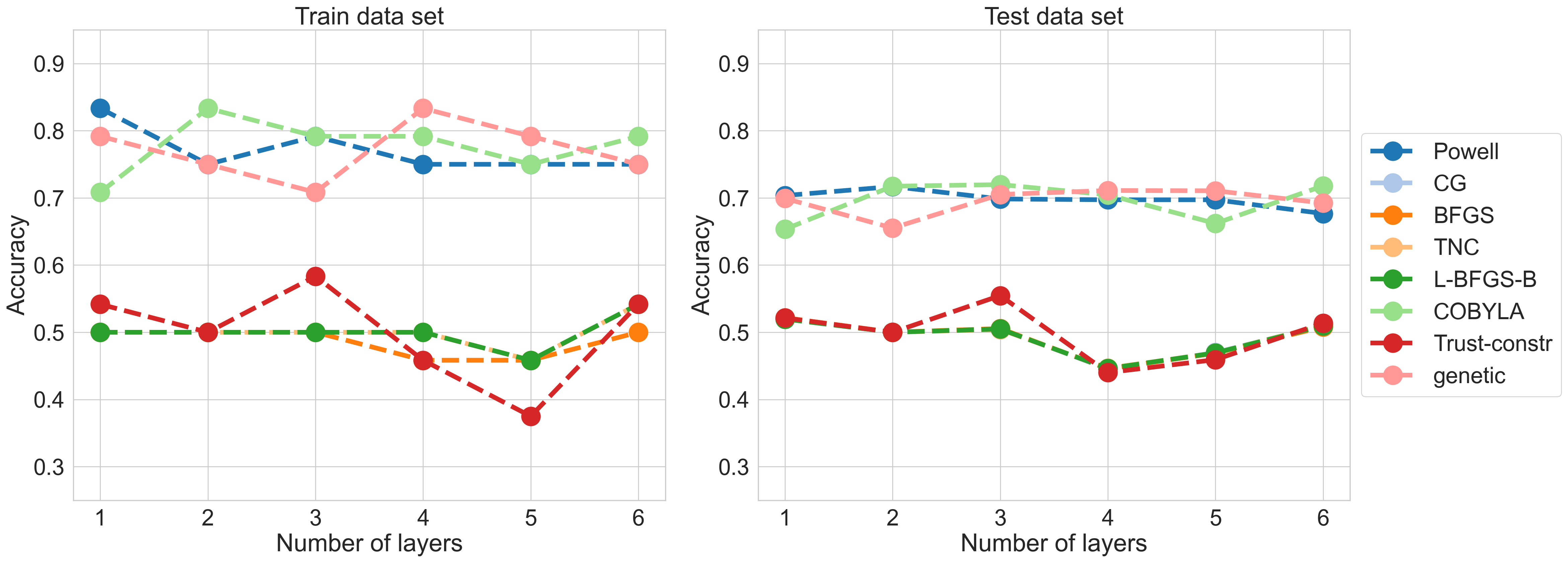}
    \caption{The plots show the accuracy of the train set on the left and the accuracy of the test set on the right as a function of the number of layers and minimizer used.
    The quantum classifier is a 10-qubits circuit for the classification of 32x32 images.
    The data set is the ``with pile-up" data set and the size of the train set is 25.
    The size of the test set is 6000.
    The loss function is ``lce". The number of shots is 2000.}
    \label{fig:noresize_lcescipy_jet_25_pileTrue}
\end{figure}

Firstly, we use for the train set a group of 25 images from the ``without pile-up" data set.
Figure~\ref{fig:noresize_squarescipy_jet_25_pileFalse} show how the accuracy changes by increasing the number of layers for different minimizers with ``square" loss function.
Figures~\ref{fig:noresize_cescipy_jet_25_pileFalse} ~\ref{fig:noresize_lcescipy_jet_25_pileFalse} show the same for ``ce" and ``lce" loss functions.
We show the same plots for the ``with pile-up" data set in Figures~\ref{fig:noresize_squarescipy_jet_25_pileTrue} ~\ref{fig:noresize_cescipy_jet_25_pileTrue} ~\ref{fig:noresize_lcescipy_jet_25_pileTrue}.

\begin{figure}
    \centering
    \includegraphics[scale=0.23]{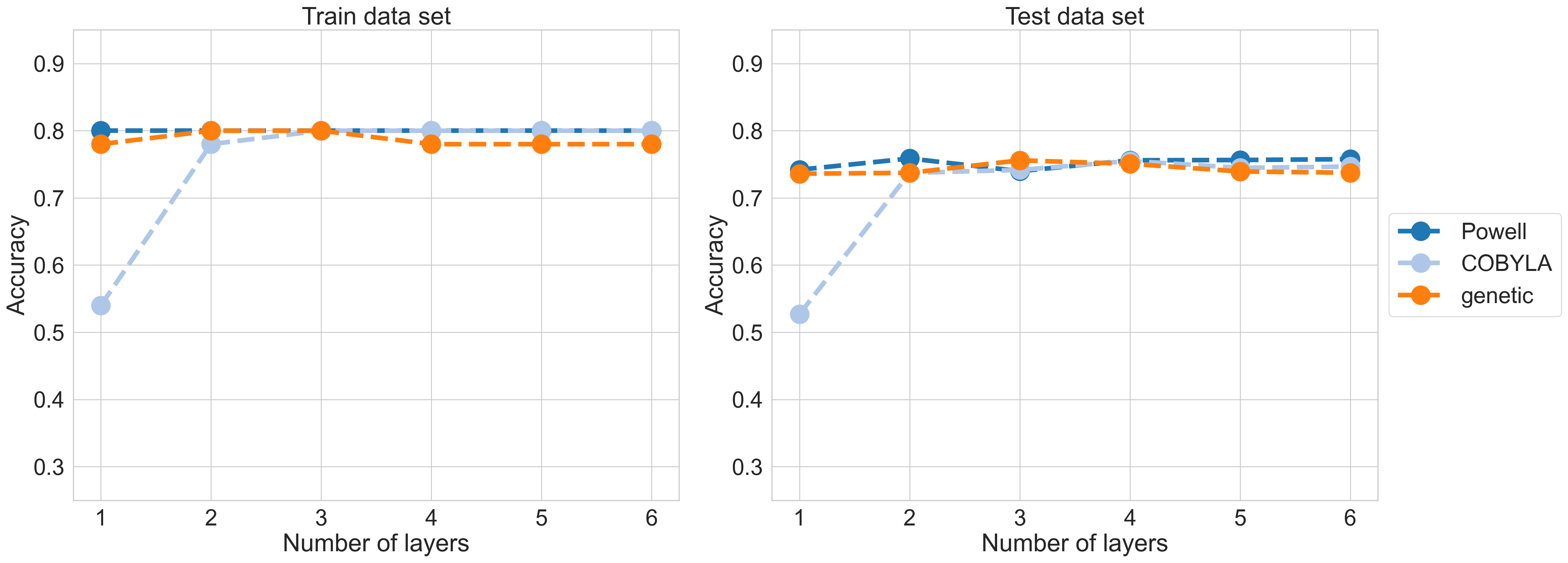}
    \caption{The plots show the accuracy of the train set on the left and the accuracy of the test set on the right as a function of the number of layers and minimizer used.
    The quantum classifier is a 10-qubits circuit for the classification of 32x32 images.
    The data set is the ``without pile-up" data set and the size of the train set is 50.
    The size of the test set is 6000.
    The loss function is ``square". The number of shots is 2000.}
    \label{fig:noresize_squarescipy_jet_50_pileFalse}
\end{figure}

\begin{figure}
    \centering
    \includegraphics[scale=0.23]{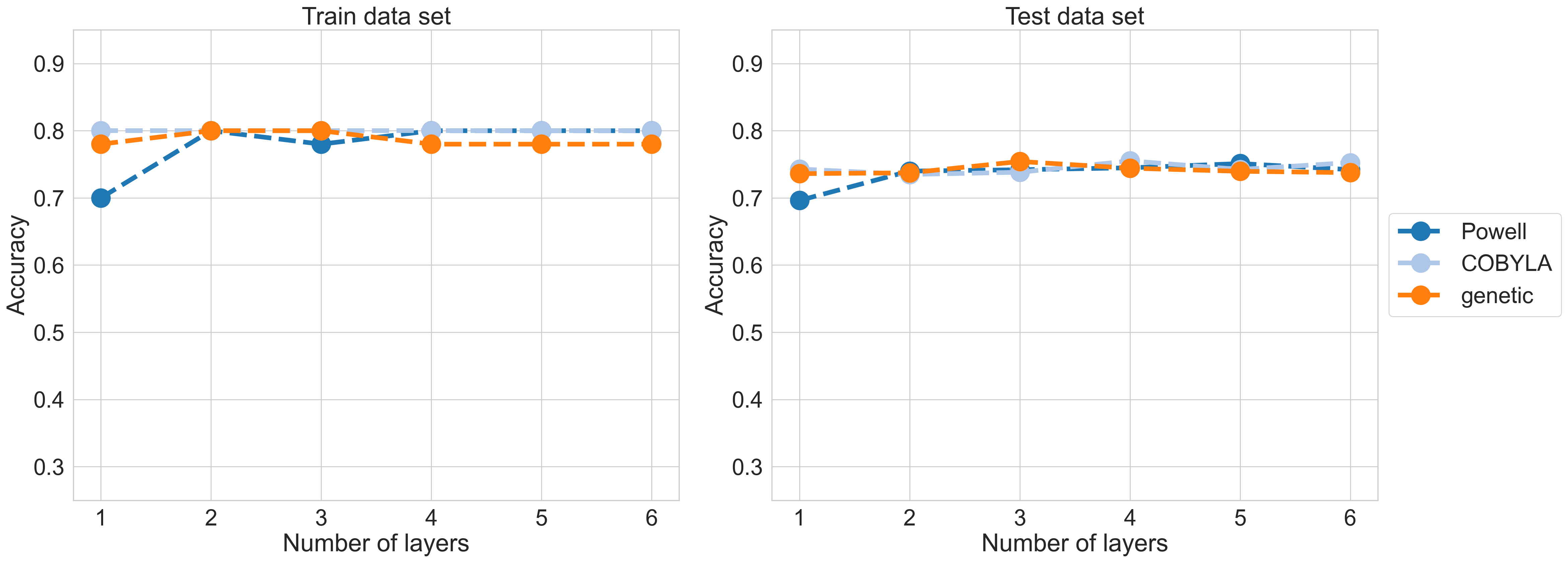}
    \caption{The plots show the accuracy of the train set on the left and the accuracy of the test set on the right as a function of the number of layers and minimizer used.
    The quantum classifier is a 10-qubits circuit for the classification of 32x32 images.
    The data set is the ``without pile-up" data set and the size of the train set is 50.
    The size of the test set is 6000.
    The loss function is ``ce". The number of shots is 2000.}
    \label{fig:noresize_cescipy_jet_50_pileFalse}
\end{figure}

\begin{figure}
    \centering
    \includegraphics[scale=0.23]{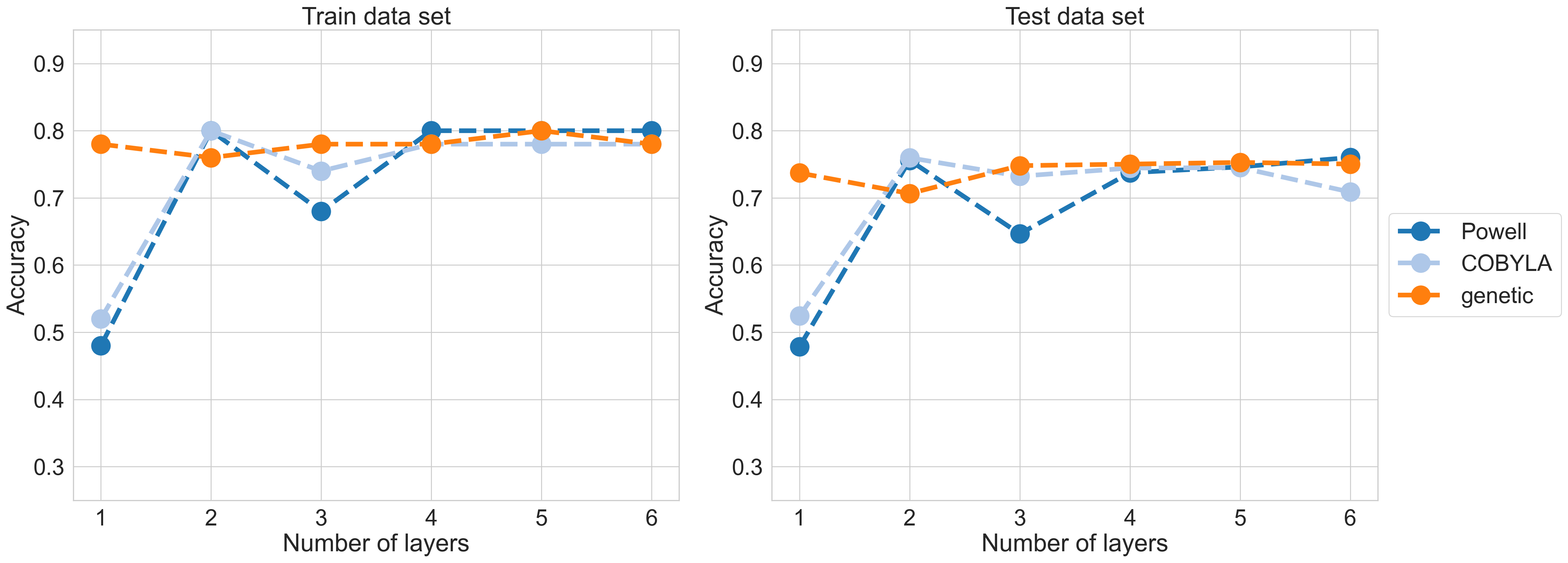}
    \caption{The plots show the accuracy of the train set on the left and the accuracy of the test set on the right as a function of the number of layers and minimizer used.
    The quantum classifier is a 10-qubits circuit for the classification of 32x32 images.
    The data set is the ``without pile-up" data set and the size of the train set is 50.
    The size of the test set is 6000.
    The loss function is ``lce". The number of shots is 2000.}
    \label{fig:noresize_lcescipy_jet_50_pileFalse}
\end{figure}

\begin{figure}
    \centering
    \includegraphics[scale=0.23]{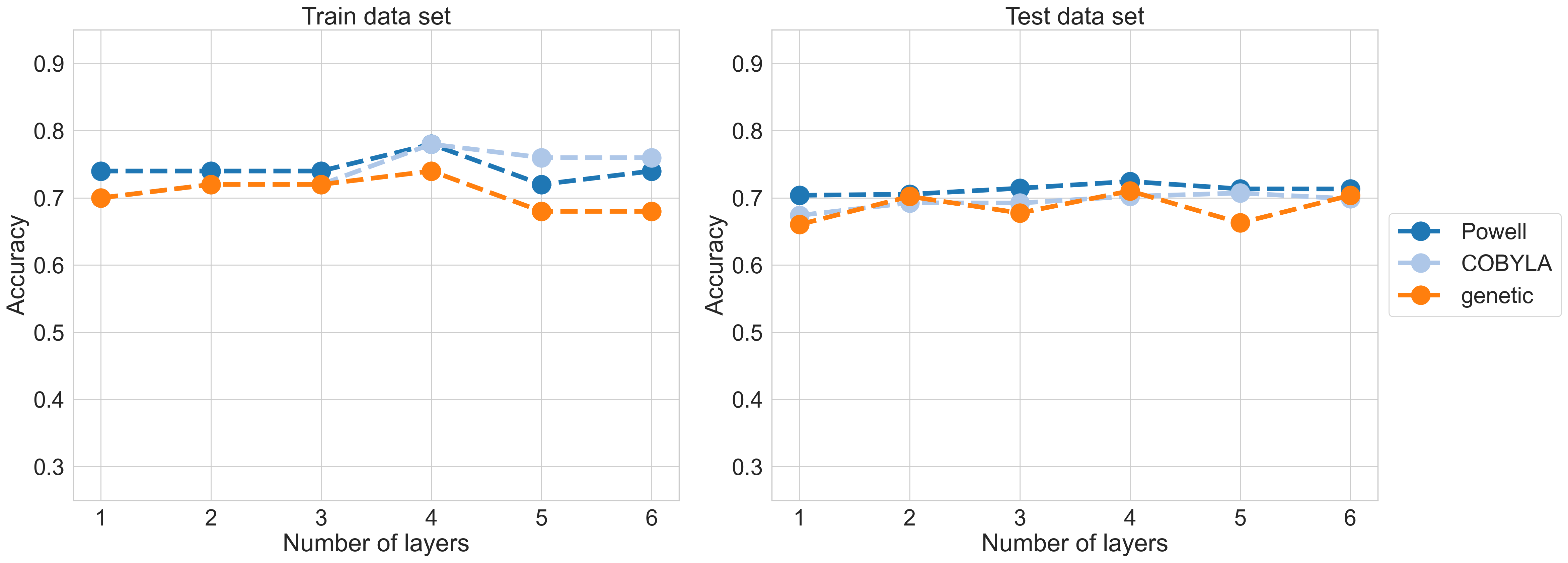}
    \caption{The plots show the accuracy of the train set on the left and the accuracy of the test set on the right as a function of the number of layers and minimizer used.
    The quantum classifier is a 10-qubits circuit for the classification of 32x32 images.
    The data set is the ``with pile-up" data set and the size of the train set is 50.
    The size of the test set is 6000.
    The loss function is ``square". The number of shots is 2000.}
    \label{fig:noresize_squarescipy_jet_50_pileTrue}
\end{figure}

\begin{figure}
    \centering
    \includegraphics[scale=0.23]{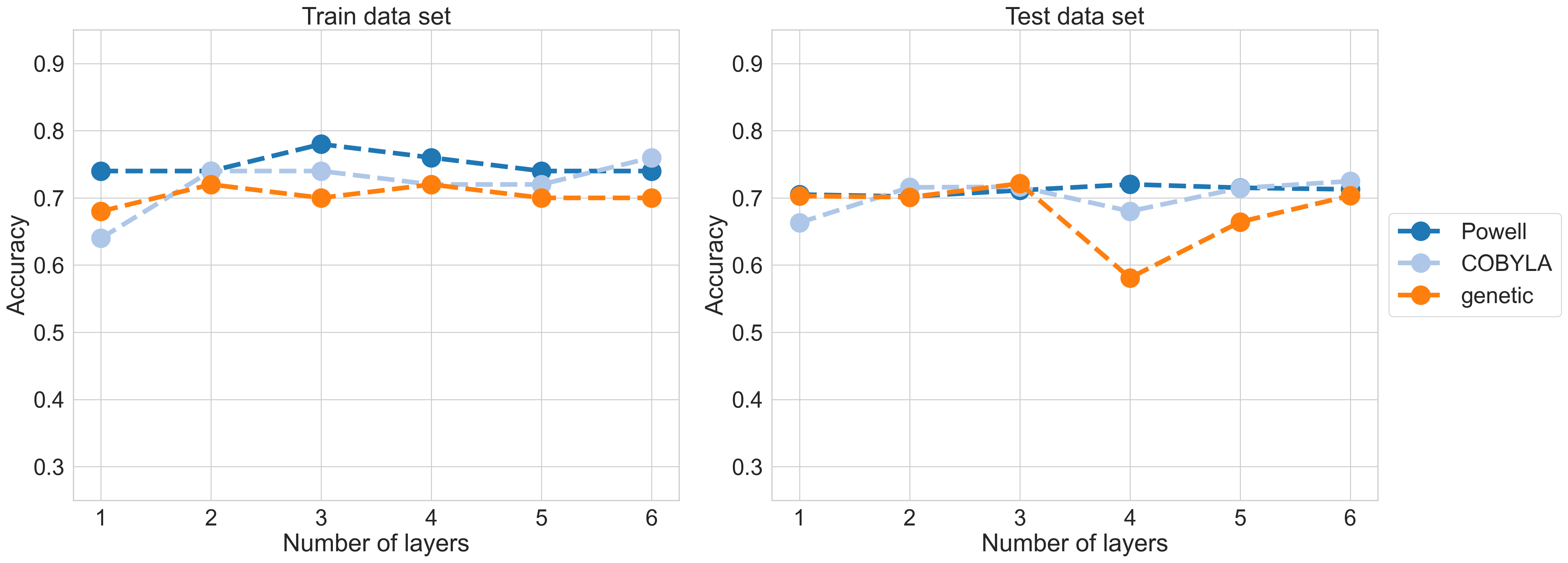}
    \caption{The plots show the accuracy of the train set on the left and the accuracy of the test set on the right as a function of the number of layers and minimizer used.
    The quantum classifier is a 10-qubits circuit for the classification of 32x32 images.
    The data set is the ``with pile-up" data set and the size of the train set is 50.
    The size of the test set is 6000.
    The loss function is ``ce". The number of shots is 2000.}
    \label{fig:noresize_cescipy_jet_50_pileTrue}
\end{figure}

\begin{figure}
    \centering
    \includegraphics[scale=0.23]{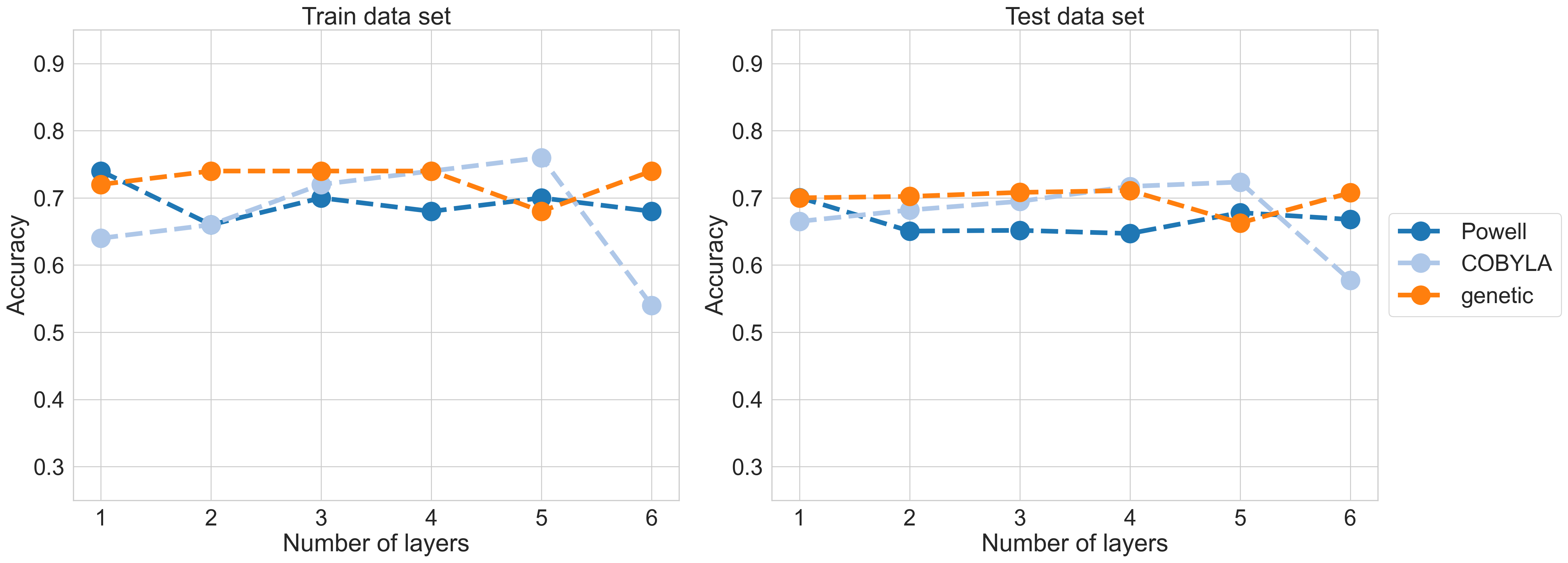}
    \caption{The plots show the accuracy of the train set on the left and the accuracy of the test set on the right as a function of the number of layers and minimizer used.
    The quantum classifier is a 10-qubits circuit for the classification of 32x32 images.
    The data set is the ``with pile-up" data set and the size of the train set is 50.
    The size of the test set is 6000.
    The loss function is ``lce". The number of shots is 2000.}
    \label{fig:noresize_lcescipy_jet_50_pileTrue}
\end{figure}

Then we increase the number of images in the train set from 25 to 50.
In these cases, we use only the best minimizers selected using 25 images as train set: Powell, COBYLA, and genetic.
Results are shown in Figures~\ref{fig:noresize_squarescipy_jet_50_pileFalse} ~\ref{fig:noresize_cescipy_jet_50_pileFalse} ~\ref{fig:noresize_lcescipy_jet_50_pileFalse} for the ``without pile-up" data set
and in Figures~\ref{fig:noresize_squarescipy_jet_50_pileTrue} ~\ref{fig:noresize_cescipy_jet_50_pileTrue} ~\ref{fig:noresize_lcescipy_jet_50_pileTrue} for the ``with pile-up" data set.

\begin{figure}
    \centering
    \includegraphics[scale=0.23]{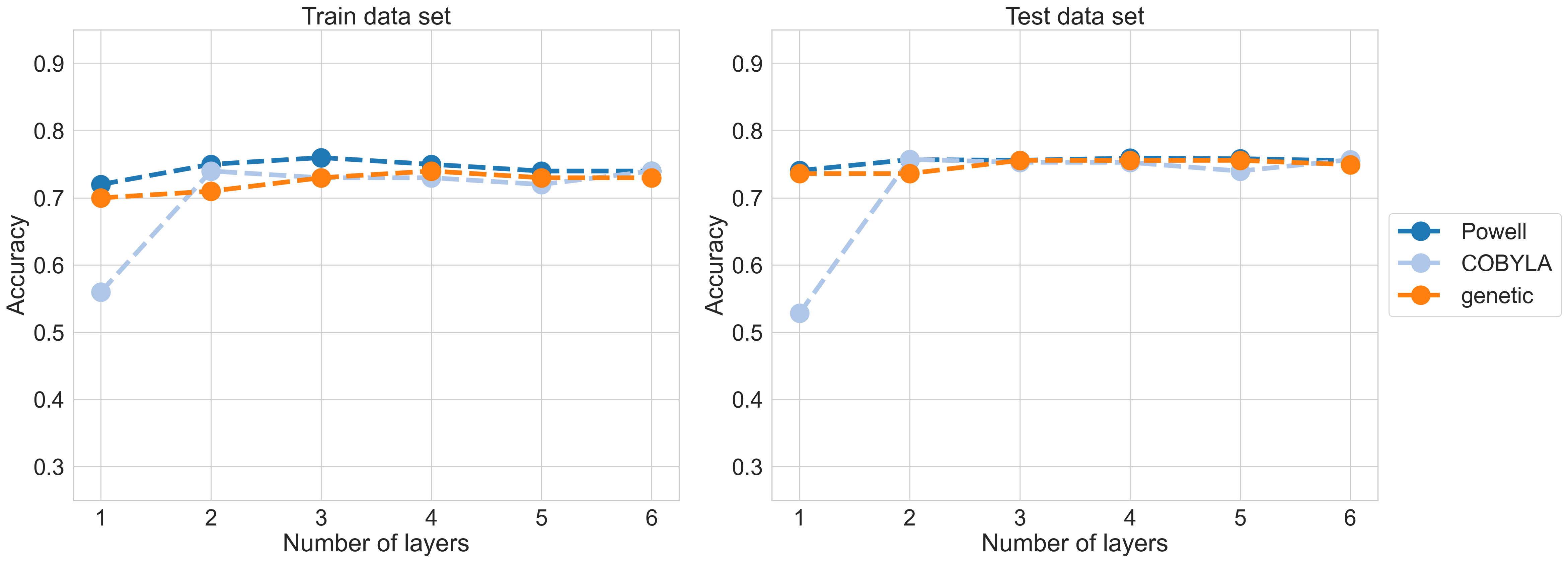}
    \caption{The plots show the accuracy of the train set on the left and the accuracy of the test set on the right as a function of the number of layers and minimizer used.
    The quantum classifier is a 10-qubits circuit for the classification of 32x32 images.
    The data set is the ``without pile-up" data set and the size of the train set is 100.
    The size of the test set is 6000.
    The loss function is ``square". The number of shots is 2000.}
    \label{fig:noresize_squarescipy_jet_100_pileFalse}
\end{figure}

\begin{figure}
    \centering
    \includegraphics[scale=0.23]{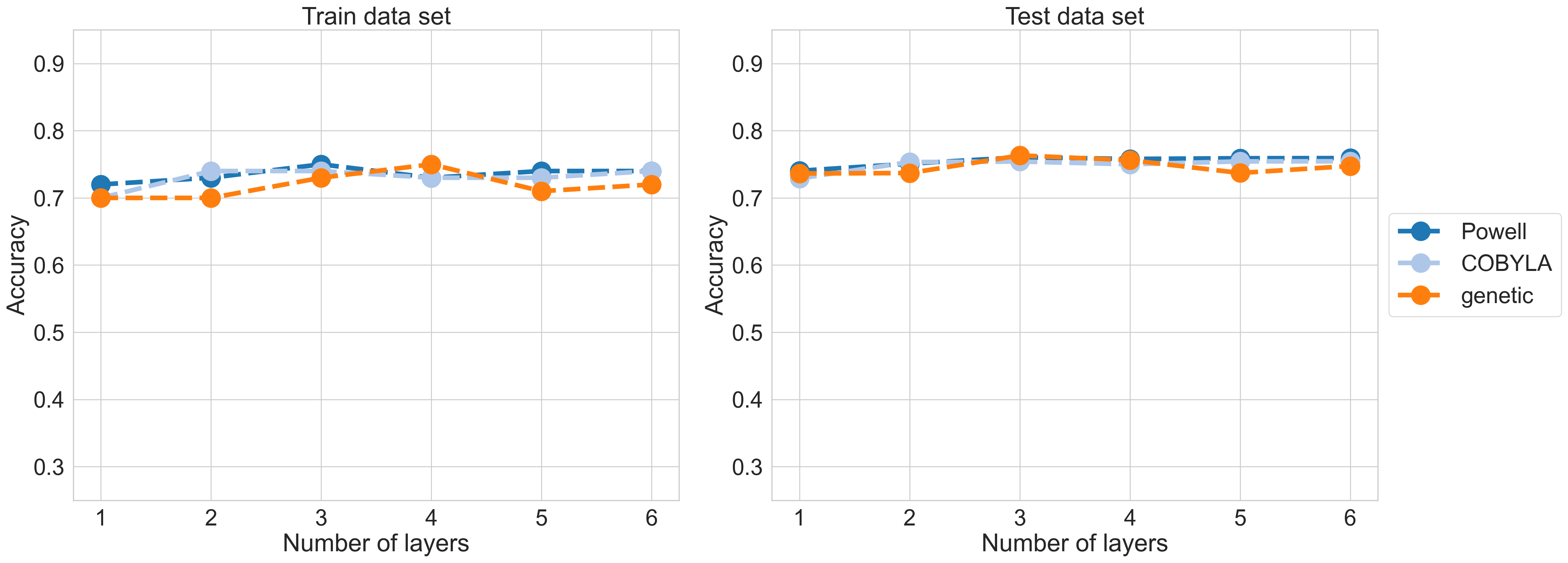}
    \caption{The plots show the accuracy of the train set on the left and the accuracy of the test set on the right as a function of the number of layers and minimizer used.
    The quantum classifier is a 10-qubits circuit for the classification of 32x32 images.
    The data set is the ``without pile-up" data set and the size of the train set is 100.
    The size of the test set is 6000.
    The loss function is ``ce". The number of shots is 2000.}
    \label{fig:noresize_cescipy_jet_100_pileFalse}
\end{figure}

\begin{figure}
    \centering
    \includegraphics[scale=0.23]{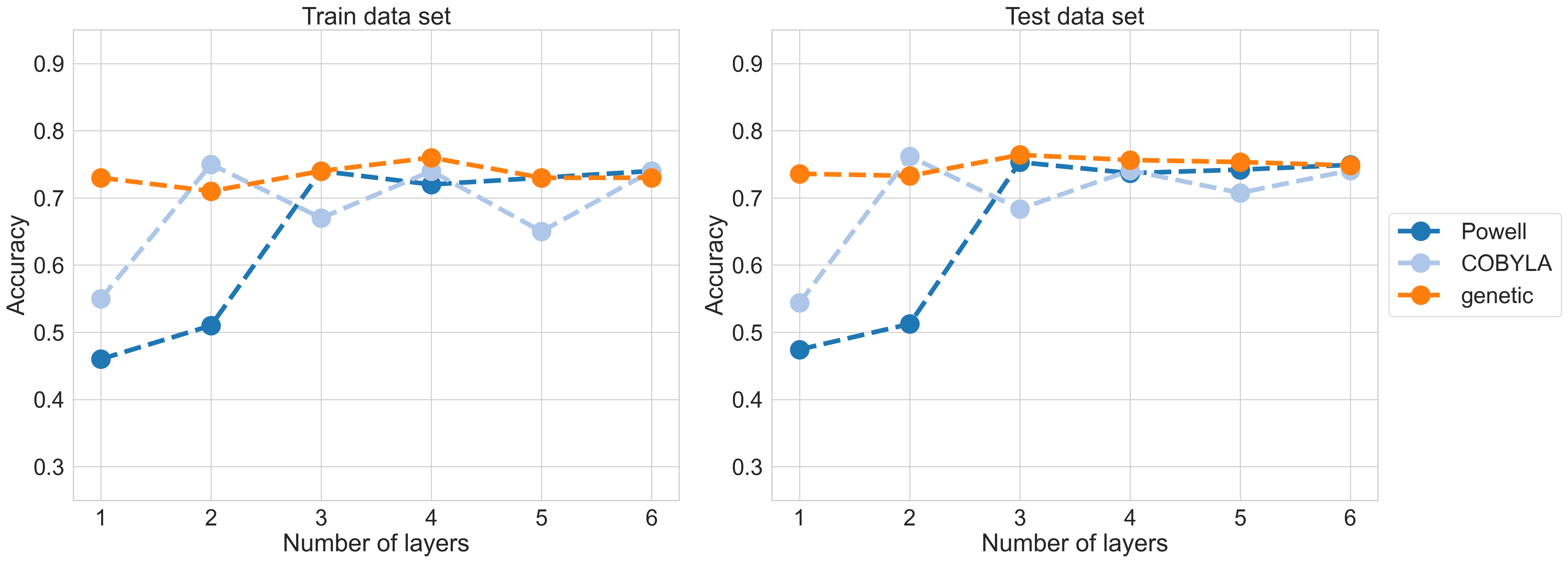}
    \caption{The plots show the accuracy of the train set on the left and the accuracy of the test set on the right as a function of the number of layers and minimizer used.
    The quantum classifier is a 10-qubits circuit for the classification of 32x32 images.
    The data set is the ``without pile-up" data set and the size of the train set is 100.
    The size of the test set is 6000.
    The loss function is ``lce". The number of shots is 2000.}
    \label{fig:noresize_lcescipy_jet_100_pileFalse}
\end{figure}

\begin{figure}
    \centering
    \includegraphics[scale=0.23]{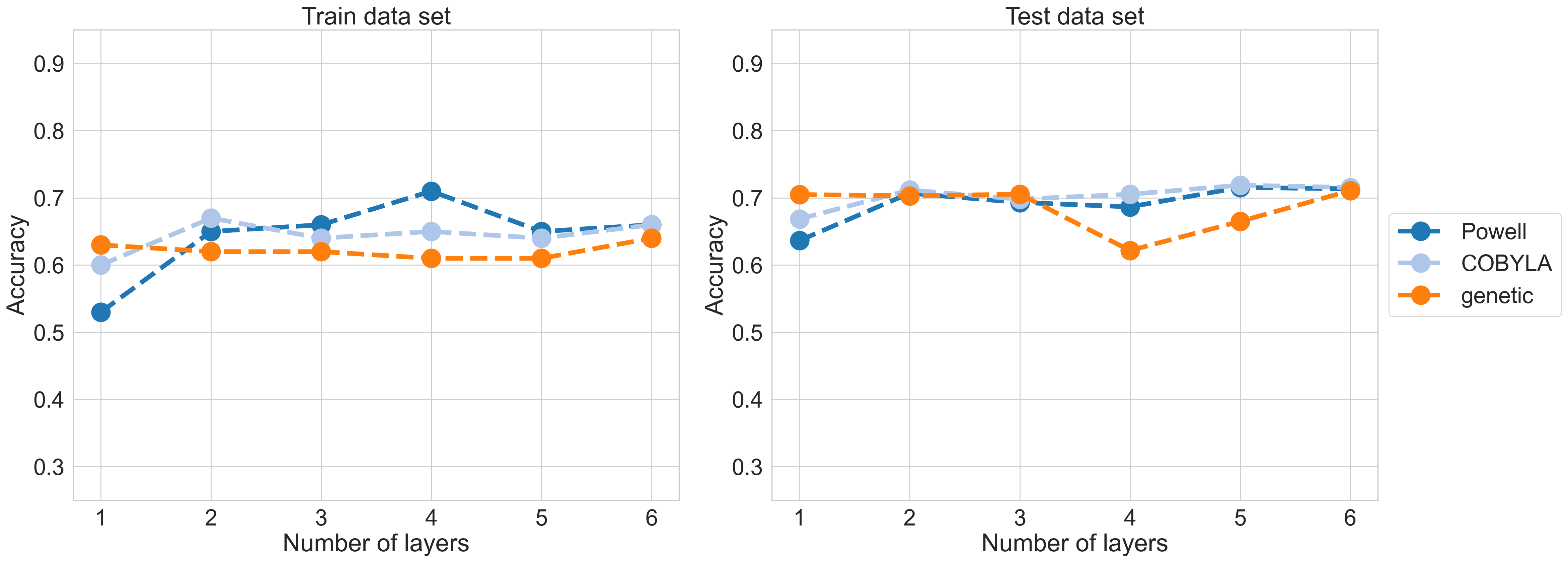}
    \caption{The plots show the accuracy of the train set on the left and the accuracy of the test set on the right as a function of the number of layers and minimizer used.
    The quantum classifier is a 10-qubits circuit for the classification of 32x32 images.
    The data set is ``with pile-up" data set and the size of the train set is 100.
    The size of the test set is 6000.
    The loss function is ``square". The number of shots is 2000.}
    \label{fig:noresize_squarescipy_jet_100_pileTrue}
\end{figure}

\begin{figure}
    \centering
    \includegraphics[scale=0.23]{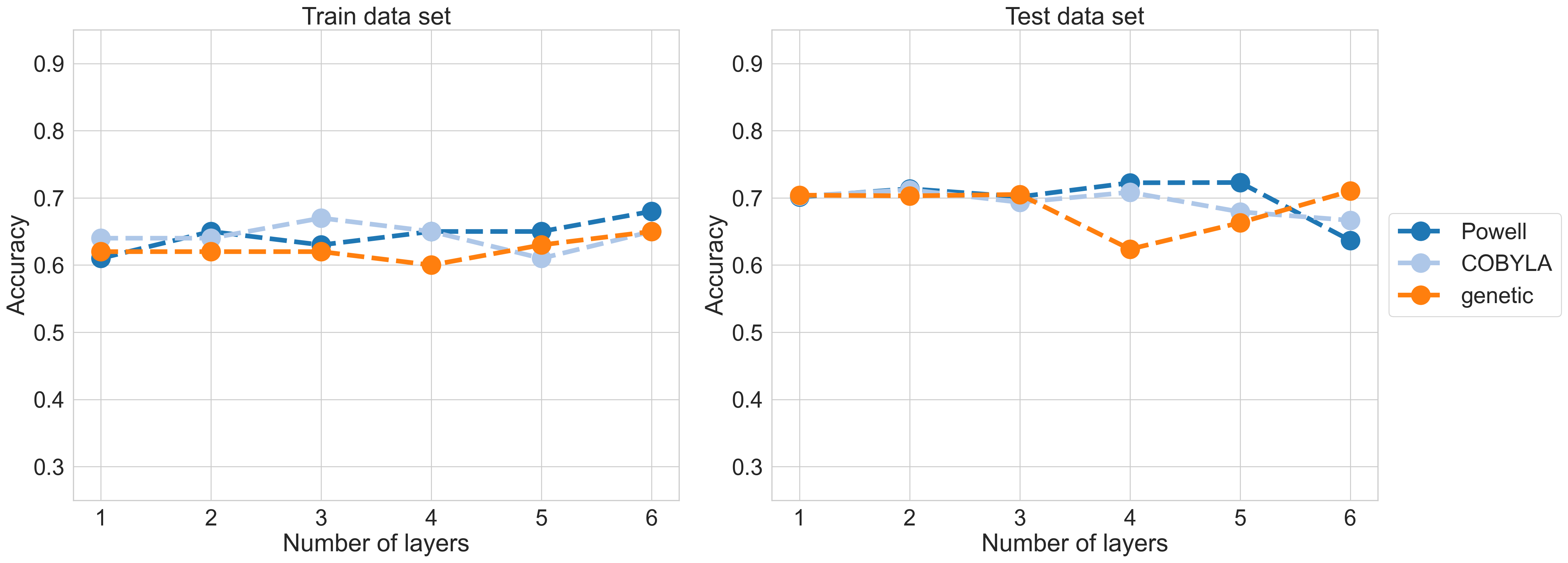}
    \caption{The plots show the accuracy of the train set on the left and the accuracy of the test set on the right as a function of the number of layers and minimizer used.
    The quantum classifier is a 10-qubits circuit for the classification of 32x32 images.
    The data set is ``with pile-up" data set and the size of the train set is 100.
    The size of the test set is 6000.
    The loss function is ``ce". The number of shots is 2000.}
    \label{fig:noresize_cescipy_jet_100_pileTrue}
\end{figure}

\begin{figure}
    \centering
    \includegraphics[scale=0.23]{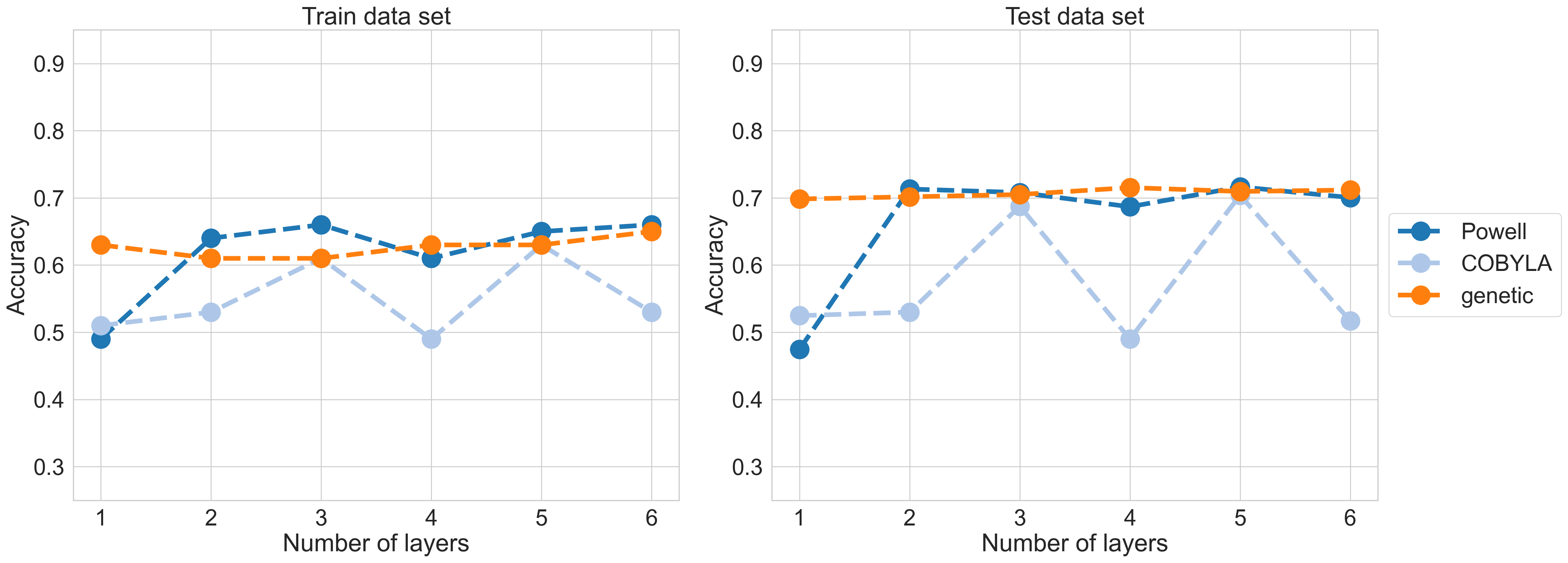}
    \caption{The plots show the accuracy of the train set on the left and the accuracy of the test set on the right as a function of the number of layers and minimizer used.
    The quantum classifier is a 10-qubits circuit for the classification of 32x32 images.
    The data set is ``with pile-up" data set and the size of the train set is 100.
    The size of the test set is 6000.
    The loss function is ``lce". The number of shots is 2000.}
    \label{fig:noresize_lcescipy_jet_100_pileTrue}
\end{figure}

Finally, we increase size of the train set to 100.
Results are shown in Figures~\ref{fig:noresize_squarescipy_jet_100_pileFalse} ~\ref{fig:noresize_cescipy_jet_100_pileFalse} ~\ref{fig:noresize_lcescipy_jet_100_pileFalse} for the ``without pile-up" data set
and in Figures~\ref{fig:noresize_squarescipy_jet_100_pileTrue} ~\ref{fig:noresize_cescipy_jet_100_pileTrue} ~\ref{fig:noresize_lcescipy_jet_100_pileTrue} for the ``with pile-up" data set.







\chapter{Results from training of jet images of size 16x16}
\label{chap:pic_jet_images16}

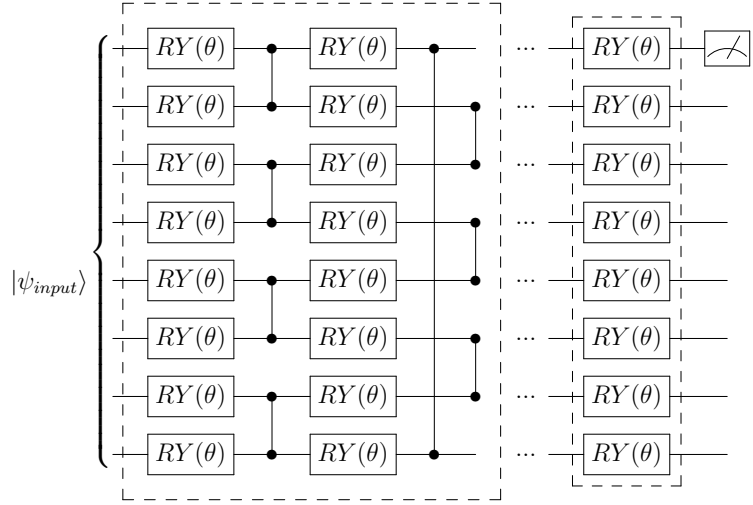
\begin{figure}
    \centering
    \adjustbox{scale=0.80}{
        \subfloat{
            \Qcircuit @C=1.4em @R=.7em {
                & \gate{RY(\theta)} & \ctrl{1} & \gate{RY(\theta)}    & \ctrl{7} & \qw & \rstick{   } & \lstick{...} & \gate{RY(\theta)} & \meter\\
                & \gate{RY(\theta)} & \ctrl{0} & \gate{RY(\theta)}    & \qw & \ctrl{1} & \rstick{   } & \lstick{...} & \gate{RY(\theta)} & \qw\\
                & \gate{RY(\theta)} & \ctrl{1} & \gate{RY(\theta)}    & \qw & \ctrl{0} & \rstick{   } & \lstick{...} & \gate{RY(\theta)} & \qw\\
                & \gate{RY(\theta)} & \ctrl{0} & \gate{RY(\theta)}    & \qw & \ctrl{1} & \rstick{   } & \lstick{...} & \gate{RY(\theta)} & \qw\\
                & \gate{RY(\theta)} & \ctrl{1} & \gate{RY(\theta)} & \qw & \ctrl{0} & \rstick{   } & \lstick{...} & \gate{RY(\theta)} & \qw\\
                & \gate{RY(\theta)} & \ctrl{0} & \gate{RY(\theta)}    & \qw & \ctrl{1} & \rstick{   } & \lstick{...} & \gate{RY(\theta)} & \qw\\
                & \gate{RY(\theta)} & \ctrl{1} & \gate{RY(\theta)}    & \qw  & \ctrl{0} & \rstick{   } & \lstick{...} & \gate{RY(\theta)} & \qw\\
                & \gate{RY(\theta)} & \ctrl{0} & \gate{RY(\theta)}    & \ctrl{0}  & \qw & \rstick{   } & \lstick{...} & \gate{RY(\theta)} & \qw
                \gategroup{1}{2}{8}{6}{2.em}{--}
                \gategroup{1}{9}{8}{9}{.9em}{--}
                \inputgroupv{1}{8}{1em}{9.4em}{|\psi_{input}\rangle \hspace{7mm}}
            }
        }
    }
    \caption{Quantum classifier circuit for classification of 16x16 images.
    The first dashed box is the Ansatz and the second one represents the final rotations.
    Dashed horizontal lines between Ansatz and final rotations indicate the possibility to concatenate multiple Ansätze.
    Rotations RY$(\theta)$ are trained during the minimization process.}
    \label{fig:circuit_jet_images16}
\end{figure}

In this Appendix, we show the accuracies obtained for the train set and the test set with the quantum classifier
by increasing the number of layers, and by changing the minimizer and the loss function.
For the training process, we resize the images to 16x16 from both data sets, with and without pile-up.
The quantum classifier circuit has 8 qubits and it is illustrated in Figure~\ref{fig:circuit_jet_images16}.

\begin{figure}
    \centering
    \includegraphics[scale=0.23]{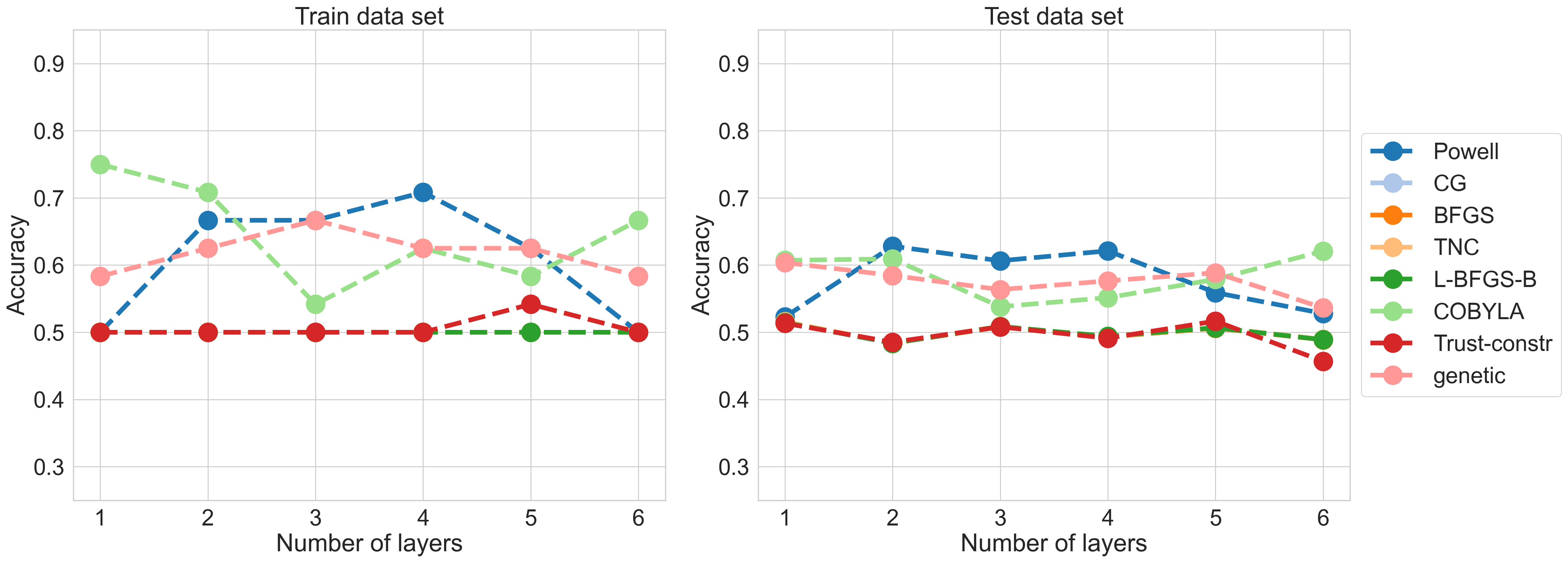}
    \caption{The plots show the accuracy of the train set on the left and the accuracy of the test set on the right as a function of the number of layers and minimizer used.
    The quantum classifier is an 8-qubits circuit for the classification of 16x16 images.
    The data set is the ``without pile-up" data set and the size of the train set is 25.
    The size of the test set is 6000.
    The loss function is ``square". The number of shots is 2000.}
    \label{fig:resize16_squarescipy_jet_25_pileFalse}
\end{figure}

\begin{figure}
    \centering
    \includegraphics[scale=0.23]{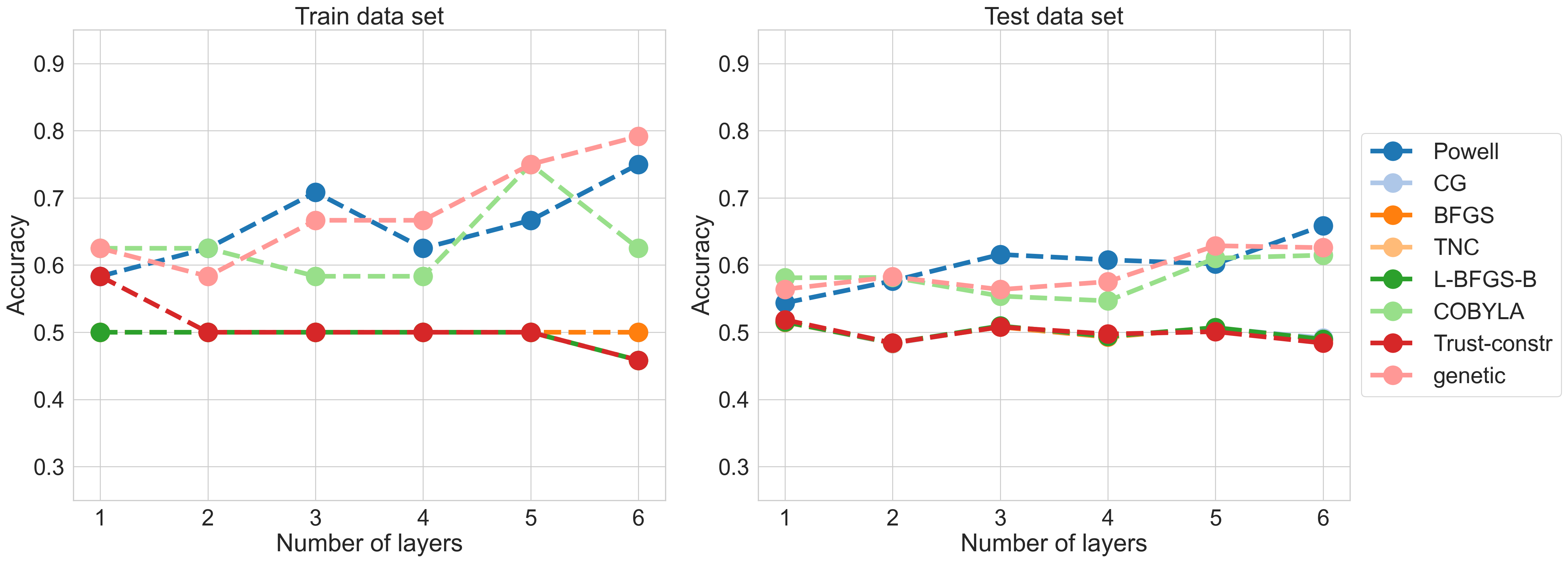}
    \caption{The plots show the accuracy of the train set on the left and the accuracy of the test set on the right as a function of the number of layers and minimizer used.
    The quantum classifier is an 8-qubits circuit for the classification of 16x16 images.
    The data set is the ``without pile-up" data set and the size of the train set is 25.
    The size of the test set is 6000.
    The loss function is ``ce". The number of shots is 2000.}
    \label{fig:resize16_cescipy_jet_25_pileFalse}
\end{figure}

\begin{figure}
    \centering
    \includegraphics[scale=0.23]{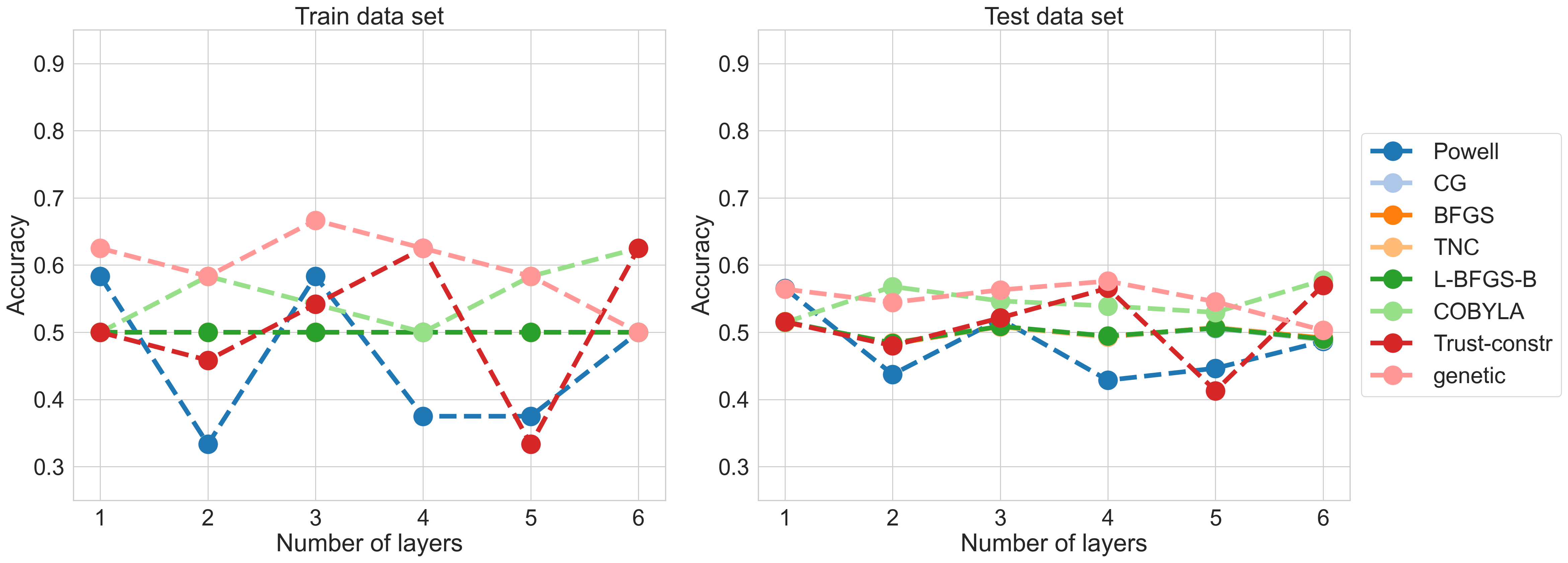}
    \caption{The plots show the accuracy of the train set on the left and the accuracy of the test set on the right as a function of the number of layers and minimizer used.
    The quantum classifier is an 8-qubits circuit for the classification of 16x16 images.
    The data set is the ``without pile-up" data set and the size of the train set is 25.
    The size of the test set is 6000.
    The loss function is ``lce". The number of shots is 2000.}
    \label{fig:resize16_lcescipy_jet_25_pileFalse}
\end{figure}

\begin{figure}
    \centering
    \includegraphics[scale=0.23]{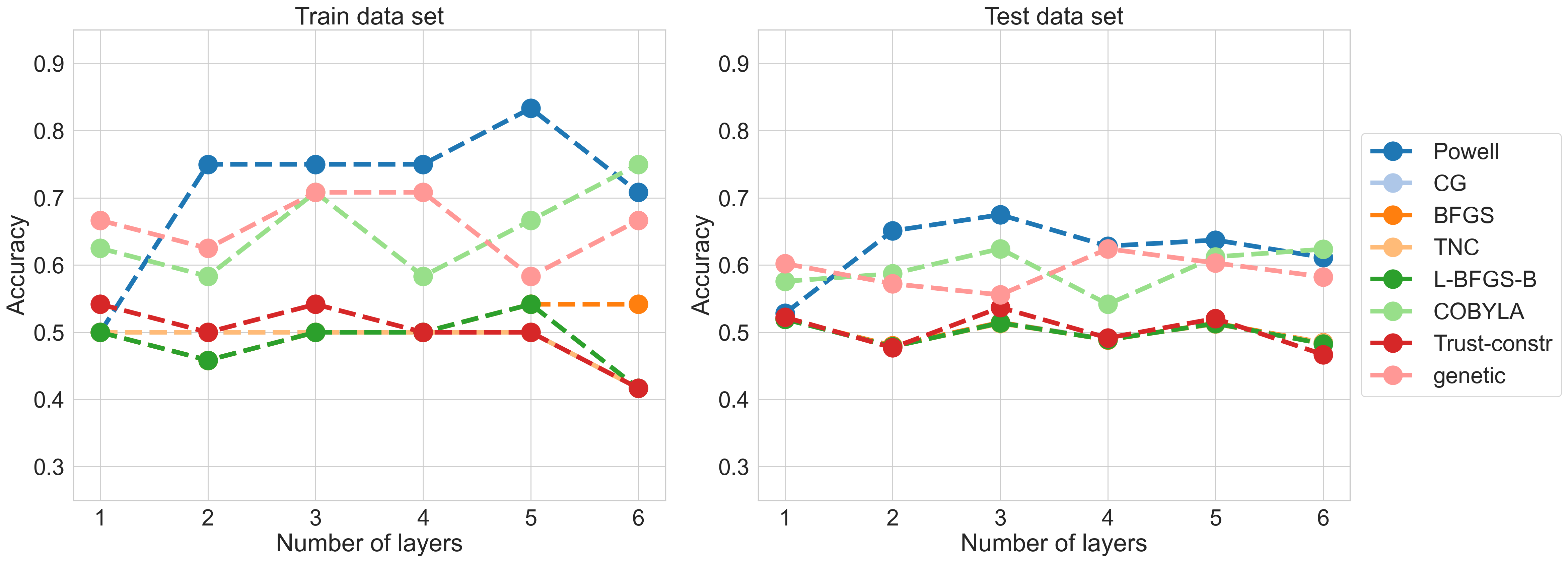}
    \caption{The plots show the accuracy of the train set on the left and the accuracy of the test set on the right as a function of the number of layers and minimizer used.
    The quantum classifier is an 8-qubits circuit for the classification of 16x16 images.
    The data set is the ``with pile-up" data set and the size of the train set is 25.
    The size of the test set is 6000.
    The loss function is ``square". The number of shots is 2000.}
    \label{fig:resize16_squarescipy_jet_25_pileTrue}
\end{figure}

\begin{figure}
    \centering
    \includegraphics[scale=0.23]{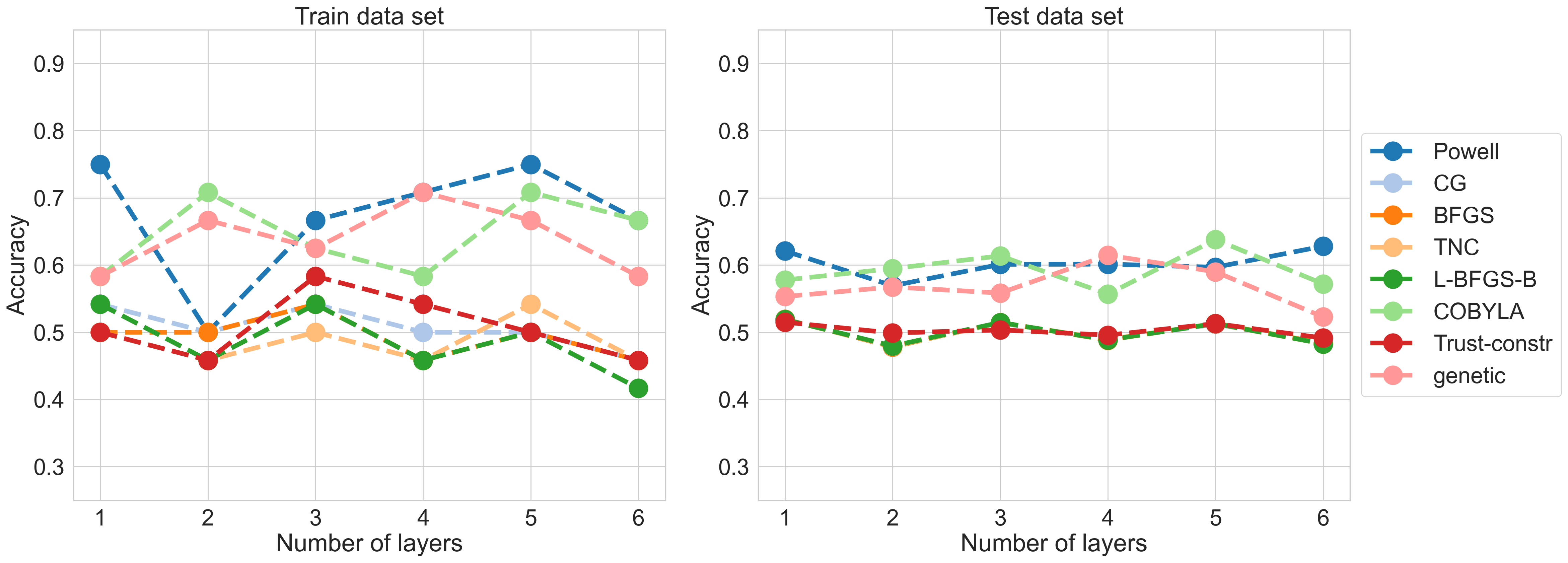}
    \caption{The plots show the accuracy of the train set on the left and the accuracy of the test set on the right as a function of the number of layers and minimizer used.
    The quantum classifier is an 8-qubits circuit for the classification of 16x16 images.
    The data set is the ``with pile-up" data set and the size of the train set is 25.
    The size of the test set is 6000.
    The loss function is ``ce". The number of shots is 2000.}
    \label{fig:resize16_cescipy_jet_25_pileTrue}
\end{figure}

\begin{figure}
    \centering
    \includegraphics[scale=0.23]{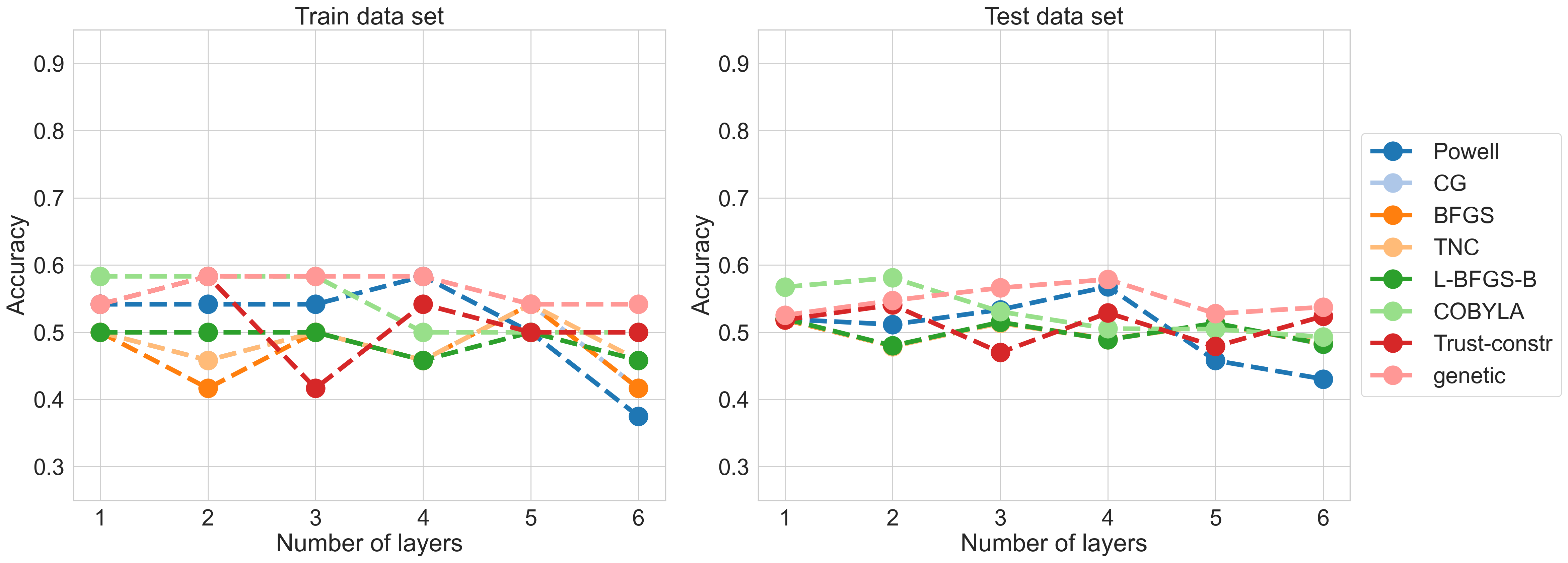}
    \caption{The plots show the accuracy of the train set on the left and the accuracy of the test set on the right as a function of the number of layers and minimizer used.
    The quantum classifier is an 8-qubits circuit for the classification of 16x16 images.
    The data set is the ``with pile-up" data set and the size of the train set is 25.
    The size of the test set is 6000.
    The loss function is ``lce". The number of shots is 2000.}
    \label{fig:resize16_lcescipy_jet_25_pileTrue}
\end{figure}

We use for the train set a group of 25 images from the ``without pile-up" data set.
Figure~\ref{fig:resize16_squarescipy_jet_25_pileFalse} show how the accuracy changes by increasing the number of layers for different minimizers with ``square" loss function.
Figures~\ref{fig:resize16_cescipy_jet_25_pileFalse} ~\ref{fig:resize16_lcescipy_jet_25_pileFalse} show the same for ``ce" and ``lce" loss functions.
We show the same plots for the ``with pile-up" data set in Figures~\ref{fig:resize16_squarescipy_jet_25_pileTrue} ~\ref{fig:resize16_cescipy_jet_25_pileTrue} ~\ref{fig:resize16_lcescipy_jet_25_pileTrue}.

\chapter{Results from training of jet features}
\label{chap:pic_jet_features}

In this Appendix, we show the accuracies obtained for the train set and the test set with the quantum classifier
by increasing the number of layers, and by changing the minimizer and the loss function.
For the training process, we use the features of jets from both data sets, with and without pile-up.
The quantum classifier circuit has 6 qubits and it is illustrated in Figure~\ref{fig:circuit_jet_features}.

\begin{figure}
    \centering
    \includegraphics[scale=0.23]{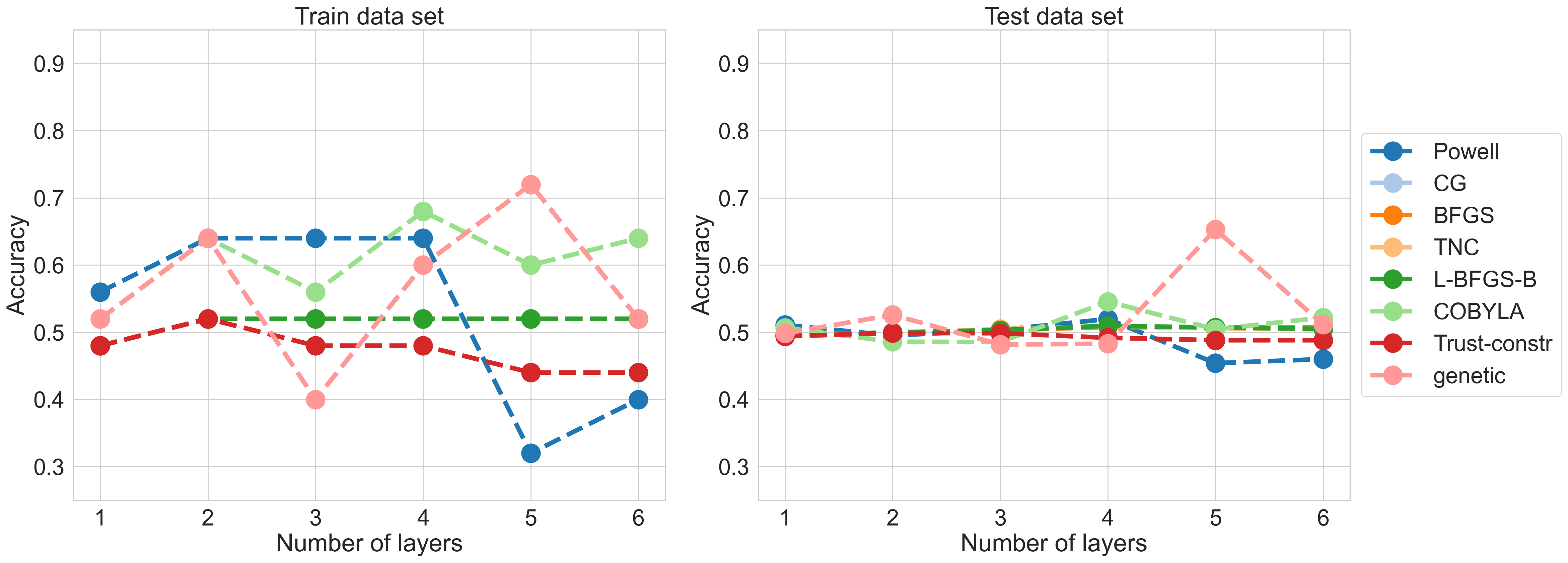}
    \caption{The plots show the accuracy of the train set on the left and the accuracy of the test set on the right as a function of the number of layers and minimizer used.
    The quantum classifier is a 6-qubits circuit for the classification of small arrays of data.
    The data set is the ``without pile-up" data set and the size of the train set is 25.
    The size of the test set is 6000.
    The loss function is ``square". The number of shots is 2000.}
    \label{fig:Fnoresize_squarescipy_jet_25_pileFalse}
\end{figure}

\begin{figure}
    \centering
    \includegraphics[scale=0.23]{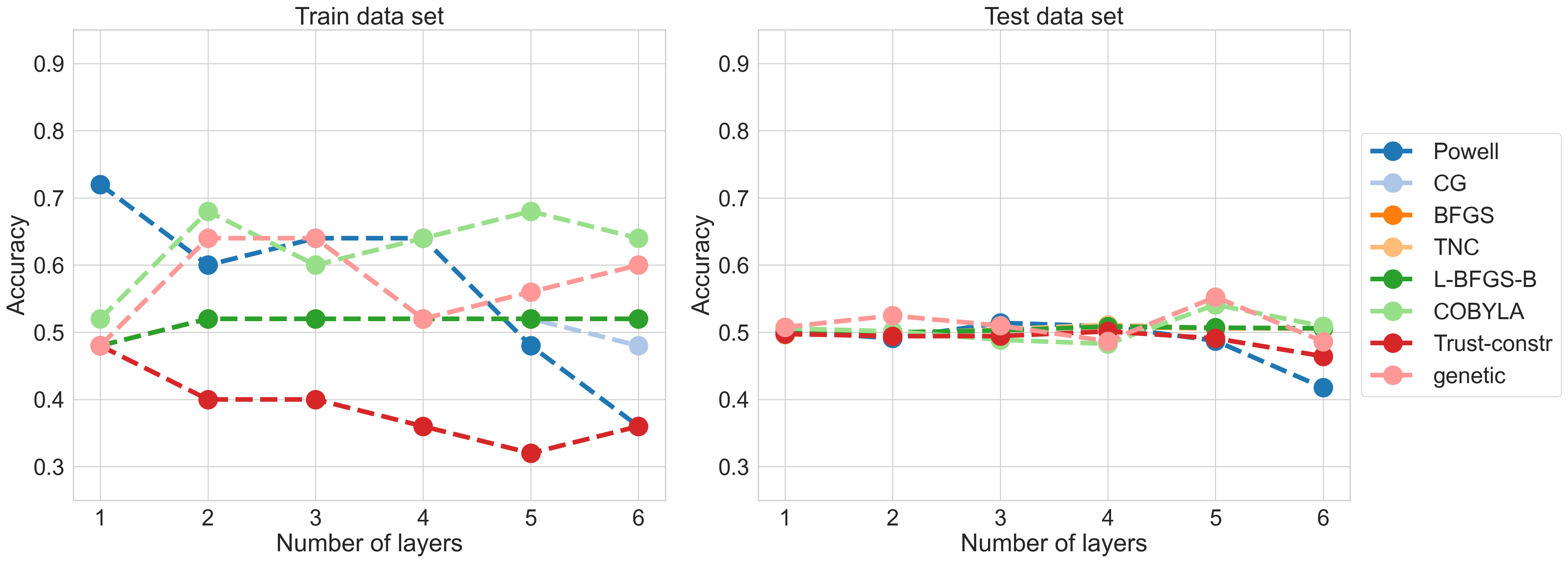}
    \caption{The plots show the accuracy of the train set on the left and the accuracy of the test set on the right as a function of the number of layers and minimizer used.
    The quantum classifier is a 6-qubits circuit for the classification of small arrays of data.
    The data set is the ``without pile-up" data set and the size of the train set is 25.
    The size of the test set is 6000.
    The loss function is ``ce". The number of shots is 2000.}
    \label{fig:Fnoresize_cescipy_jet_25_pileFalse}
\end{figure}

\begin{figure}
    \centering
    \includegraphics[scale=0.23]{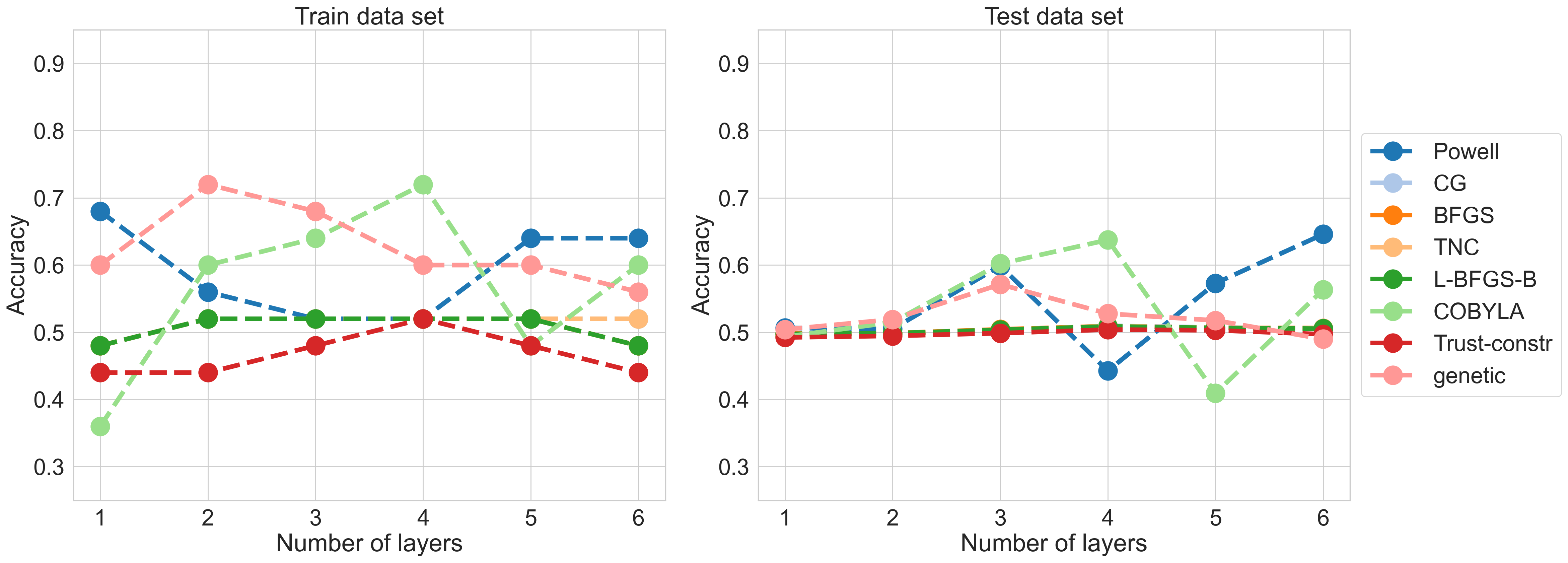}
    \caption{The plots show the accuracy of the train set on the left and the accuracy of the test set on the right as a function of the number of layers and minimizer used.
    The quantum classifier is a 6-qubits circuit for the classification of small arrays of data.
    The data set is the ``without pile-up" data set and the size of the train set is 25.
    The size of the test set is 6000.
    The loss function is ``lce". The number of shots is 2000.}
    \label{fig:Fnoresize_lcescipy_jet_25_pileFalse}
\end{figure}

\begin{figure}
    \centering
    \includegraphics[scale=0.23]{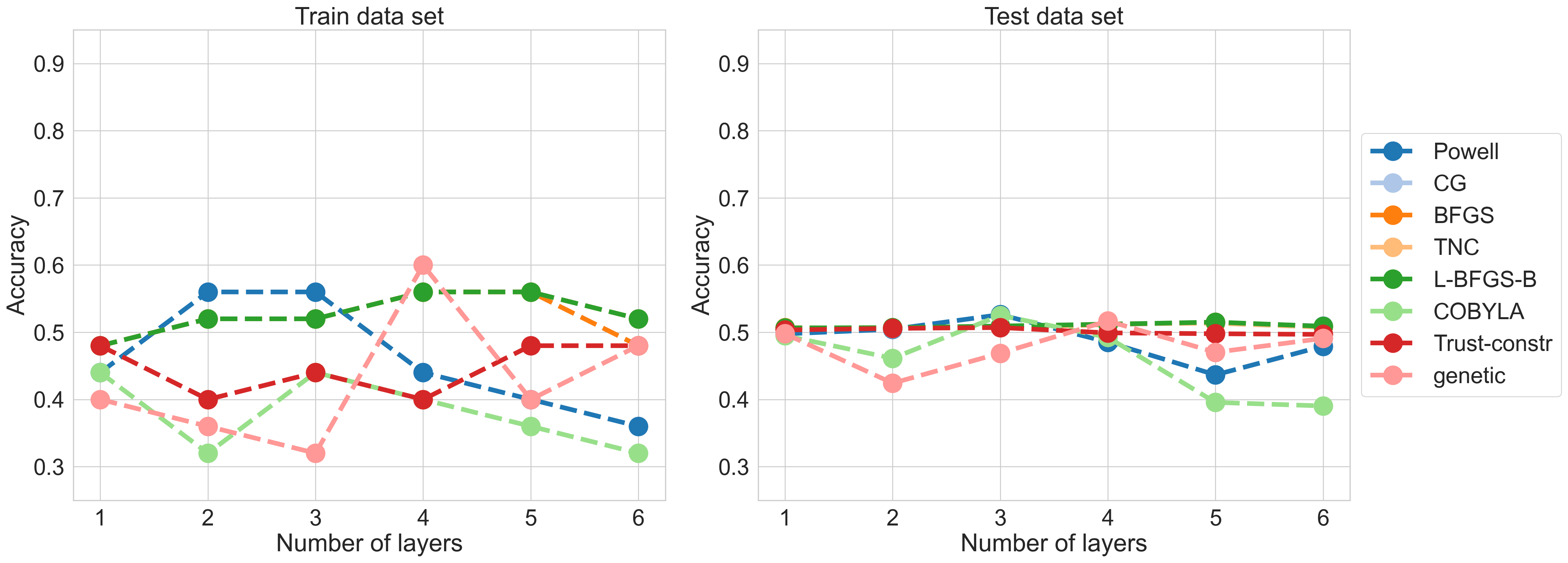}
    \caption{The plots show the accuracy of the train set on the left and the accuracy of the test set on the right as a function of the number of layers and minimizer used.
    The quantum classifier is a 6-qubits circuit for the classification of small arrays of data.
    The data set is the ``with pile-up" data set and the size of the train set is 25.
    The size of the test set is 6000.
    The loss function is ``square". The number of shots is 2000.}
    \label{fig:Fnoresize_squarescipy_jet_25_pileTrue}
\end{figure}

\begin{figure}
    \centering
    \includegraphics[scale=0.23]{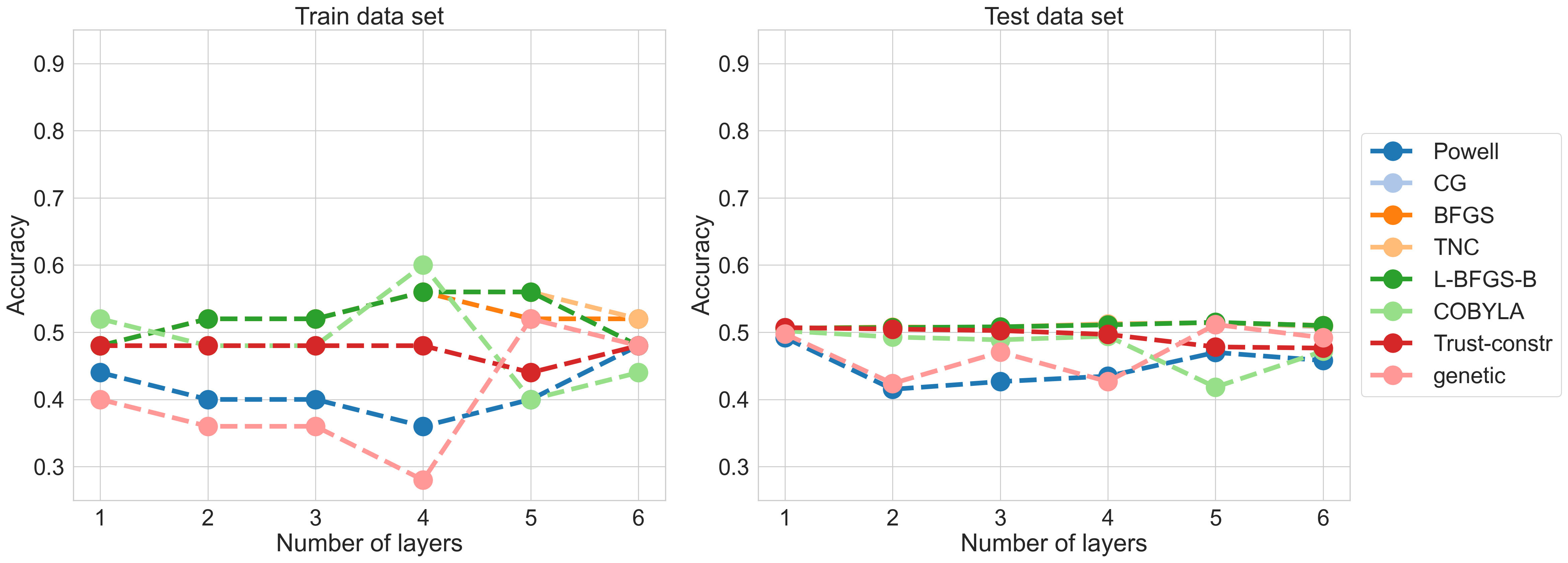}
    \caption{The plots show the accuracy of the train set on the left and the accuracy of the test set on the right as a function of the number of layers and minimizer used.
    The quantum classifier is a 6-qubits circuit for the classification of small arrays of data.
    The data set is the ``with pile-up" data set and the size of the train set is 25.
    The size of the test set is 6000.
    The loss function is ``ce". The number of shots is 2000.}
    \label{fig:Fnoresize_cescipy_jet_25_pileTrue}
\end{figure}

\begin{figure}
    \centering
    \includegraphics[scale=0.23]{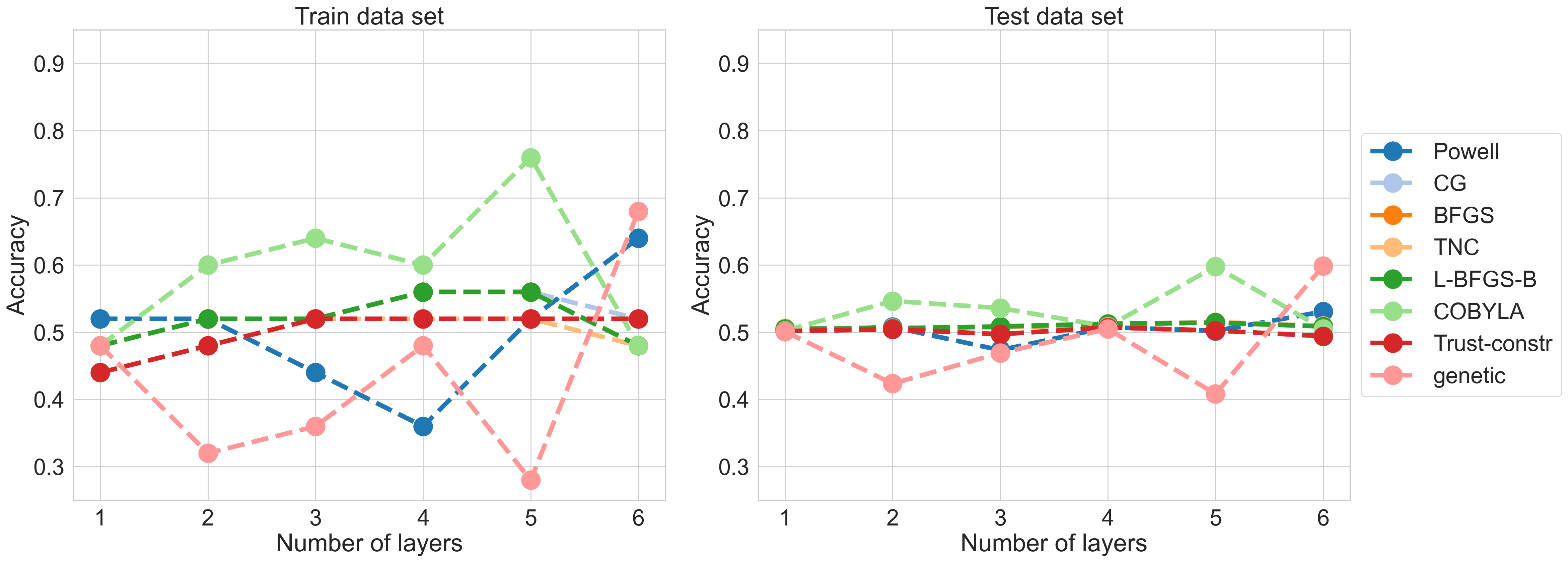}
    \caption{The plots show the accuracy of the train set on the left and the accuracy of the test set on the right as a function of the number of layers and minimizer used.
    The quantum classifier is a 6-qubits circuit for the classification of small arrays of data.
    The data set is the ``with pile-up" data set and the size of the train set is 25.
    The size of the test set is 6000.
    The loss function is ``lce". The number of shots is 2000.}
    \label{fig:Fnoresize_lcescipy_jet_25_pileTrue}
\end{figure}

We use for the train set a group of 25 elements from the ``without pile-up" data set.
Figure~\ref{fig:Fnoresize_squarescipy_jet_25_pileFalse} show how the accuracy changes by increasing the number of layers for different minimizers with ``square" loss function.
Figures~\ref{fig:Fnoresize_cescipy_jet_25_pileFalse} ~\ref{fig:Fnoresize_lcescipy_jet_25_pileFalse} show the same for ``ce" and ``lce" loss functions.
We show the same plots for the ``with pile-up" data set in Figures~\ref{fig:Fnoresize_squarescipy_jet_25_pileTrue} ~\ref{fig:Fnoresize_cescipy_jet_25_pileTrue} ~\ref{fig:Fnoresize_lcescipy_jet_25_pileTrue}.

\printbibliography[heading=bibintoc, title={Bibliography}]

\end{document}